\documentclass[twocolumn]{aa}

\usepackage{graphicx}

\usepackage{txfonts}
\usepackage{longtable}
\usepackage{color}

\begin{document}

   \title{Far-UV to mid-IR properties of nearby radio galaxies}


   \author{H. R. de Ruiter,
          \inst{}
        P. Parma\inst{},
        R. Fanti \inst{}
          \and
          C. Fanti\inst{}
          }

   \institute{INAF Istituto di  Radioastronomia (IRA),  Via Gobetti, 101, I-40129 Bologna\\
              \email{h.deruiter@ira.inaf.it, parma@ira.inaf.it, r.fanti@ira.inaf.it, c.fanti@ira.inaf.it}
             }

   \date{Received; accepted}

 
  \abstract
   {} 
   {
We investigate whether the far-UV continuum of nearby radio galaxies is due solely to the parent galaxy that passively evolves, or
if it reveals evidence for the presence of other star-forming or non-stellar components. If the UV excess is due to an additional radiation component, we compare this with other properties such as radio power, optical spectral type (e.g. high- and low-excitation galaxies), and the strength of the emission
lines. We also discuss the possible correlation between the ultra-violet flux, IR properties, and  the central black hole mass. 
}
   {
We used a sample of low-luminosity B2 radio galaxies and a small sample of higher luminosity 3C radio galaxies at comparable redshift
($z < 0.2$). Spectral energy distributions (SEDs) were constructed using a number of on-line databases that are freely available now: GALEX, SDSS, 2MASS, and
WISE.
These were compared with model SEDs of early-type galaxies with passively evolving stellar populations at various ages (typically $0.5-1.3\times 10^9$ years).
We established whether a second component was needed to obtain a satisfactory fit with the observed overall SED. 
We introduce the parameter $XUV$ , which measures the excess slope of the UV continuum between 4500 and 2000 \AA\  with respect to
the UV radiation produced by the underlying old galaxy component.
}
   {We find that the UV excess as measured by $XUV$ is usually small or absent in low-luminosity (FR I) sources, but sets in abruptly at the transition radio power,
above which  we find mostly FRII sources. $XUV$ behaves very similarly  to the strength of the
optical emission lines (in particular $H\alpha$). Below $P_{1.4 GHz} < 10^{24}$ WHz$^{-1}$ $XUV$ is close to zero. $XUV$ correlates
strongly with the $H\alpha$ line strength, but only in sources with strong $H\alpha$ emission.
We discuss whether the line emission might be due to photoionization by radiation from the parent galaxy, possibly with additional star formation, or 
if it requires the presence of a non-stellar AGN component. $XUV$ and the slope of the mid-IR are strongly correlated, as measured by the WISE bands in the interval 3.4 to 22 $\mu$m, in the sense that
sources with a strong UV excess also have stronger IR emission.
There is an inverse correlation between $XUV$ and central black hole mass: the $M_{BH}$  of objects with strong UV\ excess is on average two to three
times less massive than that of objects
without UV excess. Low-luminosity radio galaxies tend to be more massive and contain more massive black holes.

}
   {}

\authorrunning{de Ruiter et al.}
\titlerunning{}
   \maketitle
%

\section{Introduction}
\label{introduction}

Ever since the classical paper of Fanaroff \& Riley (\cite{fanaroff74}), we know that radio galaxies can be divided into two different types 
according to whether they are of low or high radio luminosity, the dividing line being at approximately $10^{24.5}$ WHz$^{-1}$ at the frequency 
1.4 GHz, but weakly dependent on the absolute magnitude of the parent galaxy (Ledlow \& Owen \cite{ledlow96}; see also Ghisellini \& Celotti 
\cite{ghisellini01}). 
This difference manifests especially in the morphology (which was the basis of the distinction of the two FR classes), 
the low-luminosity sources have a relaxed overall radio structure without hot spots at the end of the radio jets or lobes, for example, 
while this feature is usually present and even characteristic in high-luminosity sources. 
The reason for the FR division was unknown, but there is now growing evidence that both the fuelling mechanism and the merging history may
play significant roles (Evans, Hardcastle \& Croston \cite{evans08}; Hardcastle, Evans \& Croston \cite{hardcastle07}; Saripalli \cite{saripalli12}).  

High-luminosity radio sources have a significant influence on the parent radio galaxy. The radio jets plough their way through the 
interstellar medium, which may be excited or even ionized by them, and indeed the bulk kinetic energy of the jets correlates strongly with the 
luminosity of the narrow emission lines (Rawlings \& Saunders \cite{rawlings89}).   
The nuclear activity itself may also be at the origin of the emission spectrum, consisting of both lines and a continuum. 
It is much less clear if there is significant interaction (if at all) between a low-luminosity radio source and its parent galaxy.

This strong dichotomy in the optical properties of radio galaxies approximately follows the division into the two Fanaroff-Riley
(FR) classes  and is 
also related to the accretion mechanism (see for a general discussion Best \& Heckman \cite{best12}).
Emission lines in radio galaxies  have been studied by a number of authors (e.g. Zirbel \& Baum \cite{zirbel95}; for some more recent articles
 see Buttiglione et al. \cite{buttiglione09a}, 
Buttiglione et al. \cite{buttiglione09b}, Baldi \& Capetti \cite{baldi09},
and Buttiglione et al. \cite{buttiglione10}).
It has been known for a long time that strong emission lines occur
in the more powerful radio galaxies, while the weaker FRI galaxies tend to have emissionless spectra identical to those of the ordinary
early-type galaxies.
Different accretion rates for FRI and FRII sources were proposed by Ghisellini \& Celotti (\cite{ghisellini01}); they
noted that high-excitation radio galaxies (HERGs) typically accrete at about 1-10 \% of the Eddington rate, while low-excitation radio galaxies 
(LERGs) accrete at much lower rates ($\ll 1$ \%). 
At high radio luminosities, HERGs are dominant, although both types can occur at all radio luminosities  
(Best \& Heckman (\cite{best12}).
It appears that the high-excitation galaxies have on average a lower stellar mass and lower central black hole mass than the more massive galaxies 
that are associated with the low-excitation mode, while the latter are frequently associated with hot X-ray haloes.  

Allen et al. (\cite{allen06}) have suggested that accretion in LERGs in the centre of clusters may proceed directly from the hot phase 
of the intra-cluster medium. This view was supported by Hardcastle, Evans \& Croston (\cite{hardcastle07}), who suggested that LERGs 
may be powered by the hot 
X-ray emitting gas, while HERGs also need a cold gas component. However, very often circumnuclear dusty disks  can be clearly seen in 
HST images of low-luminosity B2 and nearby 3C radio galaxies (Capetti et al. \cite{capetti00}; de Ruiter et al. \cite{deruiter02}; 
Verdoes-Kleijn et al. \cite{verdoes99}; de Koff et al. \cite{dekoff00}). 
Moreover, CO has been detected in many nearby FRI radio galaxies (Prandoni et al. \cite{prandoni07}, \cite{prandoni10}; 
Lim et al. \cite{lim03}; Leon et al. \cite{leon03}; Oca{\~n}a Flaquer et al. \cite{oca10}). 
The molecular gas often appears to be in dusty disks around the nucleus.
According to de Koff et al. (\cite{dekoff00}), the dust masses found in FRI galaxies tend to be higher than those seen in
a sample of ordinary (i.e. optically selected) elliptical galaxies studied by van Dokkum \& Franx (\cite{vandok95}). 

Recent technological developments have led to the creation of large-scale optical surveys with correspondingly large data bases, 
extending from the infrared to the far-ultraviolet. This makes it possible to study the overall spectral energy distributions and  
spectral lines of many well-known sets of objects. In particular, the GALEX survey provides photometric data in the hitherto 
inaccessible far-UV 
part of the spectrum, enabling us to study the behaviour of FRI and FRII radio galaxies in that 
range of the spectrum. This is our topic here: we analyse the optical spectral energy distributions of parts of the B2 and 3C samples 
of radio galaxies and try to shed some more light on the interrelation between radio and optical properties, by fitting the spectral
energy distribution (SED; in the range $\sim 1500$\AA\ to $\sim 2\mu$) with synthetic galaxy models (Bruzual \& Charlot \cite{bruzual03}) and 
analysing the deviations in the fit.

We also compare the UV  and IR properties; in particular, we investigate the mid-IR using data collected from the 
WISE mission (Wright et al. \cite{wright10}) and in a limited number of cases also the FIR. The relevant literature for mid- and far-IR data 
is briefly discussed in Sect. \ref{data}, in which
 we describe how the optical photometry and spectroscopy were collected, 
using the on-line data bases of  GALEX,
SDSS, and 2MASS and the galaxy models of Bruzual \& Charlot (\cite{bruzual03}) to check whether any more components 
are necessary, other than some model of an early-type galaxy. We present the results in Sect. \ref{results}, where we also 
investigate the (optical) spectral line properties of the galaxies and their possible relation to the radio properties.

Finally, in Sect. \ref{discussion} we discuss the results obtained in this paper and list our conclusions.

Unless explicitly stated, all intrinsic parameters were calculated using a concordance cosmology 
with $\Omega_\Lambda = 0.7$ and $H_0 = 70$ km~s$^ {-1}$Mpc$^{-1}$.

\section{Data selection and galaxy model fitting}
\label{data}

As the basis for our analysis, we selected two samples of low-redshift ($z<0.2$) radio galaxies in the area covered by the SDSS: 
first,
B2 radio galaxies (Fanti et al. \cite{fanti87}; de Ruiter et al. \cite{deruiter90}). They are mostly low-luminosity FR I sources identified with early-type galaxies. 
Second, a sample extracted from the 3CR catalogue (Laing, Riley \& Longair \cite{laing83}). There is some overlap between the B2 and 3C samples:
the sources in common are 3C 293 (B2 1350+31), 3C 310 (B2 1502+26), 3C 315 (B2 1511+26), 3C 332 (B2 1615+32), and 3C 357 (B2 1726+32). 
The majority of the 3C sources have a higher radio luminosity and are mostly FR II. A number of the 3C sources have been observed in the ultraviolet 
as part of an HST snapshot survey of 3CR radio galaxies (Allen et al. \cite{allen02}).

Thanks to the impressive developments in optical surveys, it is now possible to obtain, for many objects at least, accurate photometric data 
ranging from the far-UV to the near-IR. Thus it is quite easy now to compare the photometry with model SEDs and 
investigate whether, apart from a "normal" SED, any signs of other components are present that might be linked to the radio activity.
Fanti et al. (\cite{fanti11}) performed a  similar study to the one presented here, in their case of compact steep spectrum (CSS) radio sources. 
We basically follow their method. 

Although information on the photometric parameters can be recovered directly from the respective GALEX, SDSS, 2MASS and WISE
sites\footnote{URLs are given in the acknowledgments.} , we summarize them here in Table \ref{tab:photpar} for easy reference. 
The SDSS does not cover the whole northern sky yet, which is why only  68 (of which 6 are FR II) B2 sources (out of the 101 in the entire sample) can
be used here, while the 3C is restricted to 14 sources, of which 8 are FR II. Virtually all of these sources have infrared data. However, not all have UV data, 
either because they are too faint or because no GALEX data are available. Nevertheless, the remaining sources constitute a sample that is 
large enough to allow a more detailed statistical investigation. The selection is random,  and the final sample used here should therefore be unbiased.

 \begin{table}
      \caption[]{Photometric parameters}
         \label{tab:photpar}
     $$
         \begin{tabular}{lrrrr}
            \hline
            \noalign{\smallskip}
           \noalign{\smallskip}
            \hline
Survey & band & $\lambda_{eff}$ & A/E(B-V) & m=0 flux density \\
            &  mag  & \AA ($\mu$m) & mag & erg~cm$^{-2}$s$^{-1}\AA^{-1}$ \\
        \hline
           \noalign{\smallskip}
GALEX & FUV & 1528\rlap{$^{\mathrm{a}}$} & 8.240\rlap{$^{\mathrm{b}}$} & $4.666\times 10^{-8}$ \\
GALEX & NUV & 2271\rlap{$^{\mathrm{a}}$} & 8.200\rlap{$^{\mathrm{b}}$} & $2.112\times 10^{-8}$ \\
SDSS & $u$ & 3553 & 5.155 & $8.269\times 10^{-9}$ \\
SDSS & $g$ & 4686 & 3.793 & $4.961\times 10^{-9}$ \\
SDSS & $r$ & 6165 & 2.751 & $2.866\times 10^{-9}$ \\
SDSS & $i$ & 7481 & 2.086 & $1.946\times 10^{-9}$ \\
SDSS & $z$ & 8931 & 1.479 & $1.366\times 10^{-9}$ \\
2MASS & J & 1.25 $\mu$m & - & $3.078\times 10^{-10}$ \\
2MASS & H & 1.65 $\mu$m & - & $1.185\times 10^{-10}$ \\
2MASS & K & 2.15 $\mu$m & - & $4.329\times 10^{-11}$ \\
WISE & 1 & 3.44 $\mu$m & - & $7.96\times 10^{-12}$ \\
WISE & 2 & 4.60 $\mu$m & - & $2.40\times 10^{-12}$ \\
WISE & 3 & 12.00 $\mu$m & - & $6.10\times 10^{-14}$ \\
WISE & 4 & 22.00 $\mu$m & - & $5.10\times 10^{-14}$ \\

 \noalign{\smallskip}
            \hline
 
         \end{tabular}
     $$ 
\begin{list}{}{}
\item[$^{\mathrm{a}}$] Budav{\'a}ri et al. (\cite{budavari09}) reported slightly different values for the central wavelengths of both FUV and NUV, 
but the difference is so small as to be negligible.
\item[$^{\mathrm{b}}$] Extinction coefficients from Wyder et al. (\cite{wyder07}).
\end{list}
   \end{table}

The photometric data were converted into luminosity using the redshift of the sources (for the B2 sources see, e.g., Fanti et al. \cite{fanti87}; 
de Ruiter et al. \cite{deruiter90}) and following the various recipes given in the on-line  web sites of the SDSS, GALEX, and 2MASS surveys. 
More details on the SDSS seventh data release (used here) can be found in  Abazijian et al. (\cite{abazijian09}). 
The GALEX mission is described in  Martin et al. (\cite{martin05}); we took the UV extinction coefficients from Wyder et al. (\cite{wyder07}). 
For an object to be included in our analysis of the UV properties, we obviously required that it is detected in at least one of the two UV bands 
of GALEX, but even if GALEX data were not available, we could in a few cases use such objects in some of the correlations, 
for instance, radio power against spectral properties.

Near-IR data were extracted from the 2MASS survey (see Skrutskie et al. \cite{skrutskie06}); 
these exist for virtually all objects in our samples.
In a minority we noticed that the 2MASS values lie systematically slightly higher (or even much higher in the case of 3C 274) than expected on the 
basis of an extrapolation of the SDSS magnitudes 
(and in one case, B2\ 1502+26, alias 3C\ 310, much lower), although the {\it \textup{slope}} of the SED is consistent with the extrapolation. 
This is expected because the galaxies are usually quite extended, and the algorithms that are used to derive the magnitudes in different surveys may 
sometimes lead to small systematic offsets. We stress, however, that the overall consistency of the photometric data from the UV to the IR is 
quite satisfactory.

Additional mid- and far-IR data were collected for a number of sources. These cover the range 3-200 $\mu$m, but 
neither IR data collected from the WISE mission nor other mid- to far-IR data were used in fitting galaxy models. The reason for this is that  other
components start to dominate the SED above $\sim5 \mu$m, which
is different from the stellar population. The MIR-FIR range can be easily explained by emission of dust. We applied some 
plausible models based on Hildebrand (\cite{hildebrand83}) and Knapp, Bies, \& van Gorkom (\cite{knapp90}). We discuss the IR in more detail below.

For almost all B2 and 3C sources considered here, data collected by the WISE mission are available (see Wright et al. 
\cite{wright10}) at 3.4, 4.6, 12 and 22 $\mu$m. The WISE mission was optimized for point sources, and it is known that there may be a 
problem with fluxes of extended objects (see Jarrett et al. \cite{jarrett13}). As discussed below, a few galaxies at low 
redshift show systematically lower fluxes than in the other IR and visual bands. However, this only occurs in a small minority of cases,
and in general the WISE data are quite useful. 
We have collected the available WISE data in Table \ref{tab:wise}.
\addtocounter{table}{1}
The references for other IR data, typically in the 20-200 $\mu$m, are a) Ogle et al. (\cite{ogle06}), b) Dicken et al. (\cite{dicken10}), 
c) Lilly et al. (\cite{lilly85}),
d) Shi et al. (\cite{shi05}), e) Maiolino et al. (\cite{maiolino95}), f) Hardcastle et al. (\cite{hardcastle09}), g) Golombek et al. 
(\cite{golombek88}), h) Frogel et al. (\cite{frogel75}), i) Tang et al. (\cite{tang09}), j) Temi et al. (\cite{temi09}), k) Moshir et al. 
(\cite{moshir90}), l) Siebenmorgen et al. (\cite{siebenmorgen04}), m)Deo et al. (\cite{deo09}), n) Sargsyan et al. (\cite{sargsyan11}),
o) Elvis et al. (\cite{elvis84}), p) Xilouris et al. (\cite{xilouris04}), q) Temi et al. (\cite{temi04}), and r) Willmer et al. (\cite{willmer09}).
Sources for which these additional IR data are available are 3C 192 (a, b), 3C 219 (a, c, d), 3C 234 (a, c, k, m, n, o), 
3C 264 (e, f, g,), 3C 272.1 (h, i, j, k), 3C 285 (b, k), 3C 296 (l), 3C 305 (b, g, l), 3C 321 (d, f, k, l, m, n),   B2 0648+27 (g), B2 1122+39 (g,k),
B2 1217+29 (i, j, k, p, q), B2 1350+31 (b, f, g, k, l), B2 1615+32 (d, k), and B2 1621+38 (r).

Although these data are very heterogenous and taken from various instruments (e.g. IRAS, ISO, and Spitzer), they appear to give
a very good idea of the extension of the SED up to 100 or 200 $\mu$m.

We list in Table \ref{tab:phot} the GALEX, SDSS, and 2MASS photometry of the B2 and 3C objects.  Either the far-UV (FUV) or the near-UV (NUV) magnitudes
(or both) are missing for a few objects.\addtocounter{table}{1}
 In the table we list in Col. 1 the B2 or 3C name, in Cols. 2 and 3  the GALEX magnitudes FUV and NUV
as found in the GALEX data base with their respective one-$\sigma$ uncertainties, and in Cols. 4-8 we provide the SDSS magnitudes ($u$, $g$, $r$, $i,$ and $z$). 
No errors are shown, since these are, at least formally, of the order of 0.01 at most. This is quite possibly much lower than the real values because we study extended nearby galaxies; in the fitting procedure described below, we adopted a 
minimum uncertainty of 0.02 mag. Finally, in Cols. 9-11 the 2MASS J, H, and K magnitudes with their one-$\sigma$ uncertainties are listed.
To investigate the spectral energy distribution, we made use of a set of models of early-type galaxies calculated by 
Bruzual \& Charlot (\cite{bruzual03}). 
The models are characterized by i) an instantaneous starburst, with ii) an IMF as given by Chabrier (\cite{chabrier03})
and iii) solar metallicity. The oldest model galaxies have an age of $1.3\times 10^{10}$ years, while the youngest ones have an age 
of a few times $10^5$ years.
We did not try to interpolate between the models, which are based on a discrete series of ages. However, this probably only has a weak effect because any improvement on the final obtained best fits can only be marginal.

We proceeded in two steps. First we checked whether a single old elliptical galaxy model was sufficient and resulted in a satisfactory fit to the photometric data. 
A minimum reduced $\chi^2$ lower than 20 was considered to represent a good fit. This value might appear to be quite high, 
but this is undoubtedly due to the (unrealistically) small formal errors on the photometry, especially in the SDSS bands.  

Many galaxies do not need an extra YSP or AGN component, meaning
that there is no trace of either star formation or non-thermal emission coming from the nucleus; 
excellent fits were then obtained with only an old galaxy model (ages around $0.5-1.3\times 10^{10}$ years). 

Since we studied 
sometimes very extended objects, this might add some uncertainty in the magnitudes as derived by the various surveys. 
In a number of cases we also have slightly deviating 2MASS bands, which are systematically displaced compared to the other bands of 
GALEX and SDSS, and this strongly contributes to increase the value of $\chi^2$, up to values $>50$. However, about three quarters of the
one-component fits have $\chi^2 < 9$, which illustrates that the old galaxy model of Bruzual \& Charlot (\cite{bruzual03}) represents the actual SED of the inactive
galaxies quite well.

\begin{figure*}
\centering
\includegraphics[width=17cm, height=23cm]{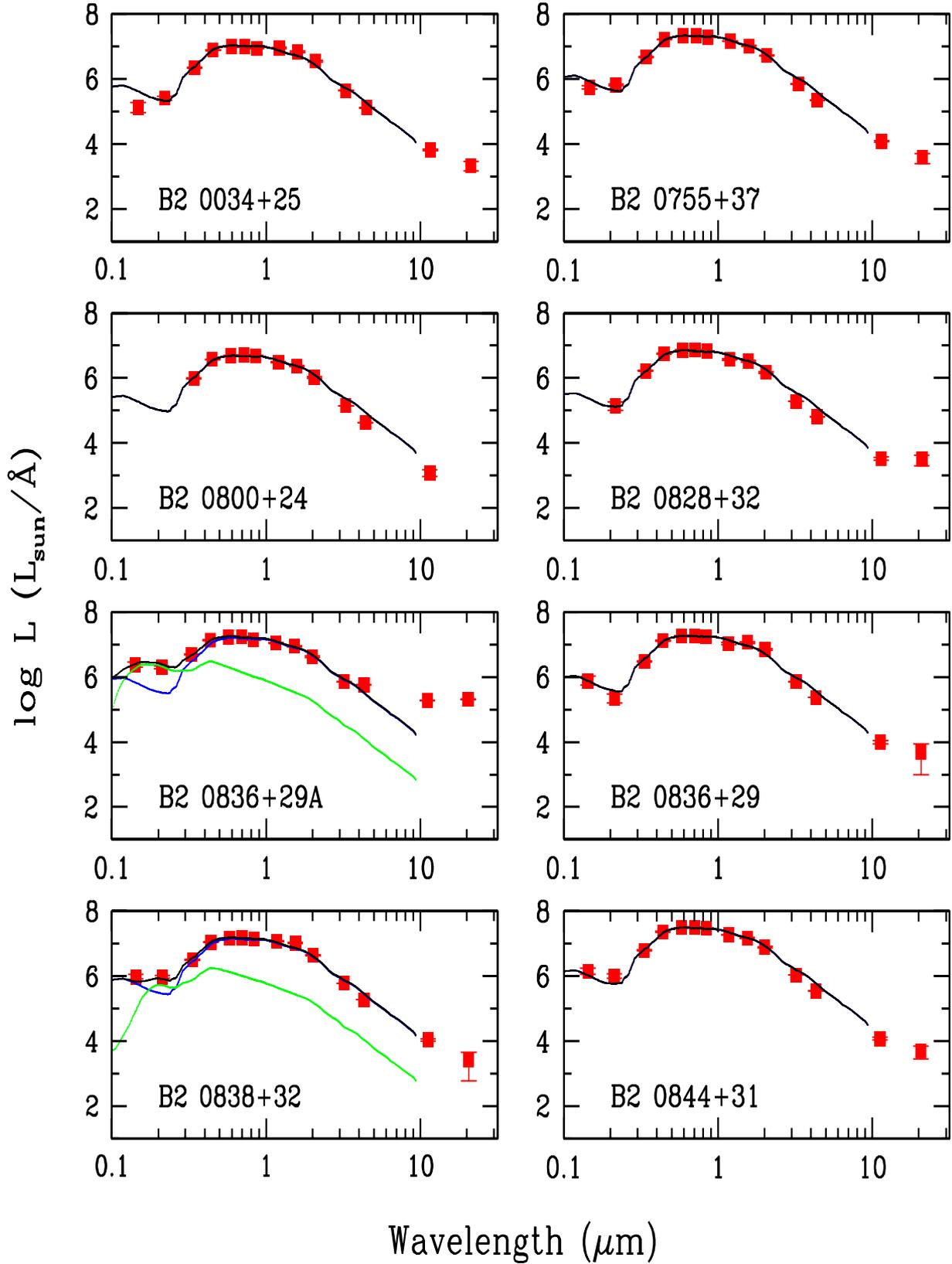}
\caption{SED fits. A solid blue line represents a Bruzual \& Charlot (\cite{bruzual03}) old elliptical galaxy model. A green line, if present, represents 
a young stellar population, and a solid black line the sum of the old and young model. In the fitting we only used 
GALEX, SDSS, and 2MASS data points.  We also show the four WISE bands, which were not used in the fit. In a number of 
cases other IR data were available, and these objects are shown at the end.}
\label{fig1}
\end{figure*}

If no satisfactory fit could be obtained with a single galaxy component, a second component was introduced, either a second younger stellar population (YSP) 
or a power-law component, due to the AGN. Of course, there might be more than one additional component, for example, two YSPs of 
different ages
or a YSP {\it \textup{and}} a power-law component (see for example Tadhunter et al. \cite{tadhunter98}; Tadhunter et al. \cite{tadhunter11}).
 However, we did not consider this possibility as the data are not sufficiently detailed to allow such a treatment.
As to the choice of a power-law or young galaxy (star-forming) component, it usually is difficult to choose which of the two gives a better fit. If 
a power-law component appears to be a serious alternative, the fitting power-law parameters are also listed in Table \ref{results}, although
we did not plot them in Fig. \ref{fig1}. A small minority of objects did not allow convergence to a satisfactory fit, and 
it therefore appears impossible to decompose the SED using two simple components. We investigate the question whether the UV excess is due to light from the AGN or from
a young stellar population in Sect. \ref{discussion}.

In Fig. \ref{fig1} we show fits with only one component, or, 
if necessary, a second younger galaxy component. Sources with additional far-IR data are given at the end.

High $\chi^2$ values may partly be due to a discrepancy between the SDSS and 2MASS, as mentioned above 
(see e.g. the cases of 3C\ 274 and B2\ 1502+26), but not always (e.g. 3C\ 234). In general, $\chi^2$ never reaches values close to unity because the formal
errors in the magnitudes of SDSS and 2MASS are much smaller than the actual errors. 
A more realistic estimate of the magnitude errors can be obtained by plotting the difference of the 2MASS magnitudes and the magnitudes of the
galaxy model used to fit the photometric data. These are shown in Fig.  \ref{fig12}.

\setcounter{figure}{0}
\begin{figure*}
\centering
\includegraphics[width=17cm, height=23cm]{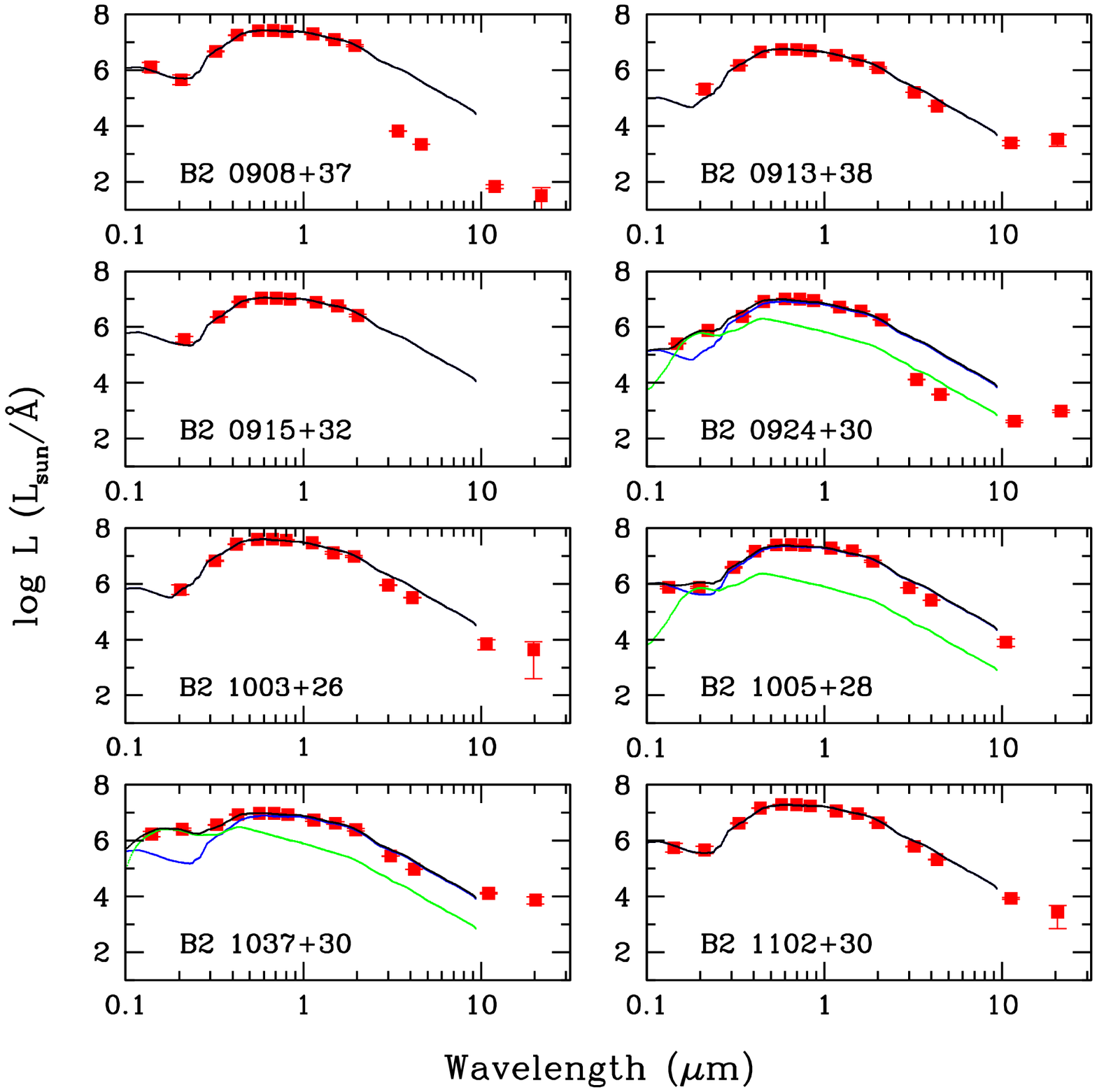}
\caption{SED fits (continued).}
\label{fig2}
\end{figure*}
\setcounter{figure}{0}
\begin{figure*}
\centering
\includegraphics[width=17cm, height=23cm]{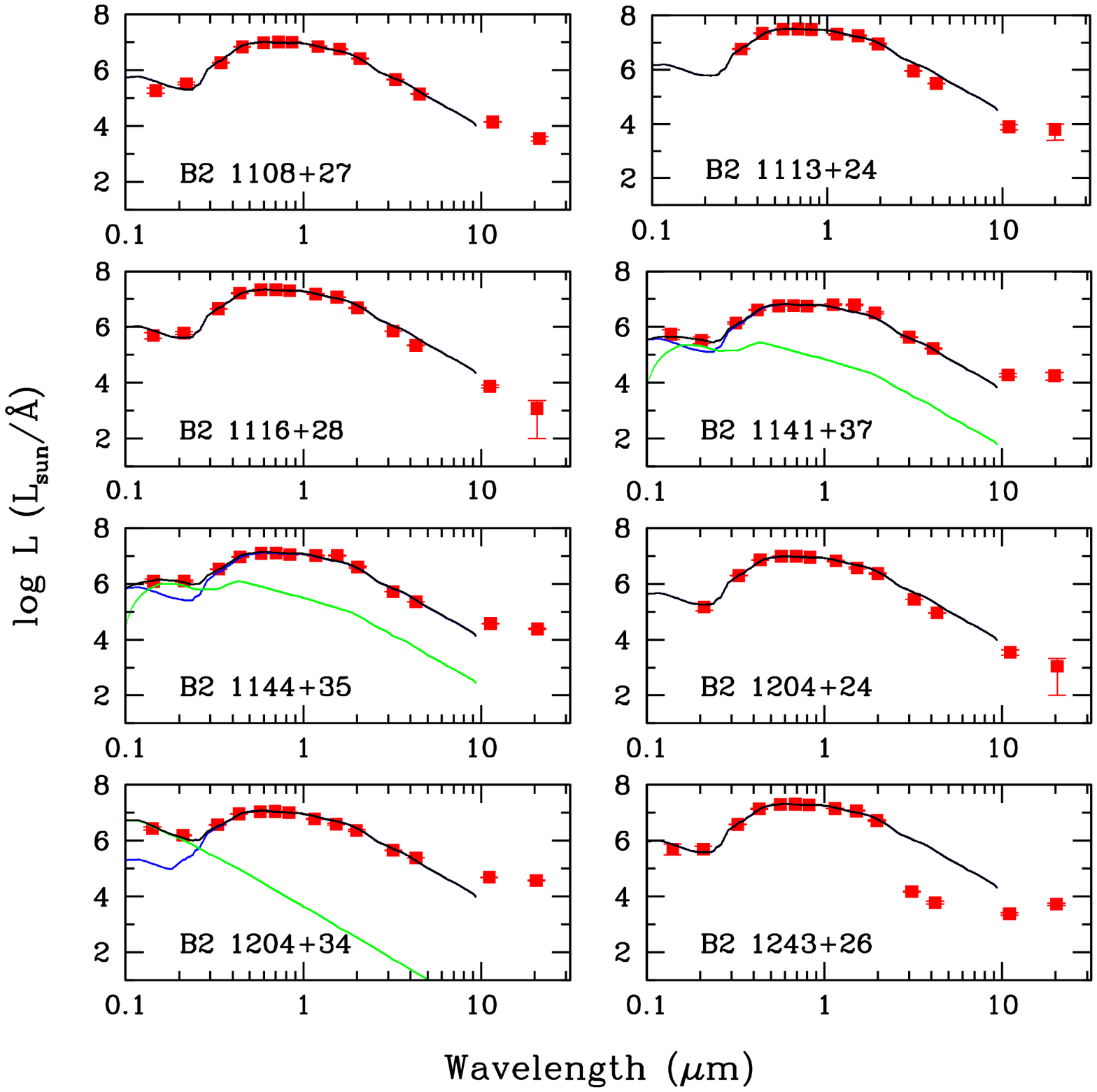}
\caption{SED fits (continued).}
\label{fig3}
\end{figure*}
\setcounter{figure}{0}
\begin{figure*}
\centering
\includegraphics[width=17cm, height=23cm]{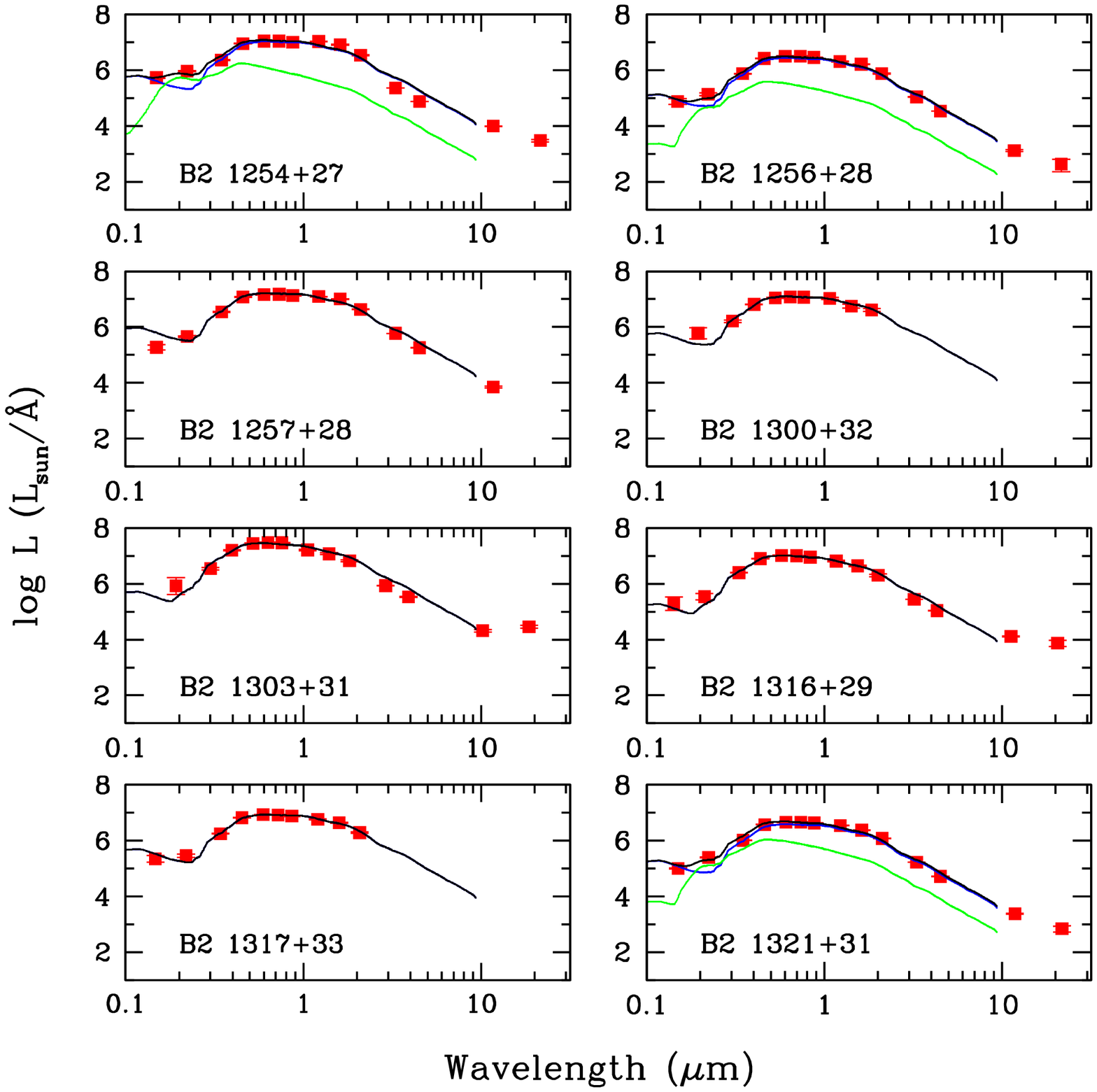}
\caption{SED fits (continued).}
\label{fig4}
\end{figure*}
\setcounter{figure}{0}
\begin{figure*}
\centering
\includegraphics[width=17cm, height=23cm]{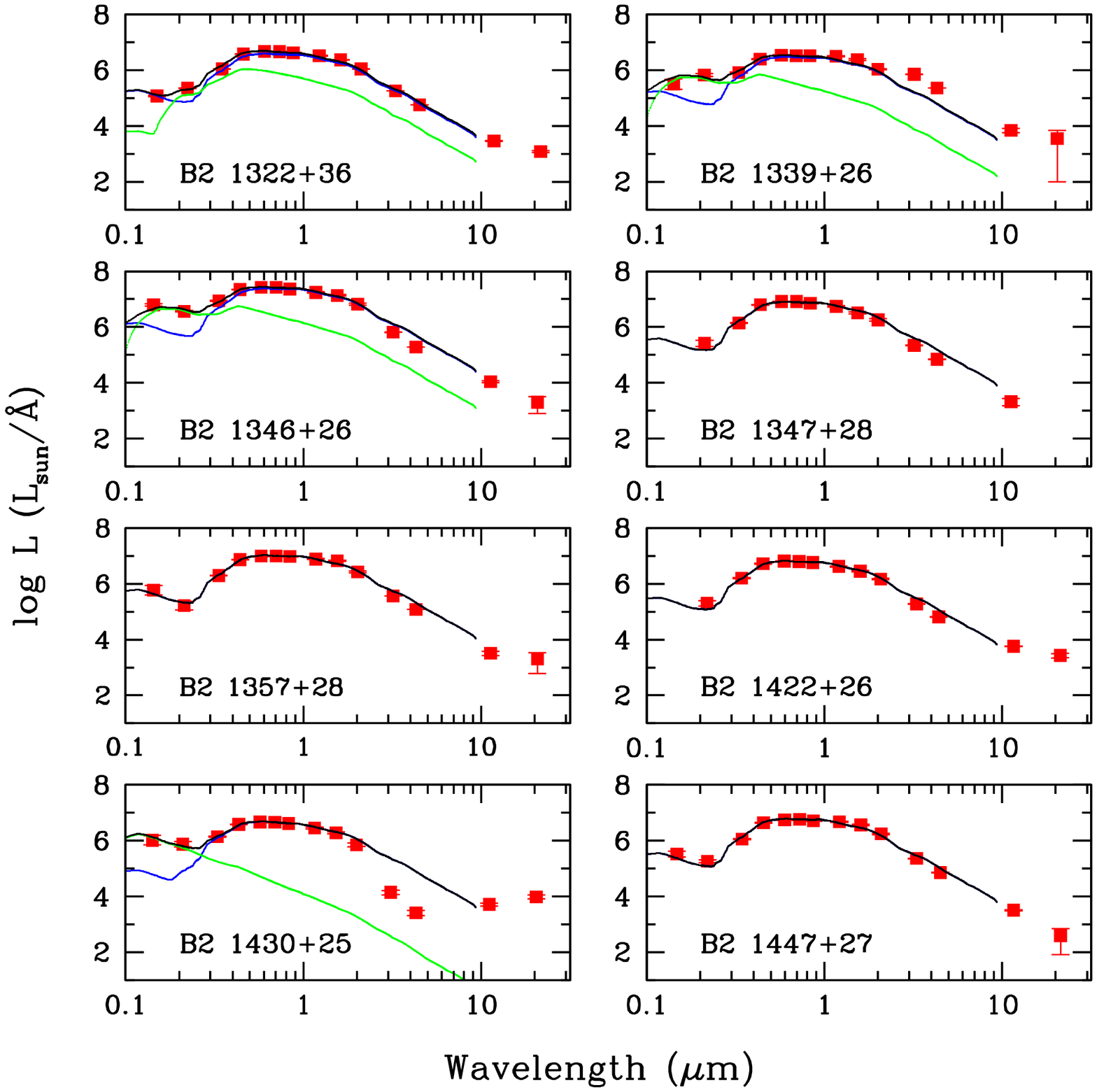}
\caption{SED fits (continued).}
\label{fig5}
\end{figure*}
\setcounter{figure}{0}
\begin{figure*}
\centering
\includegraphics[width=17cm, height=23cm]{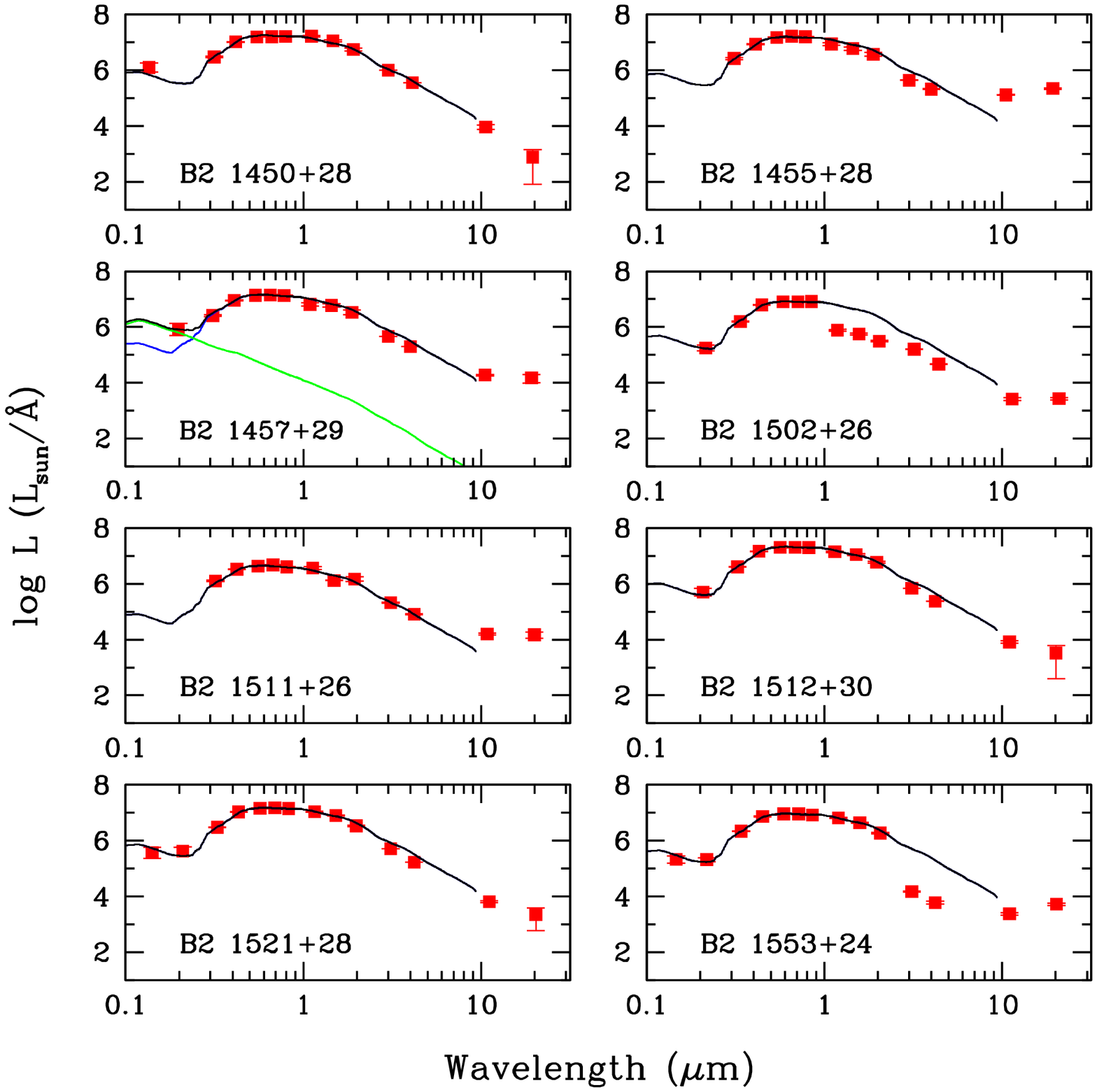}
\caption{SED fits (continued).}
\label{fig6}
\end{figure*}
\setcounter{figure}{0}
\begin{figure*}
\centering
\includegraphics[width=17cm, height=23cm]{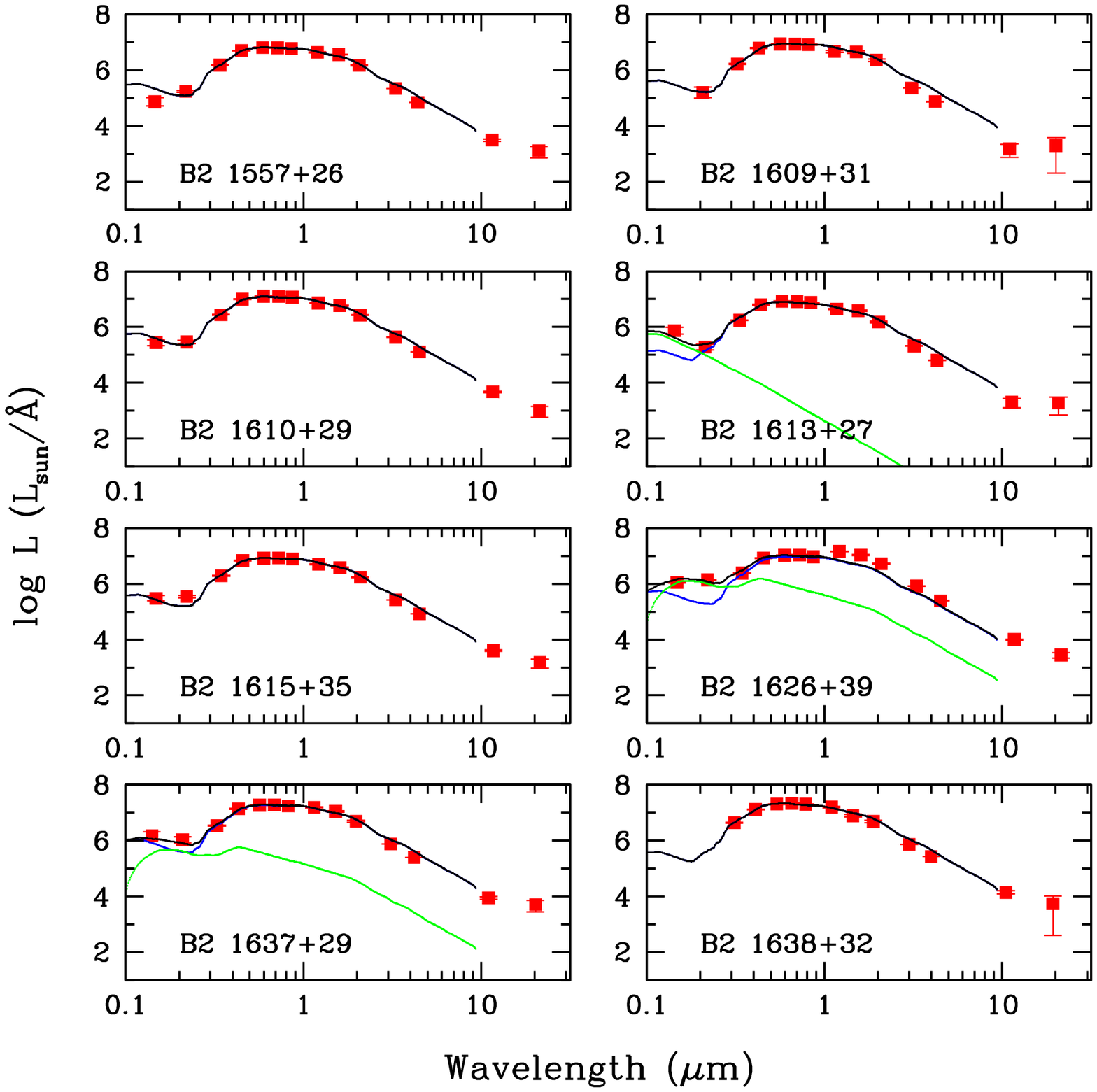}
\caption{SED fits (continued).}
\label{fig7}
\end{figure*}
\setcounter{figure}{0}
\begin{figure*}
\centering
\includegraphics[width=17cm, height=23cm]{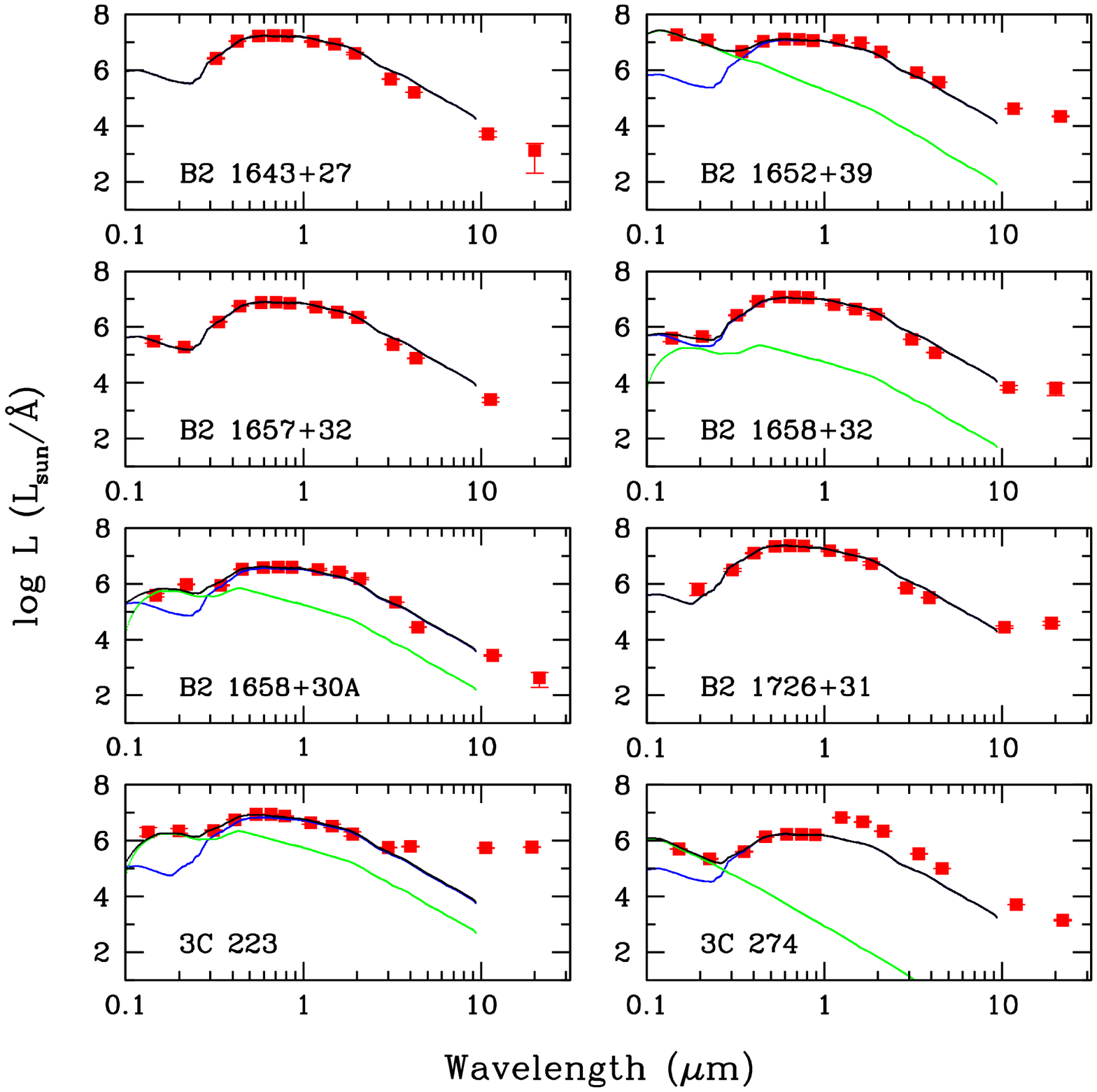}
\caption{SED fits (continued).}
\label{fig8}
\end{figure*}
\setcounter{figure}{0}
\begin{figure*}
\centering
\includegraphics[width=17cm, height=23cm]{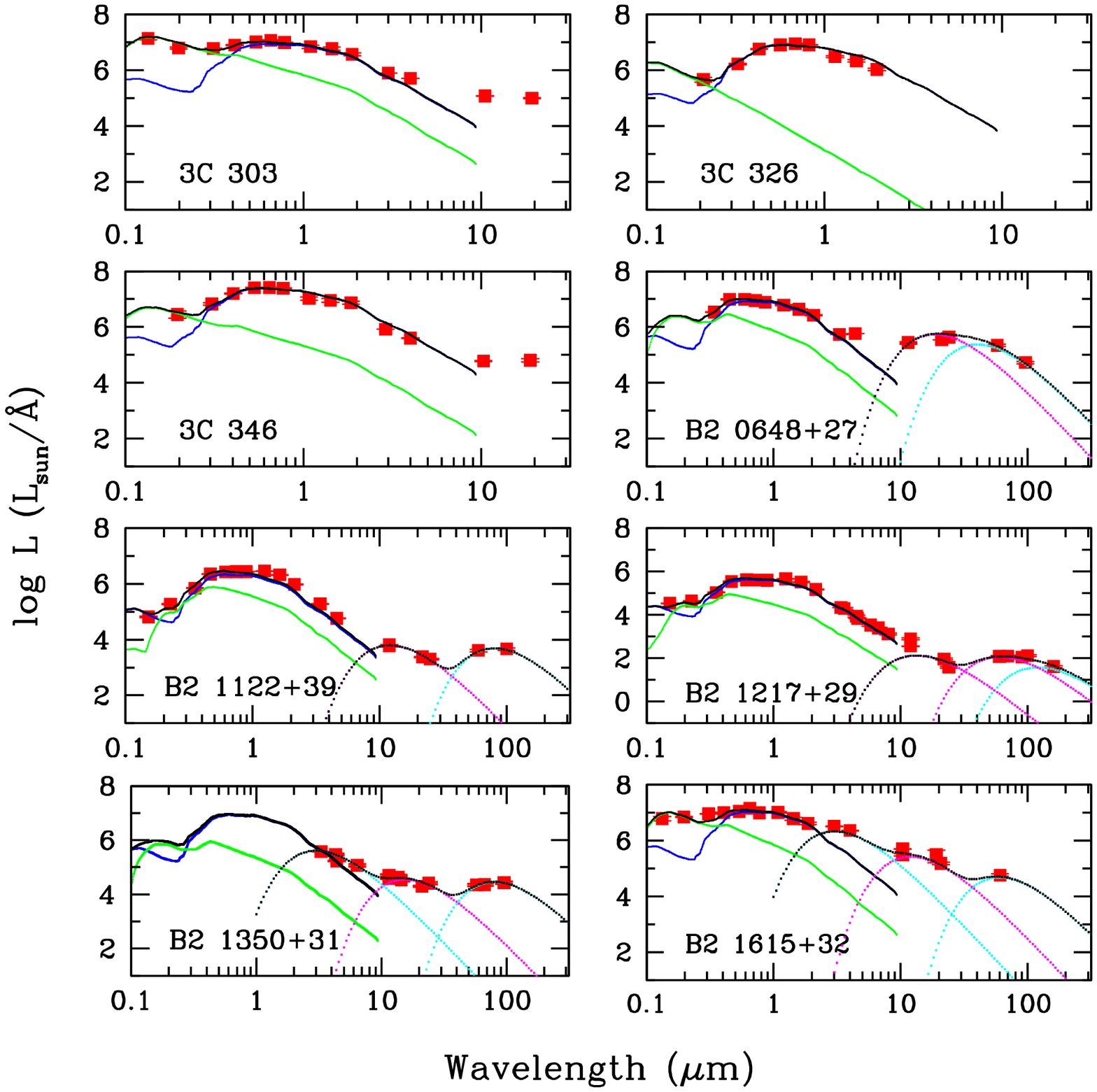}
\caption{SED fits (continued).}
\label{fig9}
\end{figure*}
\setcounter{figure}{0}
\begin{figure*}
\centering
\includegraphics[width=17cm, height=23cm]{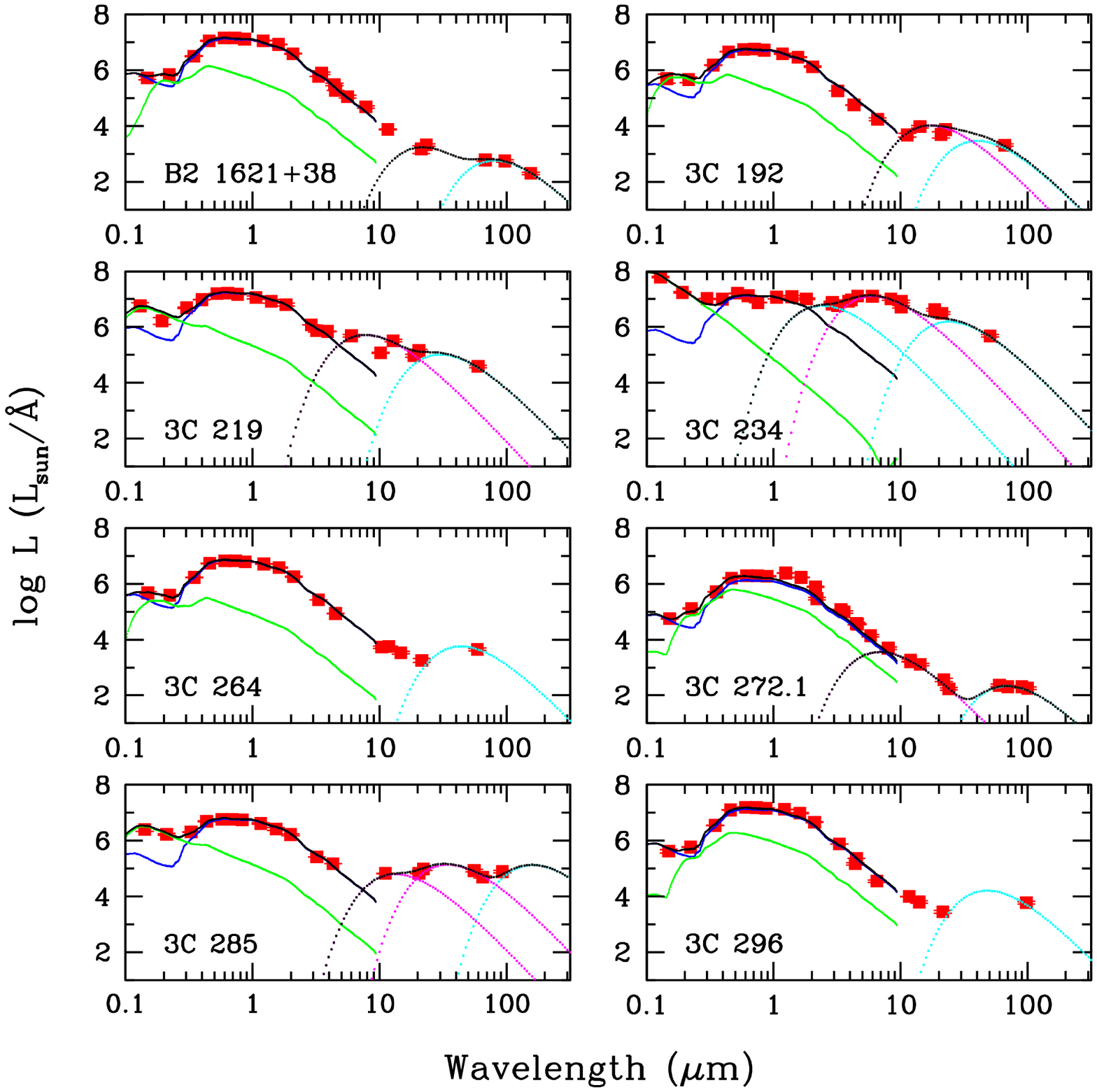}
\caption{SED fits (continued).}
\label{fig10}
\end{figure*}
\setcounter{figure}{0}
\begin{figure*}
\centering
\includegraphics[width=17cm, height=8cm]{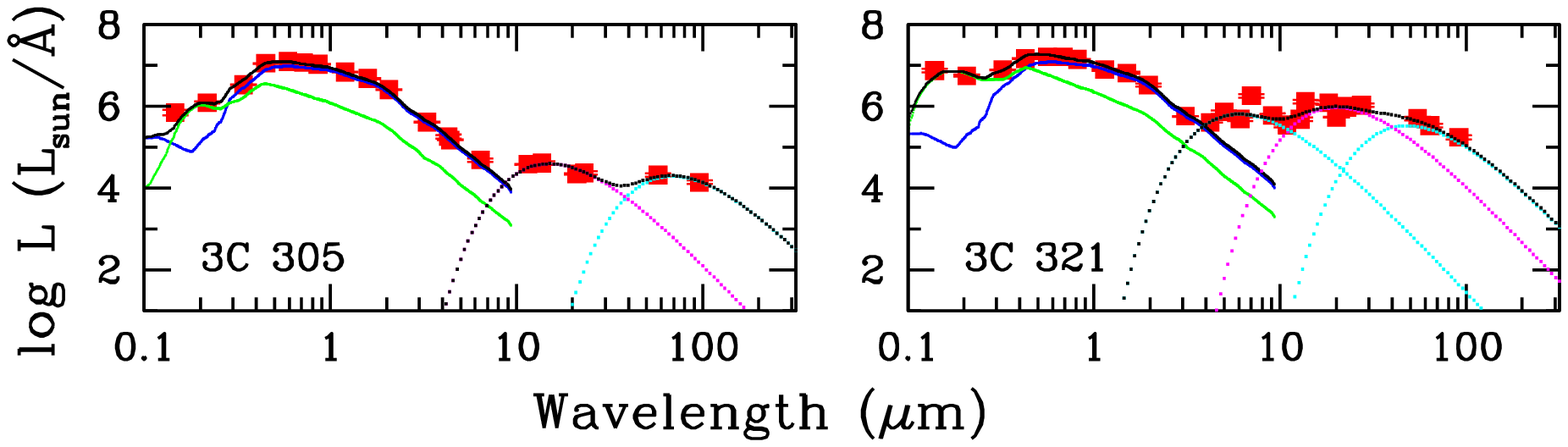}
\caption{SED fits (continued).}
\label{fig11}
\end{figure*}

From this figure one can draw four conclusions: i) only very few objects have a large difference between 2MASS and model magnitudes (the latter
being strongly constrained by the SDSS data); ii) one object, B2 1502+26 (3C 310) is quite different, the 2MASS value being as much as two magnitudes fainter
than expected from the model fit. In this case, the reason is very simple: 3C 310 is only present in the 2MASS point source catalogue and not in the
extended catalogue. Therefore the 2MASS magnitudes were determined using a point source model. Moreover, the WISE data are based on a point source
fit in determining the magnitudes, and they therefore agree well with 2MASS;
iii) with few exceptions, all other magnitudes agree within a few tenths of a magnitude, which probably tells us that a more realistic estimate of the 
magnitude errors should be of that order; iv) there appears to be a very weak correlation with redshift. 
This correlation can be safely ignored for redshifts above $\sim 0.01$, but may cause some problems for the
most nearby galaxies (e.g. 3C 274, for which 2MASS and WISE agree
well, but SDSS appears to be low). 
Leaving out the low-redshift objects and the deviating point of 3C 310, we find that $\Delta m = 0.00 \pm 0.19$.
WISE and 2MASS generally agree well, perhaps surprisingly so, because WISE was optimized for point-like sources (Jarrett et al. \cite{jarrett13}):
flux mismatches should therefore be possible considering that the objects discussed here are extended. Indeed, there are a few cases where
the WISE luminosities appear to be discrepant (B2 0908+37, B2 0924+30, B2 1243+26, B2 1430+25 and B2 1553+24) and significantly lower than the extrapolated luminosities
from the SDSS and 2MASS. For three of these five sources, the WISE luminosities were determined using a point source fit, but not B2 0908+37 and B2 1553+25, which
were fitted with an extended source model. The discrepancy of the WISE luminosities remains unaccounted for only in these two sources.
\begin{figure}
\centering
\includegraphics[width=8cm]{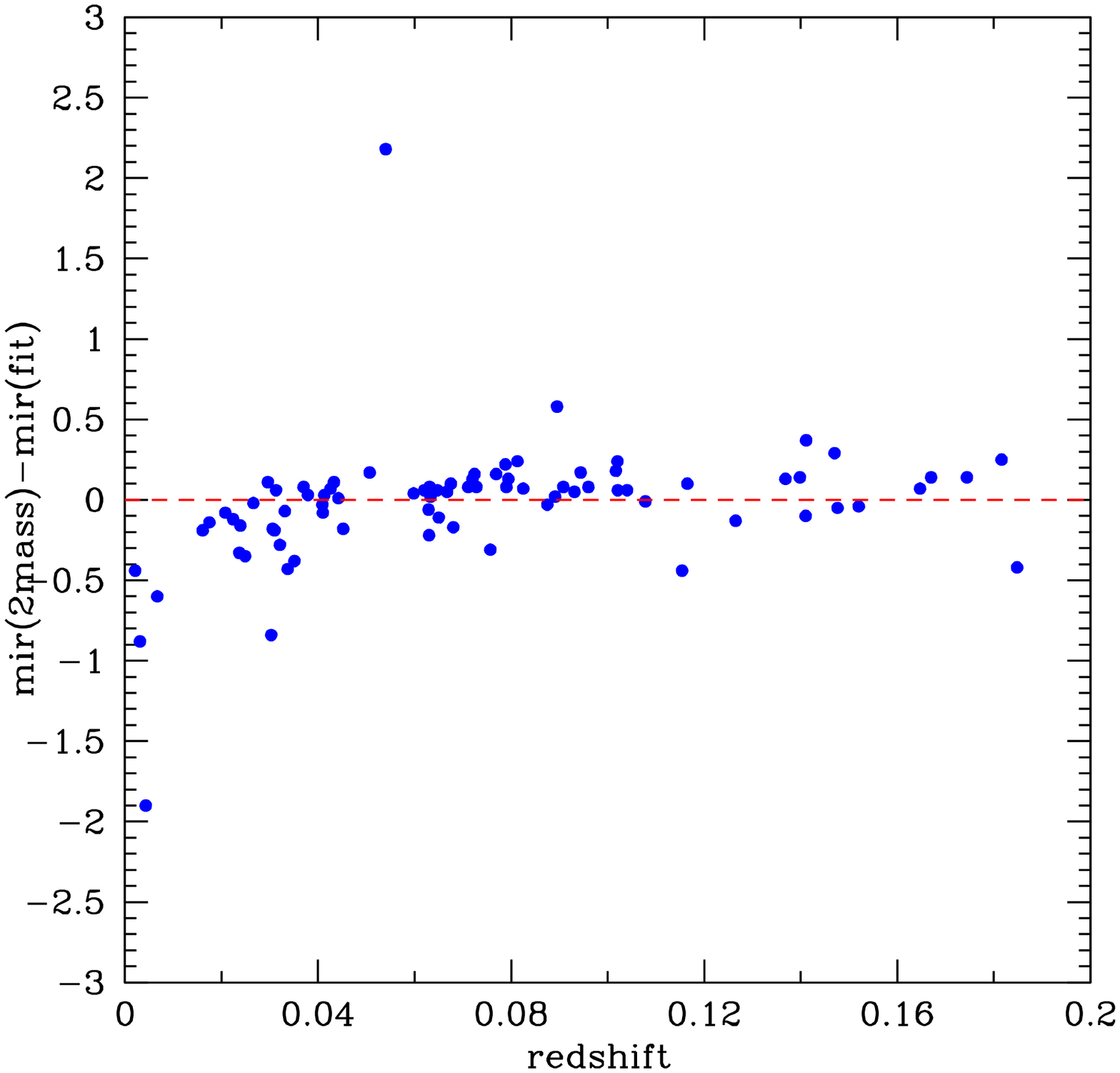}
\caption{Difference of 2MASS and model magnitudes. Each point represents the mean difference of the three 2MASS and their respective 
model magnitudes as determined by the overall SED fit.}
\label{fig12}
\end{figure}

Nevertheless, we stress that in the large majority of objects the continuity between 2MASS and the two
near-IR bands of WISE is excellent. This is different for the two mid-IR bands of WISE (at 12 and 22 $\mu$m), and this has interesting implications, as we discuss below.

\addtocounter{table}{1}
The results of the fitting are given in Table \ref{tab:results} together with  a number of parameters that are discussed in Sects. \ref{results} 
and \ref{discussion}.

\section{Results}
\label{results}

It is well known that FRI sources tend to have low-excitation spectra, often with only few weak emission lines, and these are fitted 
very well with a single-galaxy model spectrum with an age of the order 
of $5-13\times 10^ {9}$ years. This can be seen in Table \ref{tab:results}, in which we have collected a number of useful data on 
the objects studied in this paper.
The presence of a low-luminosity radio source therefore often has no influence whatsoever on the optical spectral energy distribution 
of the galaxy as a whole. 
This is, of course, in stark contrast to high-luminosity radio sources, which are usually accompanied by high-excitation spectral lines.

Several articles have recently dealt with the UV properties of early-type galaxies: Jeong et al. (\cite{jeong12}) reported 
an unexpectedly rising UV flux towards shorter wavelengths (in early-type galaxies in general), while Allen et al. (\cite{allen02}) and 
Baldi \& Capetti (\cite{baldi08}) attributed the (excess) UV flux found in 3CR radio galaxies to recent star formation. 
We here mainly concentrate on galaxies with lower radio luminosities than those discussed by Allen et al. (\cite{allen02}) 
and Baldi \& Capetti (\cite{baldi08}), and such galaxies may, at least in the optical, resemble the group of X-ray bright galaxies 
with optically "dull" AGN analysed by Trump (\cite{trump09}). We note that according to the latter authors, such galaxies do have weak but 
significant UV emission, like an unobscured AGN.

Fanti et al. (\cite{fanti11}) introduced a measure of the excess ultraviolet radiation that provides a quantitative indication of the 
amount of additional radiation in the ultraviolet region, as compared to the passive SED of an early-type galaxy. 
We use a similar but slightly different parameter, $XUV$. To be precise, 
$$ XUV = -2.5\times \log\left[\frac{(F_{4500 \AA}/F_{2000 \AA})_{\rm obs}}{(F_{4500 \AA}/F_{2000 \AA})_{\rm model}}\right]= 
\Delta m_{2000\AA}-\Delta m_{4500\AA} $$  such that $XUV$ measures the observed excess, in magnitudes, in the slope between 
4500 and 2000 \AA\ with respect to the SED of a passively evolving early-type galaxy. 
The uncertainty in $XUV$, which depends on the uncertainties in FUV and NUV 
(see Table \ref{tab:phot}), is usually $< 0.3$ mag. 
We list $XUV$ together with its uncertainty in Col. 13 of Table \ref{tab:results}.

$XUV$ measures the mean UV excess, based on both FUV and NUV. The behaviour of FUV and NUV individually does not necessarily
tell us something about the shape of an excess: even in the SED of an old elliptical galaxy there is more FUV than NUV 
(see the model SEDs in Fig. \ref{fig1}), although $XUV$ may be close to zero.
However, once the reality of a UV excess is determined, the presence or absence of an increase may tell us something about the cause of
the excess; in particular, if FUV is lower than NUV, this may be hard to explain with an AGN component in the form of a power
law.  
If there is an increase in UV that is accompanied with a UV excess, it is impossible to distinguish between a power law and a YSP with age $< 10^8$ years, and we need to turn to other methods to determine whether the prime cause is the AGN or star formation.  We discuss this in Sect. \ref{discussion}.

In Table \ref{tab:results} we also list some other parameters 
such as the galaxy model that produced the best fit to the observed SED in Cols. 2 (logarithm of the galaxy mass, in solar masses) 
and 3 (age of the galaxy model), and, if necessary, a second component in Cols. 4 and 5. 
A best fit obtained with a power law is indicated by a "p" in Col. 6: Column 4 then gives the luminosity density (in solar luminosities per $\AA$) 
at the reference point $\lambda = 2000$ \AA, and Col. 5 the slope of the power law. A 
second galaxy component is indicated by a "g" in Col. 6; in that case, we list in Col. 4 
the logarithm of the mass expressed in solar masses and the age of this galaxy component in Col. 5. 
The reduced minimum $\chi^2$ in Col. 7 refers to the final fit adopted, either with one or two components. 
We recall again that these values may in general appear to be high, which is simply due to the probably too optimistic 
estimates of the SDSS photometric uncertainties.  
Since for a number of sources the IR data do not agree very well with the SED in the optical, we have indicated these cases by footnotes to Col. 7.

For completeness we also list some older data, which can be found in Fanti et al. (\cite{fanti87}) and de Ruiter et al. (\cite{deruiter90}): 
the total power at 1.4 GHz,
(recalculated with the concordance model) in Col. 8, the FR type in Col. 9 (we note that FR 3 refers to compact sources, 
while in a few cases no FR type could be assigned). 

We determined the spectral type of the radio galaxies from the SDSS spectra (if available) from Buttiglione et al. (\cite{buttiglione09b}) or from NED.
We can distinguish the following cases: i) spectra without any sign
of emission lines. The overall  spectrum shows the typical absorption features of an inactive early-type galaxy. 
Such objects are coded by a "0" in Col. 10. There often is only very weak emission in $H\alpha$ or [N II] but not much else, and these objects have a "1". 
A discussion of weak-emission line galaxies has recently been given by Cid Fernandes et al. (\cite{cid10}; \cite{cid11}), 
aiming at distinguishing between genuine and false AGNs. 
As we here consider radio galaxies, no ambiguity should arise concerning the AGN nature (in a broad sense) of our objects.
 
Other possible spectral classifications are
"2", which stands for low-excitation galaxies (LEGs; but with enough emission lines present to allow the use of diagnostic diagrams), "3" for high-excitation galaxies (HEGs), and "4" if 
broad-line components are present as well. Since we only consider radio galaxies, LEG and HEG can be used instead of LERG and HERG 
without creating confusion. 
For the definitions of LEG and HEG, see Laing et al. (\cite{laing94}) or Buttiglione, Capetti \& Celotti (\cite{buttiglione08}), for
instance; 
an excellent discussion on the use of line ratios in the classification of optical spectra can be found in Kewley et al. (\cite{kewley06}).  
Following their analysis, we report in Col. 11 a classification of the spectral type according to the emission line ratios determined here: 
"H" for  HII, that is, star forming, "L" for liner, "S" for Seyfert, "S1" for Seyfert type 1. 
In a number of cases the information from the diagnostical line-ratio diagrams was not detailed enough to distinguish between 
liners and Seyferts, and such objects are generically classified as "A" (AGN).
A few objects fall in the zone of composite spectra, in which there is probably both star formation and an AGN; 
such objects are classified as "C" in Col. 11. We note that one object, B2 1652+39, 
which is a compact radio source with one of the highest $XUV$ values (4.05), 
is classified as a BLLac (Ulrich et al. \cite{ulrich75}); its spectrum is featureless.

In Col. 12 we list the luminosity (in $10^{33}$ Watt) of the $H\alpha$ emission.
 
In Col. 13 we list the parameter that describes the ultraviolet excess, $XUV$, as discussed above, and also its one-$\sigma$ uncertainty.
The logarithm of the ratio of the two near-IR (W1 plus W2) and the two mid-IR (W3 and W4)  WISE luminosities is given in Col. 14 (for a detailed
discussion of this parameter, see Sect. \ref{XUVIR}).
Finally, in Col. 15 we list the central black hole masses wherever available. 
They were taken from Bettoni et al. (\cite{bettoni03}, \cite{bettoni09}), 
the HYPERLEDA\footnote{URL: http://leda.univ-lyon1.fr} database (Paturel et al. \cite{paturel03}), or Snellen et al. (\cite{snellen03}), as indicated, and
using the relation between central velocity dispersion and black hole mass  of Tremaine et al. (\cite{tremaine02}).

\subsection{Comments on individual sources}
\label{comments}

B2 0648+27. Unfortunately, no GALEX data are available, but an additional young population does improve the fit (see Fig. \ref{fig1}). Moreover, spectral data
show that this galaxy is a HEG (see Emonts et al. \cite{emonts06}). MIR and FIR data are available, which can be fitted by a mixture of dust with temperatures
of 60 K (dust mass $M_d = 1.3\times 10^6 M_{\sun} $) and 130 K ($M_d = 2.9\times 10^4 M_{\sun}$).

B2 0836+29A. A young stellar component is necessary to explain the UV excess (both in FUV and NUV). It is hard to fit the excess using a power-law component coming from the AGN, but we note that this source is a HEG and also rather powerful in radio (FR type in the transition range between types I
and II). 

B2 0838+32. The moderate UV excess can be very well explained by a young stellar component. Unfortunately, no optical spectra are available (to our
knowledge). We note that the WISE data follow the old stellar population model quite well, up to and including the 22 $\mu$m point.

B2 0908+37. The WISE data fall well below the values expected from GALEX, SDSS, and 2MASS data, although the slope appears to be consistent. 
This is one of the few cases in which the WISE flux determination has failed to account correctly for the extendedness of the object.

B2 0924+30. A young stellar component gives a good fit to the spectrum, including NUV and FUV. The WISE data are different by a large factor, but again,
as in B2 0908+37, the slope appears to be consistent. We note that the WISE luminosities were derived using a point-source model.

B2 1005+28. A young stellar component improves the fit, but there is no trace of a UV excess. The optical spectrum is devoid of emission features.
The WISE data follow the old stellar component very well.

B2 1037+30. The blue and UV data can only be fitted (and well) by a young galaxy component. This object has a typical (post-)starburst spectrum, with the
higher Balmer lines present in absorption. In the diagnostic diagrams it falls in the liner region.

B2 1122+39. A young stellar component is needed to fit the UV excess. No plausible power-law component would be able to do this.
MIR and FIR data are available, which can be fitted by a mixture of dust with temperatures
of 30 K ($M_d = 1.7\times 10^6 M_{\sun} $) and 200 K ($M_d = 25 M_{\sun}$).

B2 1141+37. A young stellar population yields a good fit to the UV data. In principle, a power-law component might produce an equally plausible fit, 
but we note that the spectrum has clear star burst characteristics (higher Balmer lines present in absorption). The optical spectrum is
otherwise lacking emission features.

B2 1144+35. This source has strong Seyfert 1-like broad emission lines. No signs of star formation are present.

B2 1204+34. The UV excess may be caused by a young stellar or power-law component. As discussed below
(Sect. \ref{discussion}),  both may be present.

B2 1217+29 (NGC 4278). The fit with a YSP has a rather high $\chi^2$ value that is due to the systematically higher 2MASS points, as
compared to the GALEX and SDSS data. We note that NGC 4278 is at very low redshift (z=0.0021), which probably causes this discrepancy.
The decrease in luminosity from SDSS via NUV to FUV can be reproduced well by a YSP, but not by a power law. 
MIR and FIR data are available, which can be fitted by a mixture of dust with temperatures
of 20 K ($M_d = 1.5\times 10^5 M_{\sun} $), 40 K ($M_d = 7.0\times 10^3 M_{\sun} $) and 180 K ($M_d = 1.0 M_{\sun}$).

B2 1243+26. The WISE luminosities are low because they were determined with a point-source model.

B2 1254+27. As in B2 1217+29, the decrease in luminosity reaching from SDSS via NUV to FUV is easily reproducible by a YSP, but not by a power law.

B2 1256+28. As in B2 1217+29 and B2 1254+27, only a YSP produces a satisfactory fit.

B2 1317+33. One of the very few sources for which no WISE data are available.

B2 1321+31 and B2 1322+36. Only a YSP can explain the UV behaviour of the spectral energy distribution.

B2 1339+26. The UV excess ($XUV=2.26$ mag) is high and clearly significant. A YSP reproduces the
UV well, but there are no other signs of star formation in the optical spectrum; emission lines are absent. 

B2 1350+31 (3C 293). According to Tadhunter et al. (\cite{tadhunter11}), a YSP is present.
MIR and FIR data are available, which can be fitted by a mix of dust with temperatures
of 60 K ($M_d = 1.0\times 10^7 M_{\sun} $), 155 K ($M_d = 6.5\times 10^2 M_{\sun} $)  and 800 K ($M_d = 0.4 M_{\sun}$).

B2 1430+25. The mismatch of the WISE data is due to their being based on a point source model. The UV excess
($XUV= 2.08$) and the distribution the SDSS and GALEX points allow both a YSP or a power law as a plausible candidate.
We note, however, that the spectrum shows some moderately strong Balmer lines in absorption.

B2 1455+28. Only a one-component fit was possible because UV information is lacking. However, the source is a HEG and also
a strong FRII radio source. It is therefore likely that this source has a significant UV excess.

B2 1457+29. FUV is lacking, therefore fitting with a YSP or a power law is equally possible. No spectral data are available.

B2 1502+26 (3C 310). In this case, both the 2MASS and WISE luminosities were obtained by using a point-source model, undoubtedly 
the reason for their discrepancy with SDSS and GALEX.  

B2 1553+24. This is one of the two sources (the other being B2 0908+37) for which the mismatch of WISE luminosities remains unexplained
(an extended model was used to determine the fluxes).

B2 1613+27. Although there formally is no UV excess ($XUV=0$), there appears to be a strong increase towards FUV. The reason for this is
that $XUV$, which refers to the emitted luminosity at 2000 {\AA},  is in this case mostly determined by NUV, which still follows the old
stellar population (see Fig. \ref{fig1}). Therefore a fit that takes the increase into account can only be achieved with a very young
stellar component  or a power law. There are no obvious signs of either in the optical spectrum (there is only H$\alpha$
emission).

B2 1615+32 (3C 332). This is a Seyfert 1, with strong broad emission lines. The continuum is probably due to the AGN, as discussed in
Sect. \ref{XUVIR}. 
MIR and FIR data are available, which can be fitted by a mix of dust with temperatures
of 40 K ($M_d = 3.0\times 10^6 M_{\sun} $), 200 K ($M_d = 1.0\times 10^3 M_{\sun} $), and 800 K ($M_d = 2.1 M_{\sun}$).

B2 1621+38. 
MIR and FIR data are available, which can be fitted by a mix of dust with temperatures
of 30 K ($M_d = 2.0\times 10^5 M_{\sun} $) and 110 K ($M_d = 2.5\times 10^2 M_{\sun}$).

B2 1626+39. The UV excess can be well reproduced by a YSP.  The $\chi^2$ of the fit is high as a result of somewhat deviating 2MASS
points.
 
B2 1637+29. The slight UV excess can be explained well as being caused by a YSP. The optical spectrum shows no emission features
or signs of star formation. 

B2 1652+39. The UV excess is the second highest ($XUV=4.05$) in our sample. The increase towards the far UV means that power-law 
emission from the AGN is more probable than the YSP plotted in Fig. \ref{fig1}. This object is a well-known BL Lac (Mrk 501). It is also
worth noting that the spectral index between the red (between 7000 and 8000 \AA, from HST data) and 2000 \AA, $\alpha^{2000}_{HST} = 1.1$, 
which is virtually identical to the 
slope between FUV and NUV, $\alpha^{FUV}_{NUV} = 1.09$, thus confirming the power-law nature of the continuum in this BL Lac. 

B2 1658+32. A YSP produces a good fit to the model spectrum.

B2 1658+30A. This is a LEG, for which only a YSP gives a good fit in the UV. 

3C 192. This source has a moderate UV excess ($XUV=1.3$). The optical spectrum shows the strong emission lines typical of a HEG. 
No signs of star formation are apparent, and AGN emission producing the UV excess is more probable.
We also note that 3C 192 is an FRII radio source.
MIR and FIR data are available, which can be fitted by a mixture of dust with temperatures
of 60 K ($M_d = 1.6\times 10^4 M_{\sun} $) and 140 K ($M_d = 3.5\times 10^2 M_{\sun}$).

3C 219. Like 3C192, this is a HEG, with $XUV=1.7$. The radio source is of FR type II.  The evidence for AGN emission is
strong (see Sect. \ref{discussion}).
MIR and FIR data are available, which can be fitted by a mix of dust with temperatures
of 80 K ($M_d = 1.0\times 10^5 M_{\sun} $) and 310 K ($M_d = 1.5\times 10^2 M_{\sun}$).
 
3C 223. A YSP produces an excellent fit in the ultraviolet. This is one of the few sources with all four WISE luminosities at the same level (see
also Sect. \ref{XUVIR}).

3C 234. An unusual SED, almost flat from the far-IR to the ultraviolet (which even increases toward the far-UV). This source has the highest
$XUV$ (4.18) in our sample. Excellent fits are allowed
either by a YSP or by a power-law component. (see Sect. \ref{discussion} for a more detailed discussion).
MIR and FIR data are available, which can be fitted by a mix of dust with temperatures
of 100 K ($M_d = 4.0\times 10^5 M_{\sun} $), 400 K ($M_d = 8.0\times 10^2 M_{\sun} $), and 1000 K ($M_d = 1.5 M_{\sun}$).

3C 264. 
MIR and FIR data are available. The FIR data can be fitted by dust with a temperature
of 55 K ($M_d = 5.3\times 10^4 M_{\sun} $), while the WISE luminosities appear to follow the old stellar model spectrum out to at least 10 $\mu$m.

3C 272.1.  The FUV and NUV data suggest that a YSP is present and not a power-law component. The high $\chi^2$ value is due to the 2MASS data being
systematically higher than the model SED (see Fig. \ref{fig1}). We note that 3C 272.1 is at very low redshift (z=0.0031), which may have contributed to the
difficulty of correctly scaling all photometric bands.
MIR and FIR data are available, which can be fitted by a mix of dust with temperatures
of 35 K ($M_d = 6.0\times 10^3 M_{\sun} $) and 350 K ($M_d = 0.5 M_{\sun}$).

3C 274 (M87, Virgo A). The poor fit with a very high $\chi^2$ value is due to the systematically higher 2MASS luminosities. WISE data appear to agree better with 2MASS,
however. Like 3C 272.1, this source is at very low redshift (z=0.0043).

3C 285. Although a power law would be enough, there are good reasons to assume that significant star formation occurs (see Sect. \ref{discussion}). An excellent fit is obtained with a YSP, and according to Tadhunter et al. (\cite{tadhunter11}), a YSP is present.
MIR and FIR data are available, which can be fitted by a mix of dust with temperatures
of 15 K ($M_d = 3.0\times 10^9 M_{\sun} $), 70 K ($M_d = 3.0\times 10^5 M_{\sun} $), and 180 K ($M_d = 4.7\times 10^2 M_{\sun}$).

3C 296. A YSP provides a satisfactory fit with the model SED.
MIR and FIR data are available. As in 3C264, it can be fitted by dust, in this case with a temperature
of 50 K ($M_d = 2.6\times 10^5 M_{\sun} $), while the WISE luminosities appear to follow the old stellar model spectrum out to 20 $\mu$m.

3C 303. This source has a large UV excess ($XUV=3.4$), which can be explained equally well by a YSP or a power law.

3C 305. The far-UV can only be produced with the help of YSP, which, according to Tadhunter et al. (\cite{tadhunter11}), is indeed present.
MIR and FIR data are available, which can be fitted by a mixture of dust with temperatures
of 35 K ($M_d = 2.7\times 10^6 M_{\sun} $) and 160 K ($M_d = 6.0\times 10^2 M_{\sun}$).

3C 321. An excellent fit in the UV is obtained with a YSP, in agreement with Tadhunter et al. (\cite{tadhunter11}).
There is an abundance of
MIR and FIR data, which can be fitted by a mixture of dust with temperatures
of 50 K ($M_d = 5.5\times 10^6 M_{\sun} $), 120 K ($M_d = 8.0\times 10^4 M_{\sun} $), and 400 K ($M_d = 40 M_{\sun}$).

3C 326 and 3C 346. Unfortunately, FUV is lacking
for both objects, so it is impossible to distinguish between a YSP and a power law.

\subsection{Distribution of XUV}
\label{XUV}

\begin{figure}
\centering
\includegraphics[width=8cm]{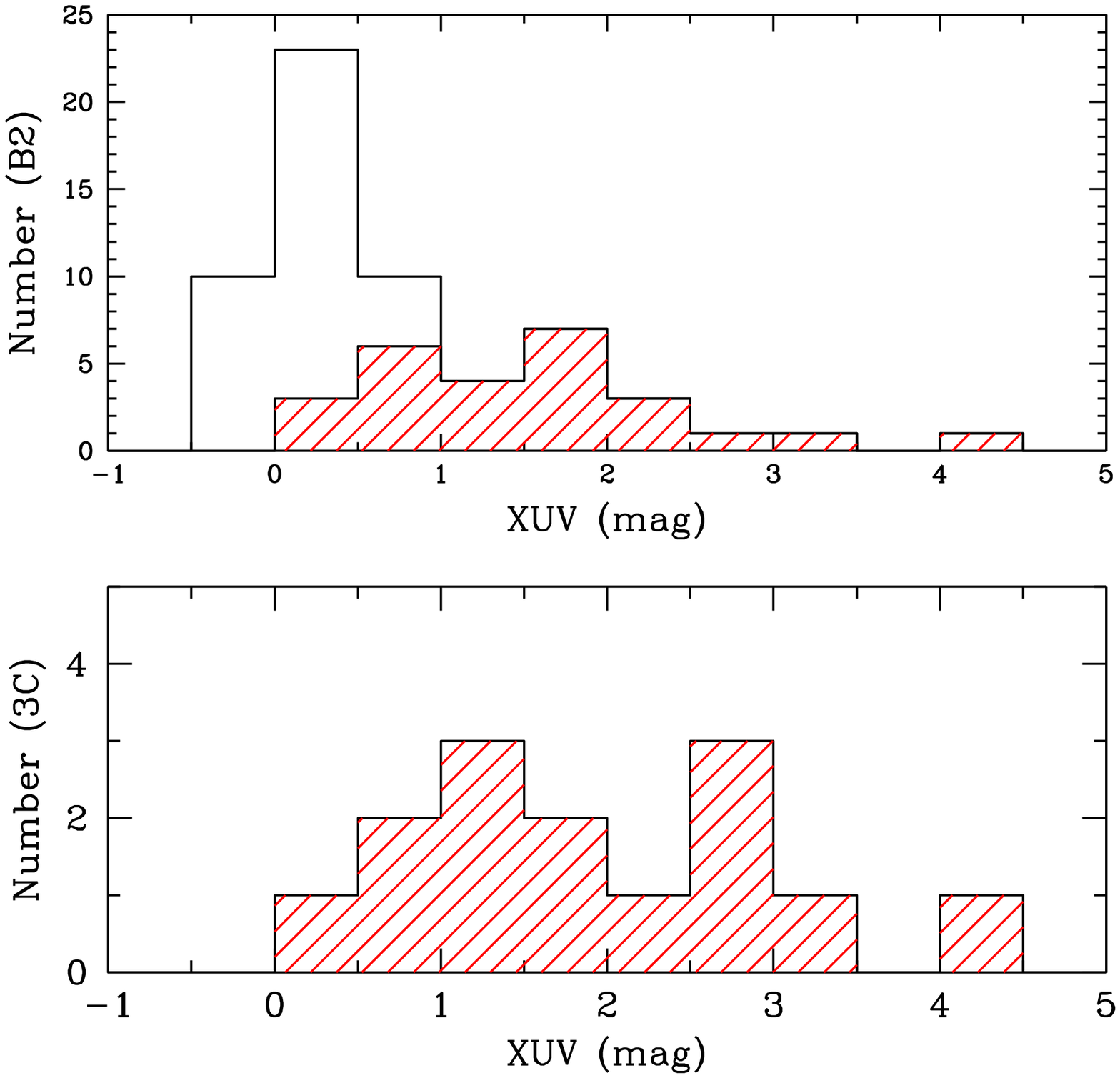}
\caption{Histograms of $XUV$ values. Upper panel: B2 radio galaxies; lower panel: 3C sources. 
We show all objects with measured $XUV$; shaded red histograms represent objects that were fitted with an 
old galaxy component plus a second one (power law or younger galaxy). All 3C galaxies were fitted with two components.}
\label{fig13}
\end{figure}

In Fig. \ref{fig13} the histograms of $XUV$ clearly show that the ultraviolet excess may be as high as 4-5 magnitudes, and such objects all need a 
second component to be satisfactorily fitted. 
As several galaxies do not need a second component, or, to be more precise, a second component does not improve the fit significantly, 
the histogram of these objects is centered on $XUV\sim 0$, as of course it should. The accuracy 
of $XUV$ is determined by NUV and FUV. We mention in passing that we 
never found a significant ultraviolet \textup{\textup{{\it deficit}}}.
The 3C sources, which are more often of the FRII type than B2 sources, do have in general more extreme values of $XUV$: about one half of 
them have $XUV > 2$, while this is true for only $\sim 10$\% of the B2 sources. This is a first indication that high radio luminosity comes
together with an ultraviolet excess.

It is well known that high-luminosity radio sources have optical spectra that usually exhibit strong emission lines like 
$H\alpha$, $[NII]\lambda\lambda 6548,6584$, $[OIII]\lambda\lambda 4959, 5007,$ and $H\beta$ (see e.g. Zirbel \& Baum \cite{zirbel95}; 
Buttiglione et al. \cite{buttiglione09a}, \cite{buttiglione09b}; Hardcastle, Evans \& Croston \cite{hardcastle09}; Capetti et al. \cite{capetti11}).

\begin{figure}
\centering
\includegraphics[width=9cm]{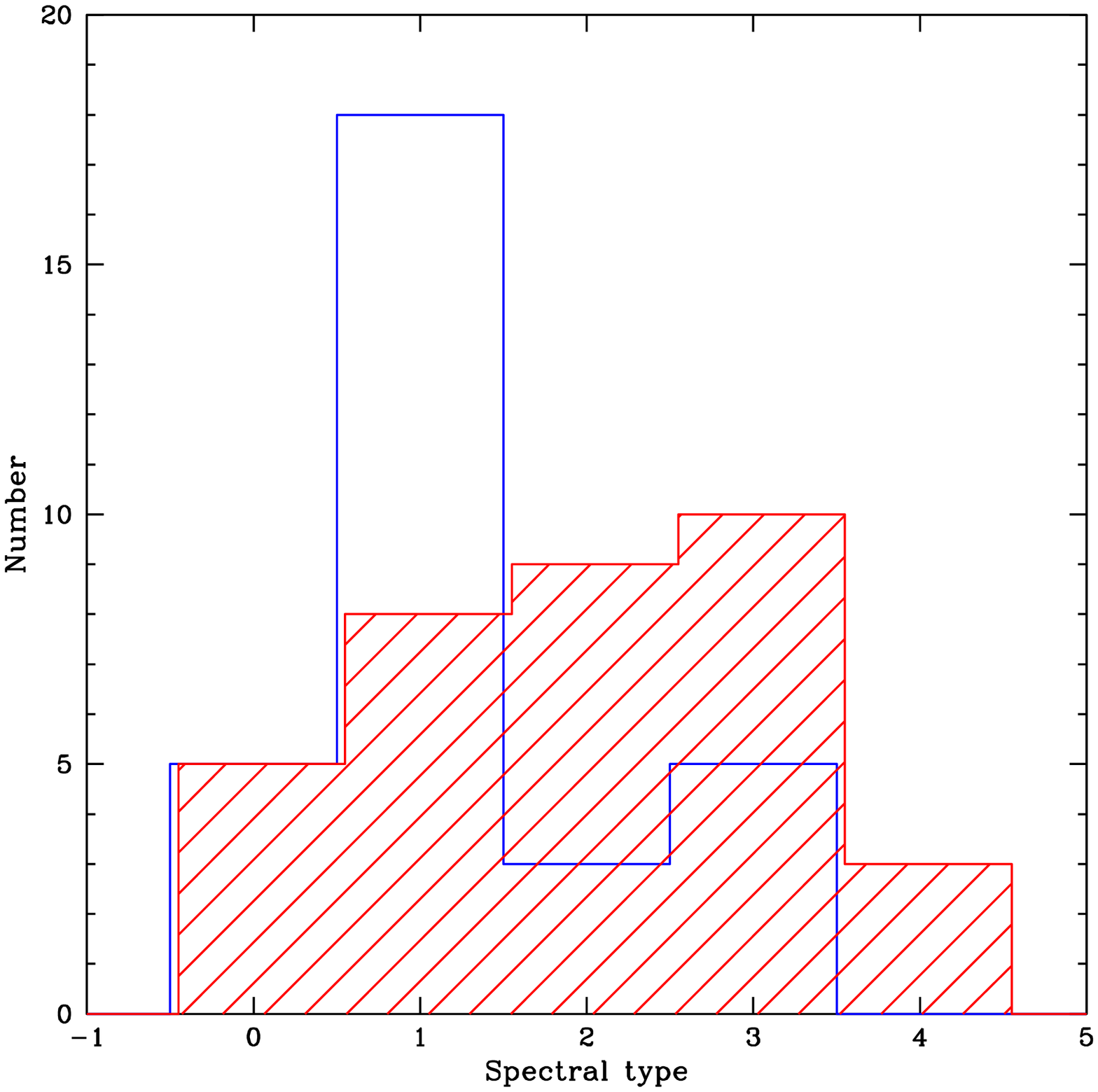}
\caption{Histograms of spectral types.  Blue histogram: objects that needed only a fit with a single old galaxy model. 
Red shaded histogram: a second component was needed. The spectral types are 
no emission lines (0), only weak emission - usually only $H\alpha$
- (1), LEG (2), HEG (3), HEG
with broad emission components (4). }
\label{fig14}
\end{figure}

\subsection{Distribution of spectral type and line luminosity}
\label{sptype}

One might expect that the majority of one-component fits corresponds to objects without significant emission features in their spectra, as can be seen 
in Fig. \ref{fig14}. 
However, this is no one-to-one correspondence because some one-component galaxies are of the HEG type. 
Therefore highly excited emission lines do not always guarantee a strong UV continuum component, 
although high $XUV$ values are seen 
more frequently in HEGs. Conversely, some of the objects that needed two components have no emission lines at all.

\begin{figure}
\centering
\includegraphics[width=9cm]{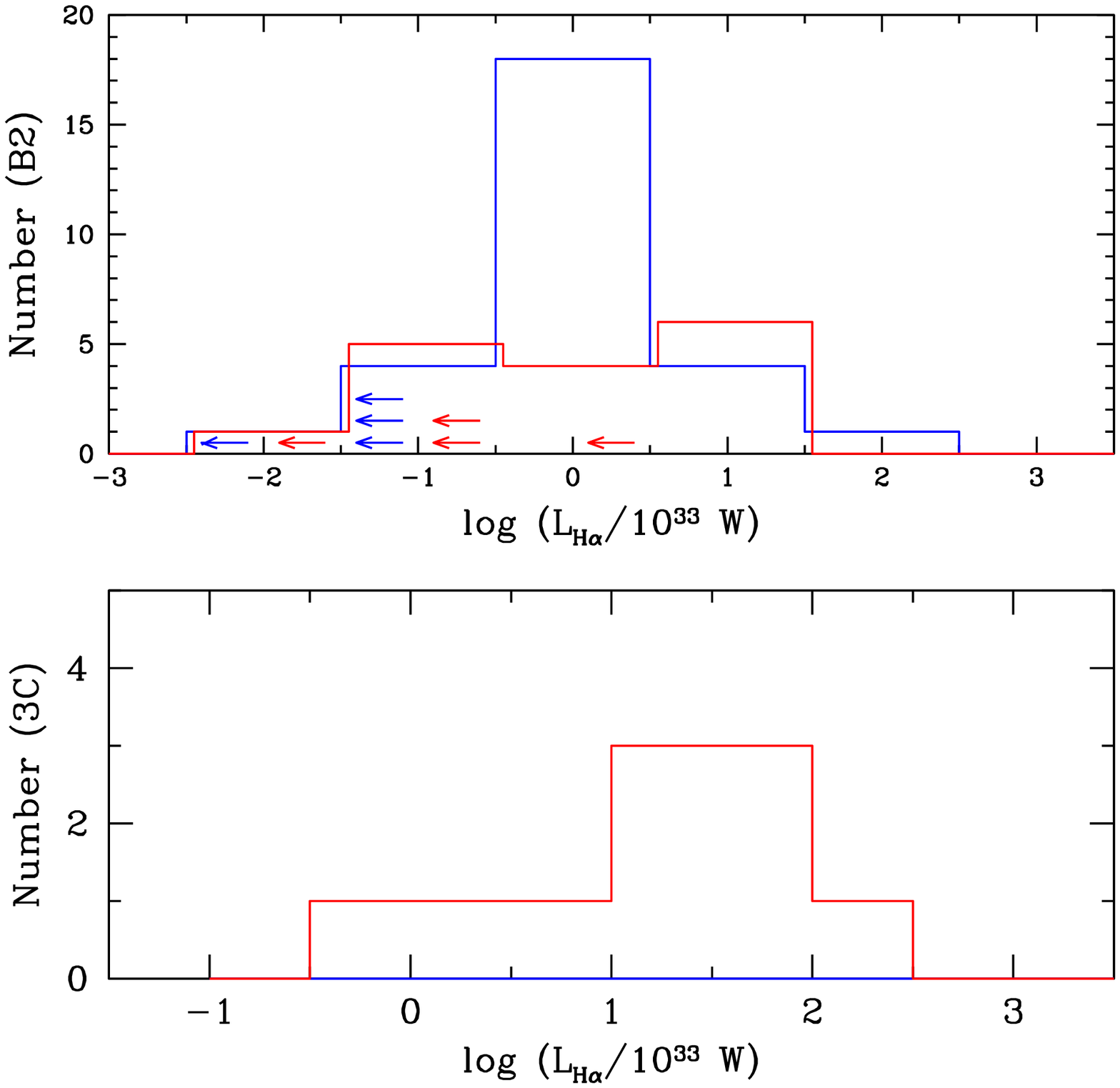}
\caption{Histograms of $H\alpha$ line luminosities. Upper panel: B2 galaxies; lower panel: 3C sources. Blue histogram: one-model fits. 
Red histogram: two-model fits.}
\label{fig15}
\end{figure}

We collected the data on the line luminosities of our sample of radio galaxies from the SDSS whenever possible. 
If necessary, the luminosities were measured directly from the SDSS spectra, with an accuracy that we estimate to be of the order of 5\%. 
In addition, there may be calibration errors, but these are hard to estimate. We used the strengths of a number of lines 
($H\alpha$, $[NII]$, $H\beta$ and $[OIII]$)  to determine the spectral type, but considered only the $H\alpha$ line luminosity in the further analysis.

\begin{figure}
\centering
\includegraphics[width=9cm]{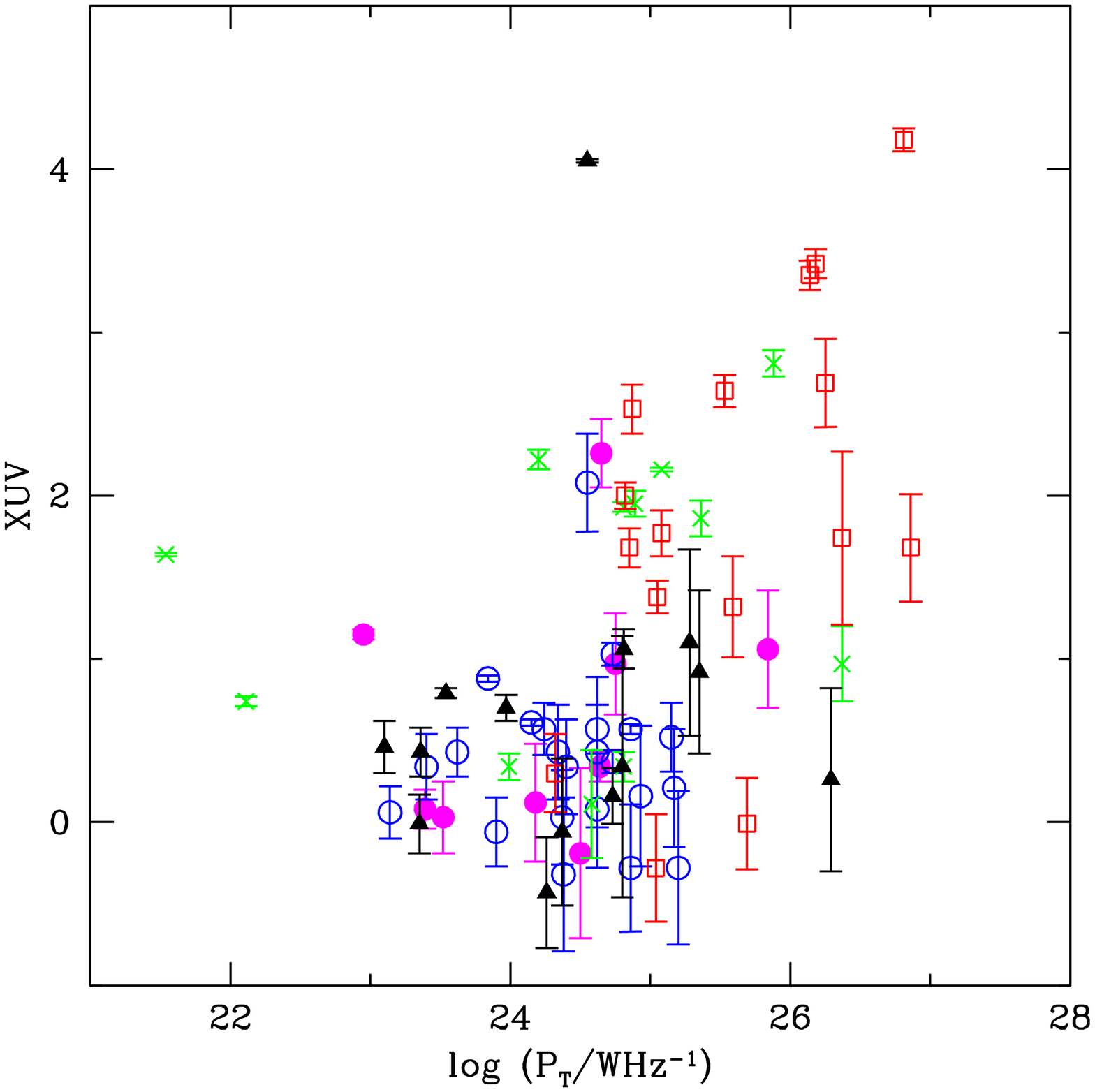}
\caption{Radio power at 1.4 GHz against UV excess ($XUV$). Points are (colour-)coded according to spectral type: (black) filled triangles if the 
spectral type is not known; (magenta) filled circles for objects without detected emission lines, (blue) open circles for objects with 
a weak emission line (usually only $H\alpha$), (green) crosses for LEGs, and (red) open squares for HEGs.}
\label{fig16}
\end{figure}

We show the histograms of the $H\alpha$ line luminosity for B2 and 3C galaxies in Fig. \ref{fig15}. 
As expected, two-component galaxies tend to have significant emission lines with $L(H\alpha)\sim 10^{34}-10^{35}$ W, 
particularly so the 3C radio galaxies. Therefore both the emission luminosity and the UV excess appear to behave (roughly) in a similar way, 
and both are somehow related to radio power and consequently FRI type. That this is true is shown in Figs. \ref{fig16} and \ref{fig17}. 
\subsection{Correlation of XUV with radio power}
\label{XUV-P}

In Fig. \ref{fig16} we plot $XUV$ as a function of radio power. 
The scatter is rather large, but there appears to be a clear dichotomy between low- and high-luminosity sources. Although there is a dividing line separating
FR I and FR II (see Ledlow and Owen \cite{ledlow96}), at the typical optical magnitude of the B2 sources this separation occurs at approximately 
$P=10^{24.5}$ WHz$^{-1}$: 
the sources below this value (indeed mostly FRI sources) have no significant UV excess, and the large majority have no or only weak line emission.
In contrast, the higher luminosity sources (mostly FRII, but also some transition types) may have a UV excess, but not necessarily very prominent. 
It rather looks as if  $XUV\approx 0$ in the FRI range (with perhaps a few rare exceptions). The $XUV$ distribution 
abruptly widens at the division of FRI and FRII. The correlation between radio power and $XUV$ is thus not simply linear over the whole power range, but more complex. 
As shown in Fig. \ref{fig16}, the high-luminosity galaxies predominantly are HEGs. 
We note, however, that the distinction is not sharp, because LEGs are in reality present over the whole range of radio power. 
Nevertheless, the LEGs follow the general trend that they mostly do not have an UV excess below $10^{24.5}$ WHz$^{-1}$, 
while they do at higher luminosities.
If we take only HEG sources, a linear regression gives $XUV = (0.85 \pm 0.36)\times (\log P_T -23.3)$, which means that the excess UV luminosity depends
only weakly on total radio power ($\propto P_T^{0.33}$), but is absent below $\log P_T=23.3$. The correlation is also not very significant, 
which is expected considering the large scatter in Fig. \ref{fig16}.

\begin{figure}
\centering
\includegraphics[width=9cm]{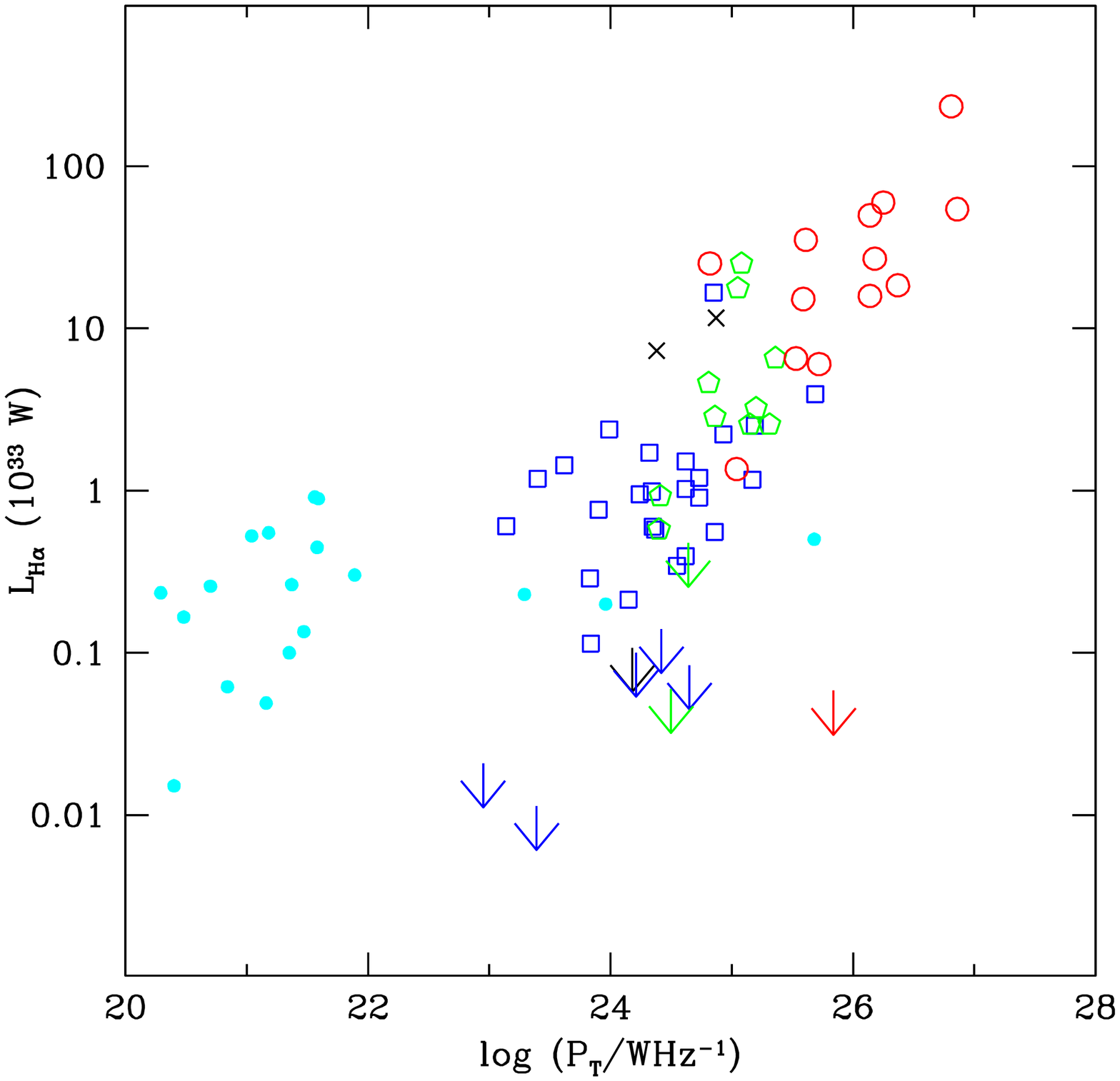}
\caption{Radio power at 1.4 GHz against $H\alpha$ line luminosity. We also
show data from the Sauron sample (see text) as (cyan) dots. Points are (colour-)coded according to FR type.
(Blue) open squares for FRI sources, (green) open pentagons for transition types (FRI-II), and red circles for FRII sources.
}
\label{fig17}
\end{figure}

The correlation between the $H\alpha$ luminosity and radio power is clearly more pronounced, as can be seen in Fig. \ref{fig17}. 
This well-known correlation is part of a more general correlation between the strength of various emission lines and radio strength 
(be it total power or jet luminosity); see e.g. Zirbel \& Baum (\cite{zirbel95}), Rawling \& Saunders (\cite{rawlings89}, Hine \& Longair (\cite{hine79}). 
In this figure  we have also plotted $SAURON$ sources (Bureau et al. \cite{bureau11}) that  have been detected in the radio. Here we also see
a clear separation in the properties of high- and low-luminosity sources.

\subsection{Correlation of XUV with $H\alpha$ luminosity}
\label{XUVHalpha}

\begin{figure}
\centering
\includegraphics[width=9cm]{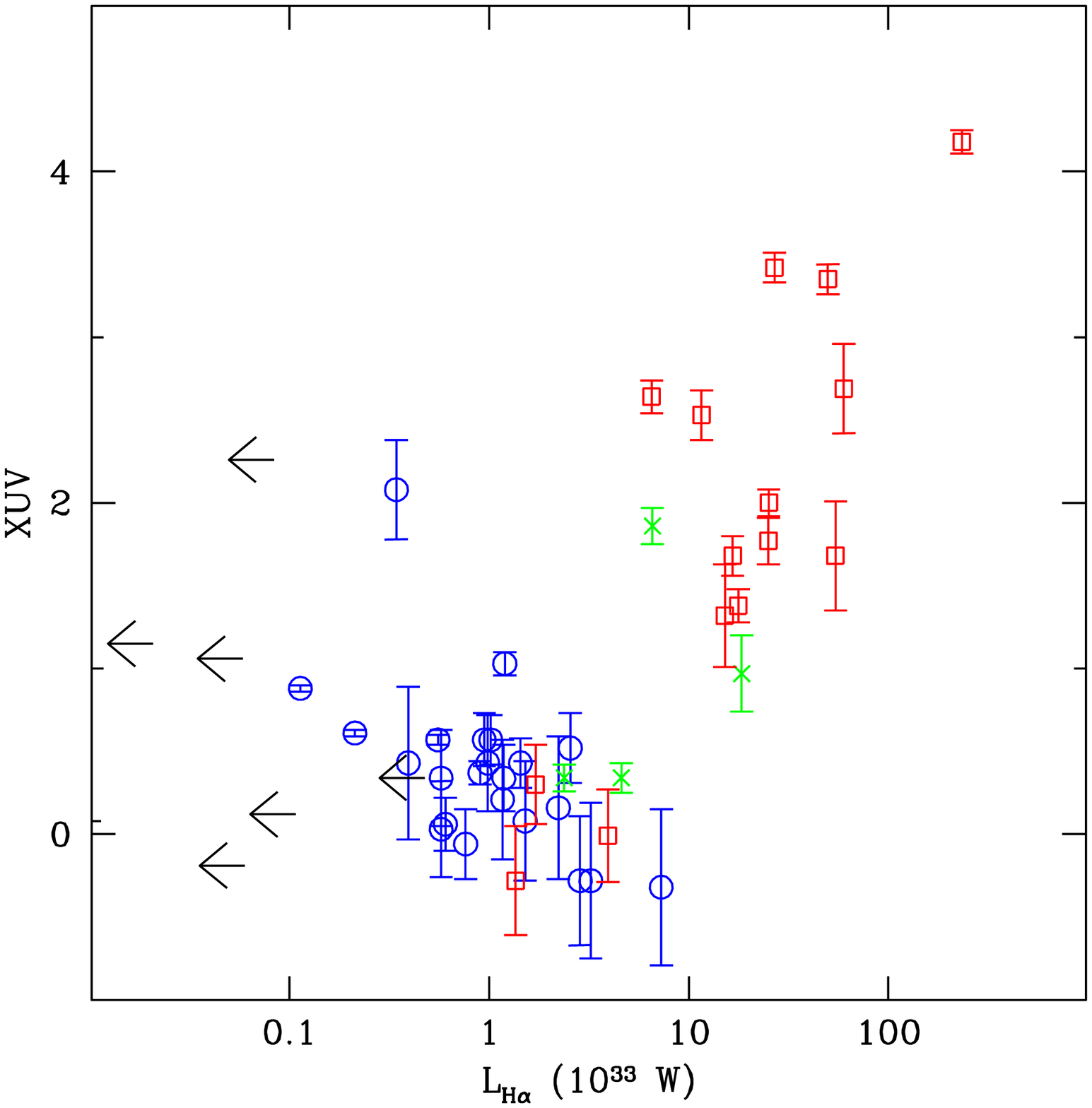}
\caption{$XUV$ as a function of $H\alpha$ line luminosity. Points are coded according to spectral type, as in Fig. \ref{fig16}.}
\label{fig18}
\end{figure}

The similarity in the behaviour of $XUV$ and $L(H\alpha)$ with respect to radio luminosity suggests that there may be a strong 
correlation between them. We show this in Fig. \ref{fig18}. 
To be more precise, there is a strong correlation for galaxies with $L(H\alpha) > 5\times 10^{33}$ W, but no correlation at all below that value. 
Therefore our data suggest that the dichotomy in $XUV$ between FRI and FRII is genuine and more profound than has been suspected up till now. 
The correlation between $XUV$ and $H\alpha$ line luminosity  is strictly true for HEGs and LEGs only (see  Fig. \ref{fig18}); 
below an $H\alpha$\ luminosity of $\sim 5\times 10^{33}$ W no significant $XUV$ appears to be present, although there is, admittedly, 
a large scatter in $XUV$, which is due to a few objects that have $XUV > 1,$ but very weak or even absent $H\alpha$\ emission.  

\begin{figure}
\centering
\includegraphics[width=9cm]{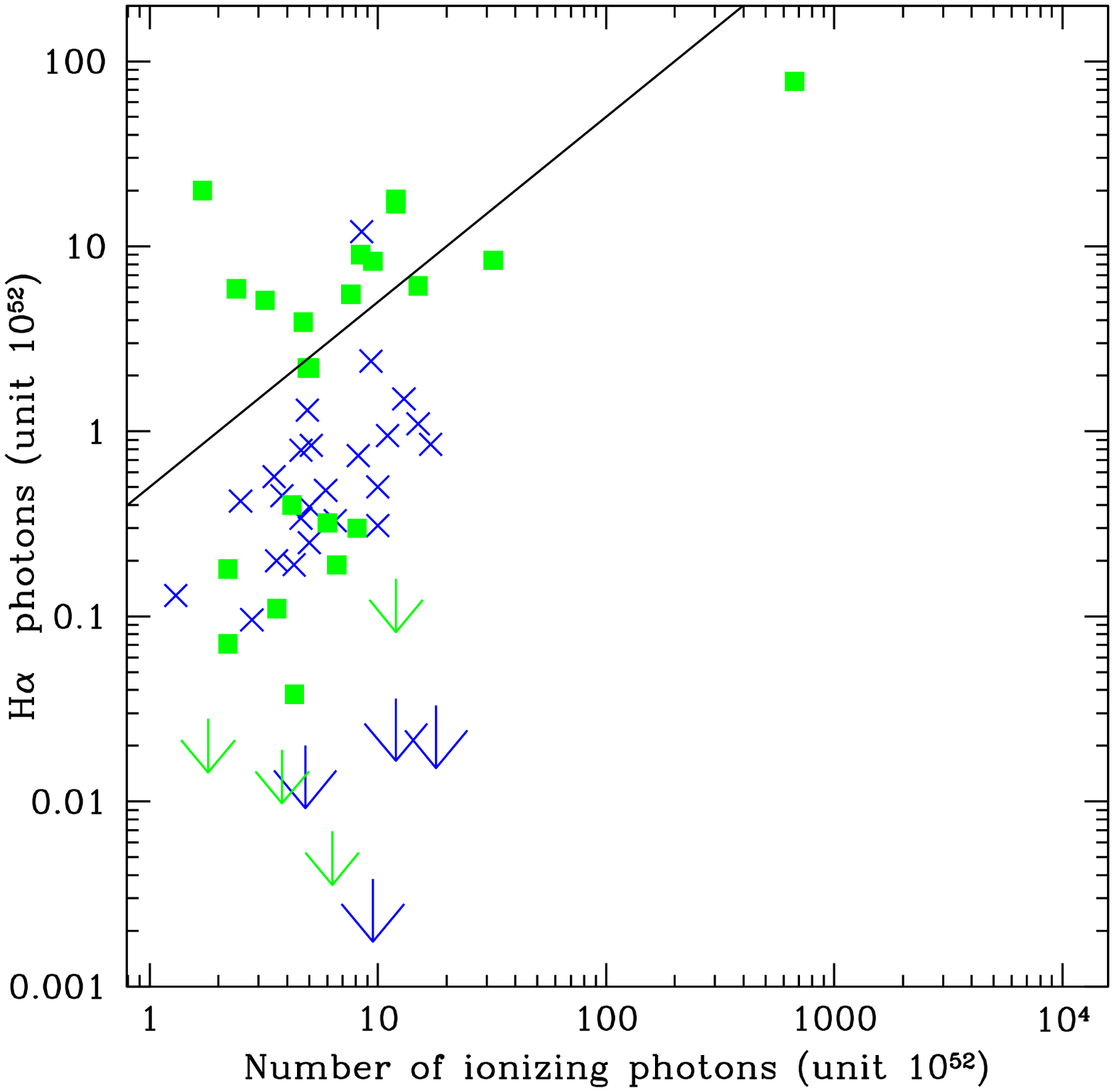}
\caption{Number of $H\alpha$ photons as derived from the line 
strength plotted against the total number of available ionizing photons.
As a rule of thumb, about two ionizing photons are needed to produce one $H\alpha$
photon (the black line). If a point is located above this line, there are not
enough ionizing photons to produce the $H\alpha$ emission. Blue crosses: galaxies
for which only one (old) galaxy model was needed to fit the observed SED. Green
squares: galaxies with a second younger galaxy component used in the model fit.
Only one blue cross (B2 1455+28) is above the line, 
but it is a single-component galaxy only due to the lack of UV information (see Sect. \ref{comments}). 
 }
\label{fig19}
\end{figure}

\subsection{Are there enough ionizing photons?}
\label{photons}

Baum, Zirbel \& O'Dea (\cite{baum95}) have suggested that the emission lines (including $H\alpha$) in FR I radio galaxies are produced directly by the parent
galaxy itself and not by the AGN. This possibility (albeit in non-radio elliptical galaxies) was discussed by Binette et al. (\cite{binette94}); therefore
we verify first that the emission lines observed in our FR I objects may be produced by the galaxy itself.

Is the ultraviolet radiation sufficiently strong to produce  $H\alpha$ in emission? We investigated this by calculating the amount of UV radiation capable 
of ionizing the gas, which can then recombine to produce $H\alpha$ photons. 
As a rule of thumb, there should be roughly two ionizing photons available to produce one $H\alpha$ photon 
(see e.g. Osterbrock \& Ferland \cite{osterbrock06}). 
Using the fitted galaxy models given in Table \ref{tab:results}, we calculated the number of available ionizing photons.
 
For objects that could be fitted by an old elliptical galaxy SED without the need to invoke any other component, 
we computed from the old
galaxy model the number of such photons  and compared this directly with the number of $H\alpha$ photons present. 
These objects are plotted in Fig. \ref{fig19} as blue crosses. 
 None of these galaxies except B2 1455+28 has very strong $H\alpha$ emission, and for that reason, 
they lie well below the line
$N_{\rm photons}(H\alpha) = 0.5\cdot N_{\rm photons}(\rm ionizing)$, so that there is no contradiction between available and 
requested number of ionizing photons. B2 1455+28 was fitted with a single model because GALEX data are lacking. However, it is known to be  a FRII radio
source associated with a HEG galaxy,  and not a typical FRI galaxy without strong optical emission lines.

The situation is slightly different for galaxies that needed a two-component fit. We plot these objects as green squares  in Fig. \ref{fig19}.   
It is clear that in a number of cases the old galaxy component plus a young stellar component alone are not able to produce a sufficient number of ionizing photons.

Of the 26 two-component galaxies, 9 exceed the ratio $N_{\rm photons}(H\alpha) = 0.5\cdot N_{\rm photons}(\rm ionizing)$. These are HEGs or broad-line
radio galaxies. In these cases it is plausible that part of the UV output  is due to the AGN and not to young stars. 
If we attribute even a minor fraction (in the range of 10 to 30 \%, for example)
of this radiation in the form of a power law, a realistic power-law slope ($1.3 \pm 0.3$) then results in a sufficiently high amount of $UV$ photons to explain the
H$\alpha$ strength.

\subsection{XUV and the mid- to far-IR}
\label{XUVIR}

Some articles concerning WISE data of radio galaxies have appeared recently. For example, Xiao-hong, Pei-sheng, \& Yan (\cite{xiao15}) discussed a large
sample of radio-luminous galaxies and concluded that galaxies in which star-formation is dominant are redder (in the 4.6-12 $\mu$m colour) than AGN 
dominated radio galaxies. There also is a separation of LEGs and HEGs in the WISE colour-colour diagram. G{\"u}rkan, Hardcastle, \& Jarvis (\cite{gurkan14})
also used WISE data. They found that quasars are stronger in the mid-IR than radio galaxies.
Is there any evidence in our data of a difference in IR properties between the various types of radio galaxies in our sample, and can this be related to the UV 
properties? 

Inspection of Fig. \ref{fig1}  reveals that the two shorter wavelength bands (3.4 and 4.6 $\mu$m, W1 and W2) of WISE 
closely follow the profile of the old stellar population, at least in the large majority of galaxies, while the two longer wavelength bands at 12 and 22 $\mu$m 
(W3 and W4) do not: W3 and W4 tend to flatten the SED beyond 10 $\mu$m. The WISE bands were not used in the model
fitting, so that W1 and W2 confirm that the SED follows that of an old stellar population out to about 10 $\mu$m.

\begin{figure}
\centering
\includegraphics[width=9cm]{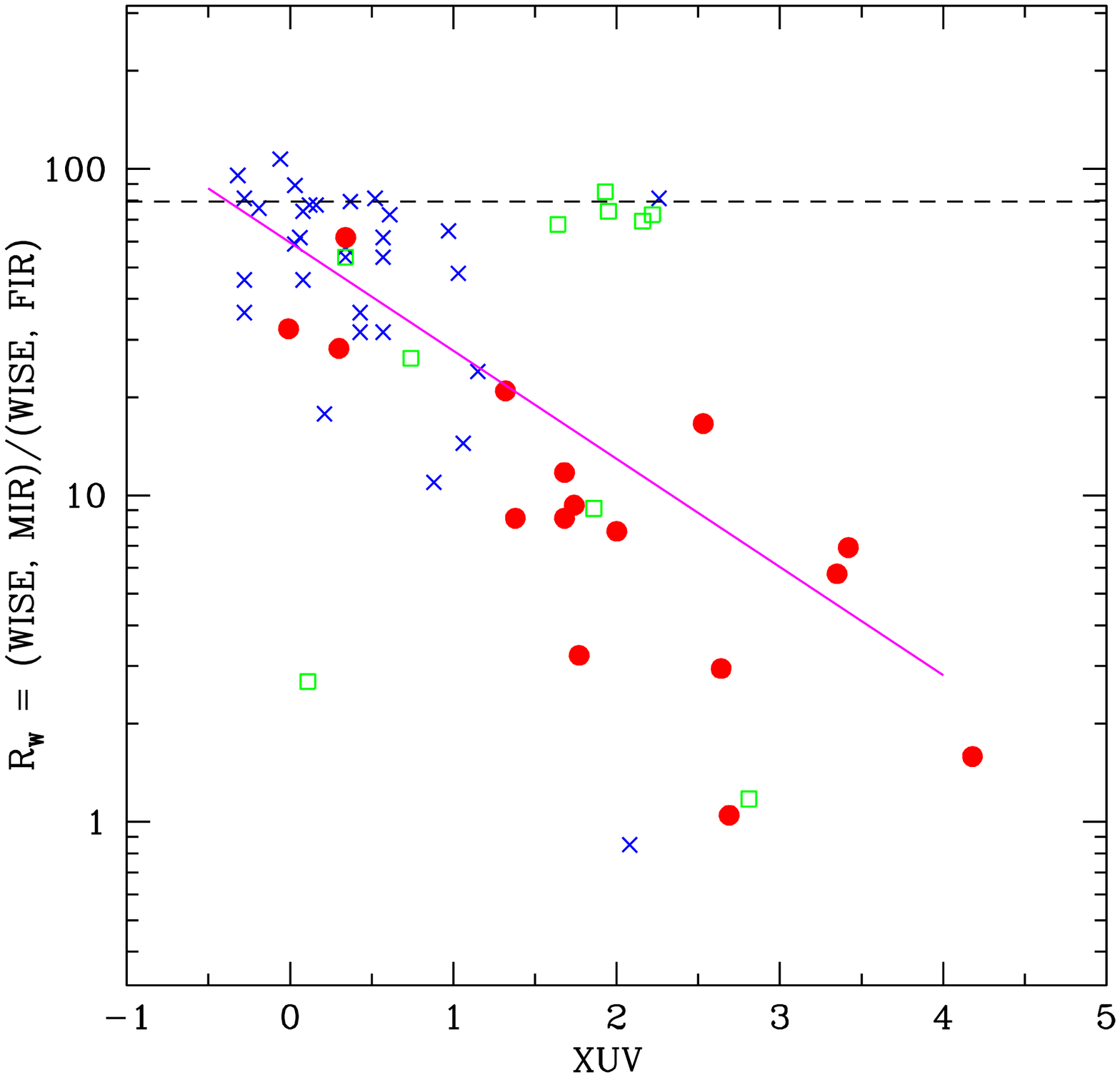}
\caption{Ratio $R_W$ of the shorter two and longer two wavelength bands of WISE plotted against $XUV$. Blue crosses are objects 
without significant emission lines in the spectrum (spectral type 0 and 1), green open squares LEGs (spectral type 2) and red dots are HEGs
 (spectral type 3 and 4, see Table \ref{tab:results}). The dashed horizontal line is the expected
$R_W$ for elliptical galaxies with only an old stellar population. The magenta line is the linear regression of $R_W$ against $XUV$ (see text).
}
\label{fig20}
\end{figure}

Is the flattening beyond 10 $\mu$m correlated with the strength of the UV excess? To investigate this, we constructed the parameter $R_W$, 
which is the ratio of the mean
luminosity of the W1 and W2 bands, and the mean of W3 and W4, that is, $R_W = (L_{W1}+L_{W2})/(L_{W3}+L_{W4})$. We plot $R_W$ against $XUV$
 in Fig. \ref{fig20}.
   
The correlation seen in Fig. \ref{fig20} is quite striking. A linear regression based on all points gives $$\log R_W = 1.77(\pm 0.07) - 0.33(\pm 0.05)\times XUV,$$
which is very significant (at the level of almost $7\sigma$). Since $XUV$ is expressed in magnitude units, this result means that the excess IR luminosity 
is practically directly proportional to the excess UV luminosity. 
At $XUV=0, $ the W1 and W2 luminosities
are almost one hundred times higher than those of W3 and W4. This is what would be expected for an SED based purely on an old stellar population up to $\sim 20 \mu$m. 
Such objects show hardly any emission features in their spectrum, in other words, they are standard elliptical galaxies. 
However, at higher $XUV,$ the ratio $R_W$ progressively
diminishes and becomes $\sim 1$ for the objects with the highest $XUV$. Such objects are mostly of the HEG type. We can conclude that the UV excess and the
IR flattening are closely related. All objects appear to have the same behaviour: if we consider only HEGs,  the linear regression remains practically the same
and is still highly significant ($4.5\sigma$) , even if there
are fewer objects.

For some objects (shown at the end in Fig. \ref{fig1}), additional far-IR data, up to 100 or even 200 $\mu$m, are available. They can be easily 
interpreted as radiation from dust with temperatures in the range of a few tens up to almost 1000 K. According to Dicken et al. (\cite{dicken10}; 
\cite{dicken12};
\cite{dicken14}), the warm dust must be heated by the AGN and not by young stars. They mainly used powerful radio galaxies, and these also show 
conspicuous  dust emission in our sample. Dicken et al. (\cite{dicken12}) stated that only about 35 \% of their sample galaxies showed signs of recent
star formation. As we show in Sect. \ref{discussion}, this is not true for the present sample, which contains many more FR I radio galaxies.

\subsection{XUV and black hole masses}
\label{XUVBH}

\begin{figure}
\centering
\includegraphics[width=9cm]{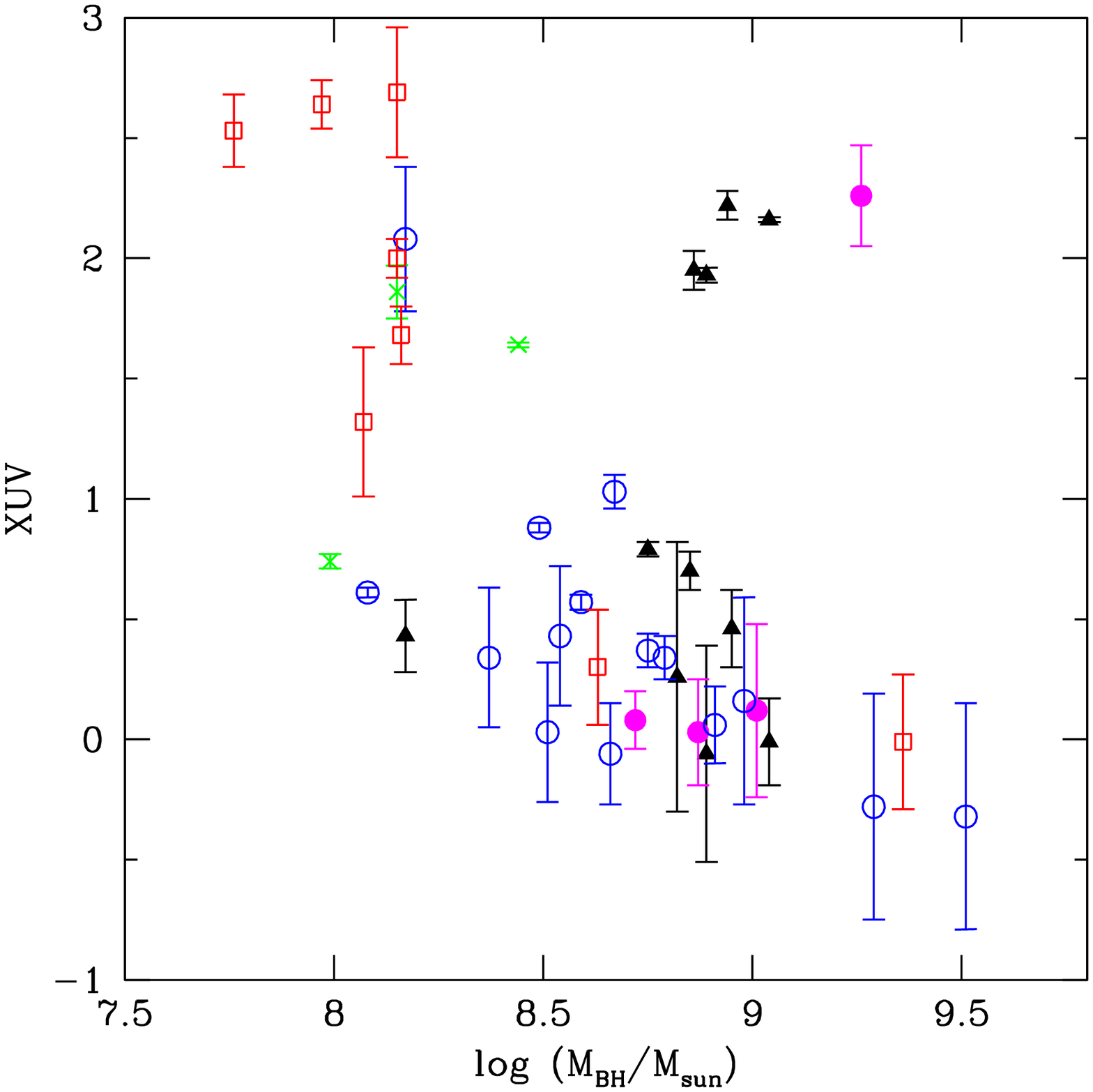}
\caption{Central black hole masses as a function of UV excess ($XUV$). Points are (colour-)coded according to spectral type, as in Fig. \ref{fig16}.}
\label{fig21}
\end{figure}

It is known that the more massive central black holes are found in low-luminosity radio sources (see for a detailed discussion Best \& Heckman \cite{best12}), 
and, considering our previous results
on the dependence of $XUV$ on radio power, we may expect that there is also a dependence of black hole mass on $XUV$. 
Using the data given in Table \ref{tab:results}, we plot the
black hole mass against $XUV$ for objects with known values of both parameters.  
Some interesting trends can be seen in Fig. \ref{fig21}. If we consider only HEGs (open squares), there is a strong anti-correlation: in galaxies without 
a UV excess ($XUV < 1$, for instance) the black hole (BH)\ mass tends to be high, of the order of $10^9 M_{\sun}$, while the galaxies with $XUV > 1$ 
have $M_{BH} \sim 10^8 M_{\sun}$. Of the galaxies with only weak emission lines (open circles), all except one
have $XUV \le 1$ and cover on average the same region as the HEGs without large UV excess. We note that a similar but much weaker anti-correlation 
between BH mass and $XUV$ may also be present for these objects.

A group of objects deviates from this trend, with black hole masses of the order of $10^9 M_{\sun}$, 
but also strong UV excess, of the order of 2  magnitudes.  Their UV excess is genuine, however.

\begin{figure}
\centering
\includegraphics[width=9cm]{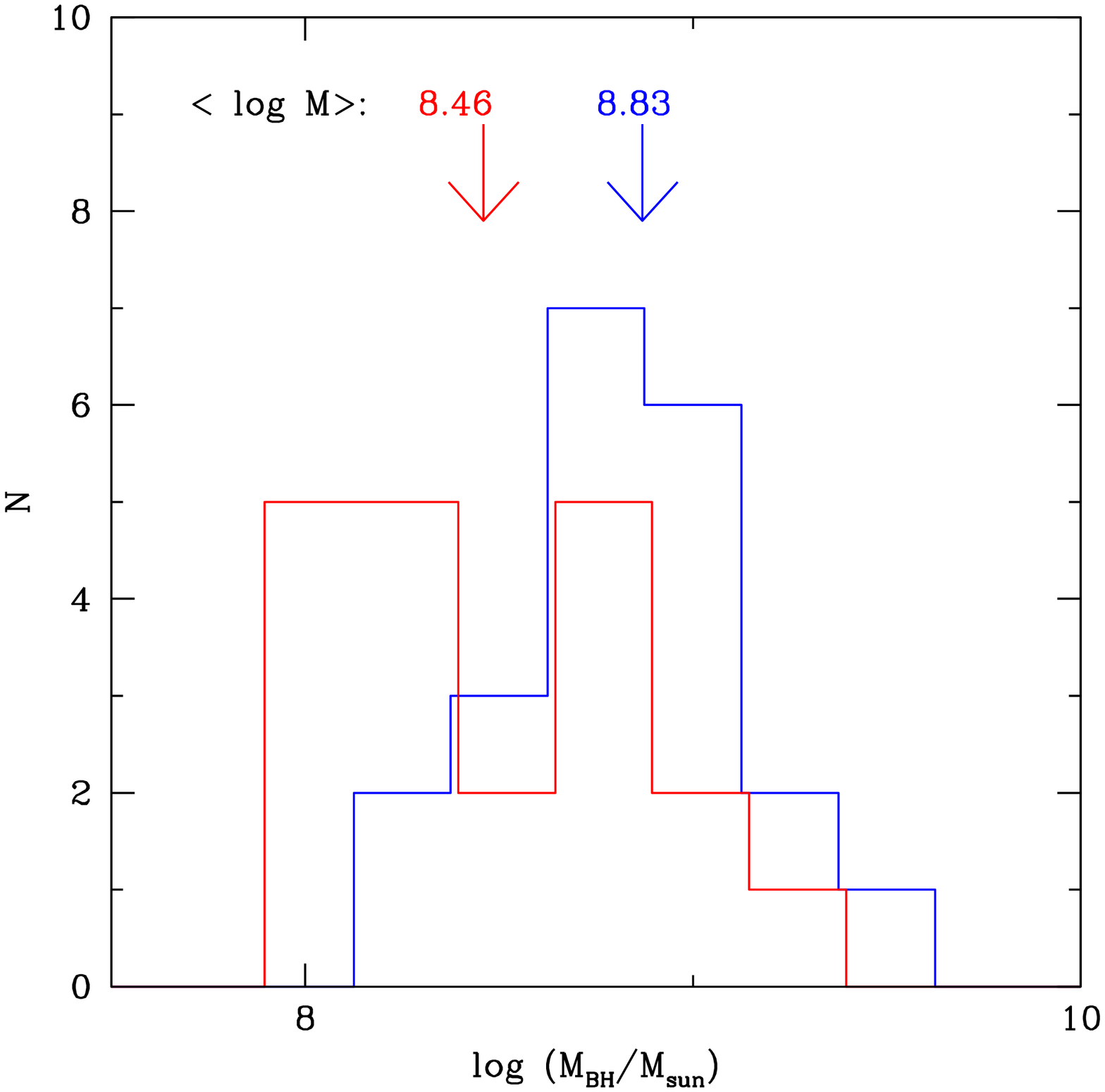}
\caption{Histogram of central BH masses, according to $XUV$ value.
Blue histogram: BH masses of galaxies with $XUV < 0.6$. Red histogram:
galaxies with $XUV \ge 0.6 $. Mean values are
$\log M_{BH}/M_{sun} = 8.83 \pm 0.07$ and $\log M_{BH}/M_{sun} = 8.46 \pm 0.09$
for galaxies without and with a UV excess, respectively. The difference is
significant at the $3.2\sigma$ level.}
\label{fig22}
\end{figure}

\begin{figure}
\centering
\includegraphics[width=9cm]{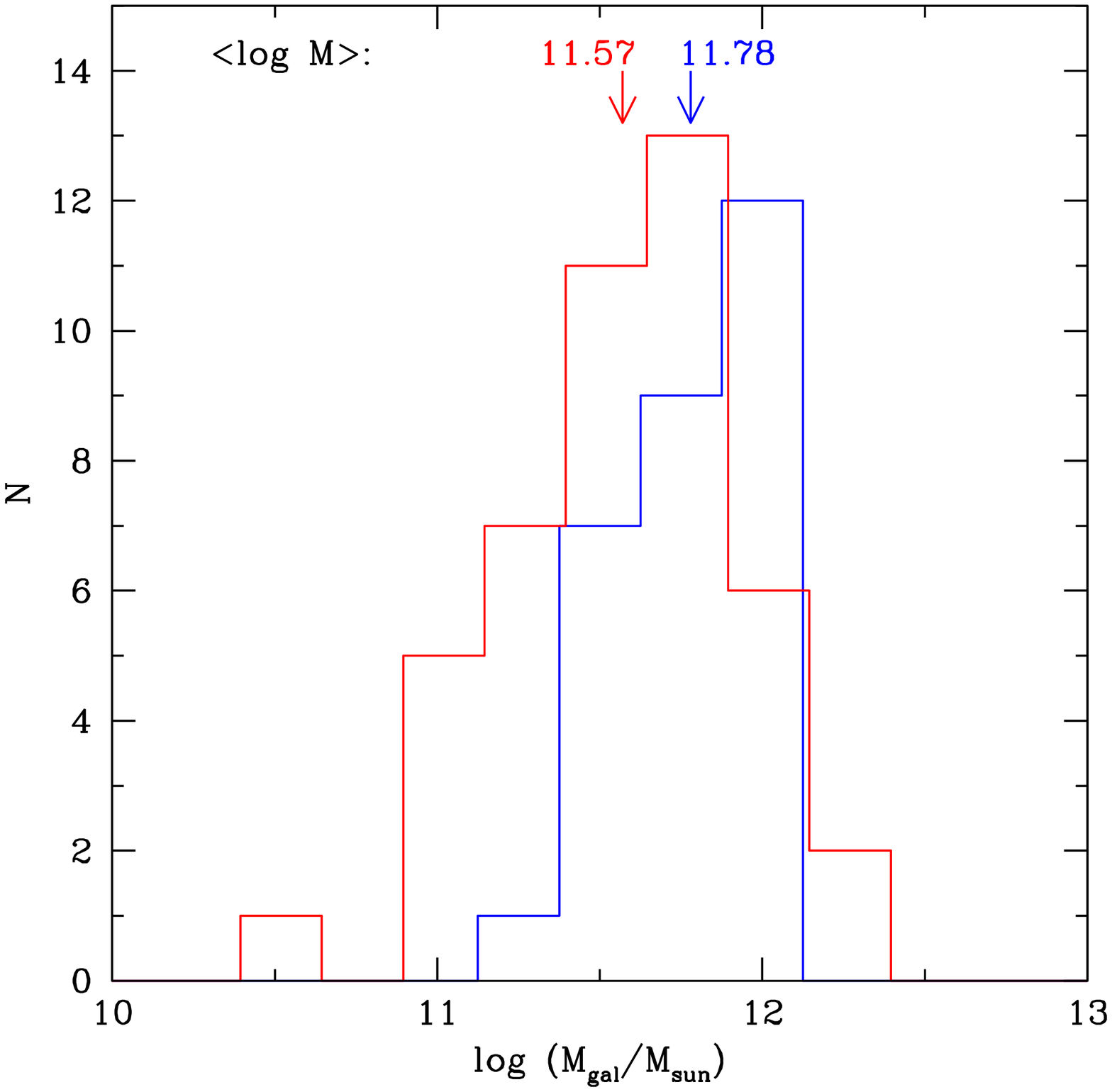}
\caption{Histogram of galaxy masses derived from SED fits (mass of main galaxy
component) according to the XUV value. Blue histogram: masses of galaxies with
$XUV < 0.4$. Red histogram: galaxies with $ XUV \ge 0.4 $. Mean
values are $\log M_{gal}/M_{sun} = 11.78 \pm 0.04$ and  $\log M_{gal}/M_{sun} = 11.57 \pm 0.05$, respectively.
The difference is significant at the $3.3\sigma $ level.}
\label{fig23}
\end{figure}

In spite of this complication, higher values of $XUV$ tend to be accompanied by lower black holes masses, as can be seen in Fig. \ref{fig22}.
Here we plot two histograms of objects with $XUV < 0.6$ and $\ge$ 0.6. The particular value of 0.6 was chosen to obtain similar numbers in
the two histograms. The distributions are clearly different. Objects with $XUV$ not very different from zero have $\log M_{BH}/M_{\sun} = 8.83 \pm 0.07$, 
while galaxies with higher $XUV$ have $\log M_{BH}/M_{\sun} = 8.46 \pm 0.09$, a difference at the $3.2\sigma$ level. 
Thus, on average, a high $XUV$  is accompanied by a black hole mass that is about 2.5 times lower than the black hole 
masses of galaxies without an additional UV component.

It is known that galaxies hosting FRII radio sources tend to be less massive (Best \& Heckman \cite{best12}; Kauffmann et al. \cite{kauffmann08}). 
More precisely, Kauffmann et al. (\cite{kauffmann08}) found lower galaxy masses for galaxies that display (strong) emission lines, that is, for HEGs.
As remarked by Best \& Heckman (\cite{best12}), the different spectral types are also related to the FRI/FRII type (see also Fig. \ref{fig17}), but the precise 
relations between all these parameters remains uncertain. Because
of the limited amount of data used in this paper, we cannot provide more detail, 
except to draw attention to the 
fact that we also found a difference in galaxy mass  between objects with high and low $XUV$ values. 
This is shown in Fig. \ref{fig23}; we have taken a slightly different separation value of $XUV$ to again obtain a similar number of objects in the two histograms. 
As in Fig. \ref{fig22}, we obtain a difference at a similar significance 
level (in this case $3.3\sigma$). This is therefore in line with the findings of Kauffmann et al. (\cite{kauffmann08}) and Best \& Heckman (\cite{best12}), if we
remember that high $XUV$ values correspond roughly to HEG and LEG type spectra, which in turn approximately correspond to FRII-type radio sources.

The histograms of Figs. \ref{fig22} and \ref{fig23} suggest that there may be a relation between $M_{gal}$ and $M_{BH}$, which would of course be expected on the
basis of the famous relation between $M_{bulge}$ and $M_{BH}$ (Ferrarese \& Merritt \cite{ferrarese00}; Marconi \& Hunt \cite{marconi03};
H\"aring \& Rix \cite{haring04}). 
Indeed, there is such a relation, but this is strictly valid only for objects with weak emission lines. Moreover, the scatter is quite large, 
and this is easy to infer from the histograms in Figs. \ref{fig22} and \ref{fig23}.

\section{Discussion}
\label{discussion}

The first far-UV detections of a few selected galaxies were made about 45 years ago, when dedicated satellites (like the Orbiting Astronomical Observatory-2; 
see Code \cite{code69}) opened up the UV window. It was discovered very soon that some elliptical galaxies surprisingly showed an increase in flux 
from the visible to
the far UV (Code \& Welch \cite{code79}; Bertola et al. \cite{bertola80}; a recent introduction on the subject can be found in Jeong et al. \cite{jeong12}). Various
explanations have been put forward, usually based on specific stellar populations or metallicity effects. It is not clear whether this is a general property of early-type 
galaxies that is
also shared by other types of ellipticals, for example those that are radio-loud or X-ray bright. Trump et al. (\cite{trump09}) found that about 30 \%  of X-ray bright, 
but optically
"dull" early-type galaxies (passive red galaxies) exhibit weak but significant blue emission, presumably originating in the (unabsorbed) nucleus. 

That UV emission and the morphology of radio sources may be closely connected has been known since the work of McCarthy et al. (\cite{mccarthy87}): 
they found that in
distant radio sources the UV emission is aligned with the radio structure. However, this result is based on very high redshift sources.   

Only very recently, far-UV  data are routinely available, thanks to the very powerful GALEX mission, and therefore different types 
of radio-loud objects can now be studied easily down to about 1000 \AA.
Fanti et al. (\cite{fanti11}) performed such a study and analysed compact steep spectrum (CSS) radio sources. They concluded that an ultraviolet excess 
(with respect to the SED of
a passively evolving elliptical galaxy) is present in most of these galaxies. This UV excess is, according to them, due to a young stellar population, which is not directly 
caused by the radio source: both the young stars and the radio source are the result of a recent wet major merger.

We here analysed the overall SED of low-redshift ($z<0.2$) B2 and 3C radio galaxies from the far-UV to the near-IR. It is immediately clear 
(Fig. \ref{fig16}) that
weaker radio galaxies, basically FR I type radio galaxies, show little evidence for a UV excess. These FRI  galaxies mainly have few, if any, emission lines 
(except for
a few LEGs) and can therefore be considered to be the radio equivalent of the X-ray bright optically dull galaxies of Trump et al. (\cite{trump09}). 
However, the UV behaviour is quite different because the weak but significant UV component seen by Trump et al. (\cite{trump09}) is absent from the "optically dull" 
FRI radio galaxies.

Several parameters appear to be closely linked, viz. the strength of the emission lines, or more specifically, the spectral type, the UV excess as measured by our 
parameter $XUV,$ and the radio power. The parameter correlation is shown in Figs. \ref{fig16} - \ref{fig18}. These figures strongly 
suggest that there is
a fundamental difference between FRI and FRII sources, as the behaviour in the various plots is very different below and above $10^{24.5}$ WHz$^{-1}$, which is the
standard luminosity (at 1.4 GHz) that separates FRI from FRII sources. Below this, there is no correlation at all in any of the figures, while above this, all plots show a 
rather abrupt
onset of UV excess and strong line luminosities. Especially remarkable is the close correlation of $XUV$ and $H\alpha$ luminosity (if $>5\times10^{33}$ W), while at
lower $H\alpha$ luminosity a correlation is entirely absent. Therefore the spectral features typical of HEGs, and less obviously LEGs, appear to be closely related
to an excess UV flux. In fact,  Baldi \& Capetti (\cite{baldi08}) concluded that only HEGs are associated with powerful AGN, probably triggered by a recent wet 
major merger. 
They ascribed the enhanced UV flux to recent star formation and stated that LEGs do not show enhanced SF with respect to quiescent galaxies. 
We note, however, that B2 1141+37, B2 1430+25, B2 1350+31 (3C 293), and 3C 321 (Holt et al. \cite{holt07}; Tadhunter et al. \cite{tadhunter11}) 
are LEGs with enhanced star formation (see Table \ref{tab:sfagn}).

So far, we have only indirectly dealt with the question of AGN vs. star formation. However, for a number of galaxies more information exists that can be
used to obtain a more definite answer. To this end, we used four different methods, the results of which are summarized in Table \ref{tab:sfagn}. 
In Cols. 1 and 7 we list the names of the radio galaxies.
We inspected the SDSS spectra, if available, to determine whether there are (post-)starburst signs. In some cases, higher Balmer lines 
in absorption are clearly present. These galaxies are denoted by "SF" in  Cols. 2 and 8 ("Balmer"). 

 \begin{table*}
      \caption[]{Star formation or AGN?}
         \label{tab:sfagn}
     $$
         \begin{tabular}{lccccclccccc}
            \hline
            \noalign{\smallskip}
            \hline
           \noalign{\smallskip}
           & \multicolumn{5}{c}{Method} & & \multicolumn{5}{c}{Method} \\     
B2/3C & Balmer & Diagn. & FIR/MIR & P.L. & Overall & B2/3C & Balmer & Diagn. & FIR/MIR & P.L. & Overall \\
     1     &     2        &     3       &       4         &   5    &       6       &    7       &     8          &     9       &     10        & 11  &     12   \\
\noalign{\smallskip}
\hline
\noalign{\smallskip}
0648+27\rlap{$^{\mathrm d}$}   & SF & AGN  &        &      & SF  & 1621+38    &   & &  AGN &        & AGN  \\
0755+37\rlap{$^{\mathrm c}$}   &       & SF &        &      & SF  & 1626+39\rlap{$^{\mathrm c}$}    & No & SF & SF      & SF & SF \\
0836+29   &       & SF &        &      & SF  & 1652+39\rlap{$^{\mathrm d}$}    & No & &            &        & AGN \\
0913+38   &       &       & Mix &      & Mix & 1658+30A\rlap{$^{\mathrm c}$} &   & SF &            & SF  & SF \\
1003+26   &       & SF &        &       & SF   & 192\rlap{$^{\mathrm d}$}          & ?   & AGN  & AGN\rlap{$^{\mathrm a}$} & SF  & AGN \\
1005+28   &       &       &        & SF & SF  &   219\rlap{$^{\mathrm d}$}           & ?   & AGN  & AGN & AGN & AGN \\
1037+30\rlap{$^{\mathrm d}$}   & SF & SF &         &       & SF  &   223\rlap{$^{\mathrm d}$}           & No & AGN  &          & SF    & Mix \\
1141+37   & SF &      &         &       & SF  &   234\rlap{$^{\mathrm d}$}           & No & AGN  & AGN & Mix & AGN \\
1144+35\rlap{$^{\mathrm d}$}   & ?    & AGN  &         &       & AGN    &   264   & ?    & SF   &           & Mix & Mix \\
1204+34\rlap{$^{\mathrm d}$}   & ?    & AGN  &         & SF & Mix &   272.1        & No &   & SF    & Mix & Mix \\ 
1217+29\rlap{$^{\mathrm c}$}   & No & SF     & Mix &  SF & Mix &  274\rlap{$^{\mathrm c}$}           & ? & SF  &          &  SF  & SF \\
1254+27   & ?    &      &         &        & ?     &  285\rlap{$^{\mathrm d}$}       & SF\rlap{$^{\mathrm b}$} & AGN   & Mix\rlap{$^{\mathrm a}$}  &  SF  & SF \\
1339+26   & ?    &      &         &        & ?     &  296          & No  & SF  &          & Mix & Mix \\
1346+26\rlap{$^{\mathrm c}$}   &       & SF &        &  SF  & SF  &  303\rlap{$^{\mathrm d}$}        & No  & AGN  &          & Mix & Mix \\
1350+31\rlap{$^{\mathrm c}$} & SF\rlap{$^{\mathrm b}$} & SF & SF\rlap{$^{\mathrm a}$}  &         & SF  &  305\rlap{$^{\mathrm d}$}          & SF\rlap{$^{\mathrm b}$} & AGN  & SF    &        & Mix \\
1430+25   & SF & SF &         &  SF & SF  &  321\rlap{$^{\mathrm c}$}        & SF\rlap{$^{\mathrm b}$}      &   & Mix\rlap{$^{\mathrm a}$}   &        & Mix \\
1613+27   &       & SF &         & SF  & SF  &  326\rlap{$^{\mathrm c}$}          & ?    & Mix   &           & SF & SF \\
1615+32\rlap{$^{\mathrm d}$}   & No & AGN & Mix  & AGN & AGN  &  346\rlap{$^{\mathrm d}$}  &        & AGN    &           & AGN & AGN \\
 \noalign{\smallskip}
 \hline
         \end{tabular}
     $$ 
\tablefoottext{a} Mid/Far-IR data from Dicken et al. (\cite{dicken10}), at 25 and 60 $\mu$m. All other data refer to 24 and 70 $\mu$m.
\tablefoottext{b} Tadhunter et al. (\cite{tadhunter11}).
\tablefoottext{c} LEG.
\tablefoottext{d} HEG.

   \end{table*}

We also used emission-line diagnostics (see Table \ref{tab:results}). If a galaxy has a typical HEG or Seyfert-like spectrum, it is classified as AGN in Cols. 3 and 9.
Some galaxies with classification "H" for an HII-like spectrum are also listed in Cols. 3 and 9 ("Diagn."). We note that  many FR I galaxies have very few if any emission lines
and thus cannot be classified. 

Another empirical test, first introduced by de Grijp et al. (\cite{degrijp85}) and revisited by Dicken et al. (\cite{dicken12}), concerns the ratio of
far- and mid-IR. If this ratio is higher than 5, it indicates star formation, below 2.5 it indicates an AGN, and in between a mixture between the two. We show the
result in Cols. 4 and 10 ("FIR/MIR").

Finally, we can make use of the strengths of the nucleus in the optical and radio, provided we have HST data, which we have taken for a number of the B2 
and 3C sources from Capetti et al. (\cite{capetti02}) and Chiaberge, Capetti, \& Celotti (\cite{chiaberge99}).  Assuming a power-law and extrapolating the observed
HST strength down to 2000 \AA\  using the mean spectral index 1.3 found between the radio and HST, we compared this with the UV excess found at 2000 \AA. If a
power law reaches $\sim 50$ \% of the observed UV excess, we conclude that the radiation from an AGN is dominant.
We note
that the choice of 1.3 for the spectral index is not critical if we assume that a non-thermal power-law slope should be in the interval 1-1.5. Any variation will 
not be larger than 30 \%. 
The results are given in Cols. 5 and 11 ("P.L."). An overall
assessment, based on these four methods, is given in Cols. 6 and 12 ("Overall").    

Most HEG sources in this table tend to have IR emission originating in AGN heated dust, while in  other types of sources (with weak emission lines) the
origin tends to be young stars.  There are a few exceptions, however, like 3C 285. In the large majority of FR I sources, there is no sign of an AGN continuum either
in the UV or in  IR, and at most some star formation may be present. 

The available data on black hole masses point to a relatively clear anti-correlation between $XUV$ and black hole mass (see Fig. \ref{fig21}): high black hole masses 
tend to have no UV excess, while the highest $XUV$, $>1$, for
example, on average have $M_{BH}$ ten times lower ($10^8$ instead of $10^9$). The only
deviation comes from a group of five galaxies (four of which have no known spectral type, while the fifth does not have emission lines), with relatively high $XUV$ 
and high
black hole mass. In any case, the correlations involving the mass of black hole or galaxy mass have a large scatter, as can be inferred from Figs. \ref{fig21} and \ref{fig22}.
This is to be expected, since in the mass (black hole or galaxy) - radio power plane the FR I and FRII sources occupy a wide range in both these parameters (see
Ghisellini \& Celotti \cite{ghisellini01}), while the dividing line between FRI and FRII is tilted in a way such that radio sources remain FR I type at high radio powers if
the galaxy mass (and by extrapolation the black hole mass) is higher (Ledlow \& Owen \cite{ledlow96}). The net result would be that FRI sources have on average a
higher galaxy and black hole mass than FR II sources, even if the scatter should be large since both types cover large parts of the mass-radio power plane. Therefore
the systematic difference in black hole mass and galaxy mass discussed by Kauffmann et al. (\cite{kauffmann08}) and Best \& Heckman (\cite{best12}) 
may be a selection effect.  

The  FRI-FRII dichotomy may be directly related to different accretion rates (Ghisellini \& Celotti \cite{ghisellini01}), which in turn may be due to
the merger history of the parent galaxy (Baldi \& Capetti \cite{baldi08}); therefore low- and high-luminosity radio sources are fundamentally different not only in their radio and optical properties, but also in the way they were formed.

\section{Summary}
\label{summary}
We summarize here the main points that have been treated in this paper.

\begin{itemize}
\item We used GALEX, SDSS, and 2MASS photometry to fit SEDs of nearby radio galaxies taken from a compound B2/3C sample. We first attempted a fit with
a model of a passively evolving old elliptical galaxy. 
In many cases, a second component
was needed to produce a satisfactory fit (we tried both a young stellar component and a power-law type). 
\item We used the parameter $XUV$  to quantify the excess in the UV over the basic, old,  passively evolving galaxy SED (usually $0.5-1.3 \times 10^{13 }$ years old).
\item The behaviour of $XUV$ is very similar to that of the emission lines strength as measured by $H\alpha$. Both correlate with radio power in the FRII regime.
\item At low radio powers no correlation is found at all, $XUV$ basically being close to zero. This same kind of behaviour is also seen for the $H\alpha$ luminosity. 
\item In contrast with X-ray bright "optically dull" galaxies (Trump \cite{trump09}), for instance, in the FRI dull galaxies (without a trace of AGN or SF activity in the optical spectrum), 
no UV excess radiation is seen.
\item In many FRII/HEG radio galaxies with strong $H\alpha$ emission, a young stellar component may result in a satisfactory fit, but this does not necessarily 
provide enough ionizing UV photons that would justify the strength the $H\alpha$ emission.
\item A  correlation exists between the UV excess, as measured by $XUV$, and the ratio of near- and far-IR WISE bands: flattening of the SED in the IR is accompanied
by an increasingly strong $XUV$. This correlation is close to linear, in the sense that the excess UV luminosity is directly proportional to the excess IR luminosity (both
measured with respect to the underlying old elliptical galaxy).
\item Analysis of a number of parameters based on the optical spectra, the ratio of far- and near-IR strengths, and of the ratio of the strength of the nucleus in
the optical and radio led to the conclusion that the HEG sources (which are in general the most powerful radio sources as well) show AGN
radiation in both the UV and IR, while sources with weaker emission lines and typically FR I sources tend to have star formation signatures and little
evidence for AGN radiation.
\item There appears to be a general (inverse) correlation between the mass of the central black hole and $XUV$ (although with the caution that a few objects 
appear to deviate
from this general trend; it is not clear if this deviation is genuine or due to problems with the derivation of $XUV$).
\item We confirm some trends that are known from the literature: there is evidence that the mass of black holes is higher (by a factor of about 2-3) for objects
 without $XUV$ 
excess, and this translates, via the correlation of $XUV$ and radio power or spectral type, into the known fact that LEGs, "dull",  and FRI radio galaxies have on average
higher black hole masses than HEGs and FRII radio galaxies (Best \& Heckman \cite{best12} and references therein). A similar effect is seen for the mass 
of the galaxy instead of the black hole mass: again the FRI radio galaxies are more massive than FRII/HEGs
by about a factor 2.
\item We conclude that all these points indicate a fundamental difference between FRI and FRII galaxies.
\end{itemize}

\begin{acknowledgements} 
 
We thank the referee, whose comments forced us to rethink many parts of this work, but were stimulating and in the end very helpful.
 
We made use of a number of on-line data bases. 
The Galactic Evolution Explorer (GALEX) is operated for NASA by the California Institute of Technology under NASA contract NAS5-98034. 
URL: http://www.galex.caltech.edu.

Funding for the SDSS and SDSS-II has been provided by the Alfred P. Sloan Foundation, the Participating Institutions, 
the National Science Foundation, the U.S. Department of Energy, the National Aeronautics and Space Administration, 
the Japanese Monbukagakusho, the Max Planck Society, and the Higher Education Funding Council for England. 
The SDSS Web Site is http://www.sdss.org/.
The SDSS is managed by the Astrophysical Research Consortium for the Participating Institutions. 
The Participating Institutions are the American Museum of Natural History, Astrophysical Institute Potsdam, University of Basel, 
University of Cambridge, Case Western Reserve University, University of Chicago, Drexel University, Fermilab, 
the Institute for Advanced Study, the Japan Participation Group, Johns Hopkins University, the Joint Institute for Nuclear Astrophysics, 
the Kavli Institute for Particle Astrophysics and Cosmology, the Korean Scientist Group, the Chinese Academy of Sciences (LAMOST), 
Los Alamos National Laboratory, the Max-Planck-Institute for Astronomy (MPIA), the Max-Planck-Institute for Astrophysics (MPA), 
New Mexico State University, Ohio State University, University of Pittsburgh, University of Portsmouth, Princeton University, 
the United States Naval Observatory, and the University of Washington.

This publication makes use of data products from the Two Micron All
Sky Survey, which is a joint project of the University of Massachusetts
and the Infrared Processing and Analysis Center/California Institute of
Technology, funded by the National Aeronautics and Space Administration
and the National Science Foundation. URL: http://www.ipac.caltech.edu/2mass/.

This publication makes use of the data products from the Wide-field Infrared Explorer, which is a joint project of the University
of California, Los Angeles, and the Jet Propulsion Laboratory/California Institute of Technology, funded by the National
Aeronautics and Space Administration. URL: http://wise2.ipac.caltech.edu/docs/release/allsky/expsup/.

We acknowledge the usage of the HyperLeda database (http://leda.univ-lyon1.fr).

This research has made use of the NASA/IPAC Extragalactic Database (NED) which is operated by the Jet Propulsion Laboratory, 
California Institute of Technology, under contract with the National Aeronautics and Space Administration. 

\end{acknowledgements}

\longtab{2}{
\begin{longtable}{lrrrrrrrr}
\caption{\label{tab:wise} WISE magnitudes of B2 and 3C radio galaxies}\\
\hline\hline
 Source & $W1$ & $\pm$ & $W2$ & $\pm$ & $W3$ & $\pm$ & $W4$ & $\pm$ \\
\hline
\endfirsthead
\caption{continued.}\\
\hline\hline
 Source & $W1$ & $\pm$ & $W2$ & $\pm$ & $W3$ & $\pm$ & $W4$ & $\pm$ \\
\hline
\endhead
\hline
\endfoot
0034+25  & 10.08 & 0.01 & 10.11 & 0.01 &  9.34 & 0.06 &  7.87 & 0.37 \\
0648+27  & 10.40 & 0.01 &  9.07 & 0.01 &  5.80 & 0.01 &  2.88 & 0.01  \\
0755+37  & 10.12 & 0.01 & 10.10 & 0.01 &  9.22 & 0.05 &  7.81 & 0.37  \\
0800+24  & 11.97 & 0.01 & 12.01 & 0.01 & 11.87 & 0.26 &  0.00 & 0.00 \\
0828+32  & 11.99 & 0.01 & 11.95 & 0.01 & 11.11 & 0.12 &  8.50 & 0.38  \\
0836+29  & 11.10 & 0.01 & 11.03 & 0.01 & 10.41 & 0.13 &  8.56 & 0.85 \\
0836+29A & 11.54 & 0.01 & 10.53 & 0.01 &  7.71 & 0.01 &  4.93 & 0.02 \\
0838+32  & 11.42 & 0.01 & 11.38 & 0.01 & 10.48 & 0.09 &  9.35 & 0.85 \\
0844+31  & 10.80 & 0.01 & 10.75 & 0.01 & 10.39 & 0.10 &  8.65 & 0.45 \\
0908+37  & 12.09 & 0.01 & 12.00 & 0.01 & 11.75 & 0.18 &  9.88 & 0.99 \\
0913+38  & 12.96 & 0.01 & 12.91 & 0.02 & 12.15 & 0.24 &  9.15 & 0.48 \\
0924+30  & 13.55 & 0.02 & 13.60 & 0.03 & 11.94 & 0.00 &  8.36 & 0.00 \\
1003+26  & 12.20 & 0.01 & 12.07 & 0.02 & 12.16 & 0.43 & 10.00 & 1.68 \\
1005+28  & 12.96 & 0.01 & 12.84 & 0.01 & 12.56 & 0.32 &   -   &  -   \\
1037+30  & 12.91 & 0.01 & 12.83 & 0.01 & 10.97 & 0.07 &  8.87 & 0.31 \\
1102+30  & 11.50 & 0.01 & 11.45 & 0.01 & 10.85 & 0.10 &  9.45 & 0.83 \\
1108+27  & 10.10 & 0.01 & 10.11 & 0.01 &  8.59 & 0.02 &  7.42 & 0.20 \\
1113+24  & 11.88 & 0.01 & 11.79 & 0.01 & 11.74 & 0.25 &  9.32 & 0.65 \\
1116+28  & 11.19 & 0.01 & 11.19 & 0.01 & 10.84 & 0.12 & 10.14 & 2.44 \\
1122+39  &  7.68 & 0.01 &  7.69 & 0.01 &  6.01 & 0.01 &  4.45 & 0.04  \\
1141+37  & 12.97 & 0.02 & 12.73 & 0.03 & 11.07 & 0.14 &  8.42 & 0.32 \\
1144+35  & 11.38 & 0.01 & 11.02 & 0.01 &  8.97 & 0.02 &  6.78 & 0.06 \\
1204+24  & 12.50 & 0.01 & 12.48 & 0.02 & 11.98 & 0.22 & 10.54 & 1.98 \\
1204+34  & 12.05 & 0.01 & 11.49 & 0.01 &  9.16 & 0.02 &  6.80 & 0.05 \\
1217+29  &  7.25 & 0.01 &  7.28 & 0.01 &  6.44 & 0.02 &  5.25 & 0.13 \\
1243+26  & 15.99 & 0.05 & 15.76 & 0.13 & 12.73 & 0.00 &  9.17 & 0.00 \\
1254+27  & 10.22 & 0.01 & 10.20 & 0.01 &  8.35 & 0.02 &  6.97 & 0.10 \\
1256+28  & 10.71 & 0.01 & 10.78 & 0.01 & 10.30 & 0.08 &  8.80 & 0.49 \\
1257+28  &  9.12 & 0.01 &  9.15 & 0.01 &  8.63 & 0.07 &   -   &  -   \\
1303+31  & 13.25 & 0.02 & 13.02 & 0.03 & 12.02 & 0.00 &  8.97 & 0.00 \\
1316+39  & 12.39 & 0.01 & 12.16 & 0.01 & 10.45 & 0.06 &  8.35 & 0.27 \\
1321+31  &  9.57 & 0.01 &  9.62 & 0.01 &  8.93 & 0.05 &  7.57 & 0.26 \\
1322+36  &  9.77 & 0.01 &  9.77 & 0.01 &  8.97 & 0.03 &  7.22 & 0.09 \\
1339+26  & 11.48 & 0.01 & 11.44 & 0.01 & 11.20 & 0.19 &  9.27 & 1.05 \\
1346+26  & 11.16 & 0.01 & 11.21 & 0.01 & 10.33 & 0.08 &  9.47 & 0.64 \\
1347+28  & 12.65 & 0.01 & 12.63 & 0.02 & 12.39 & 0.31 &   -   &  -   \\
1350+31  & 10.98 & 0.01 & 10.59 & 0.01 &  8.19 & 0.03 &  6.21 & 0.04 \\
1357+28  & 11.76 & 0.01 & 11.76 & 0.01 & 11.61 & 0.18 &  9.47 & 0.74 \\
1422+26  & 11.30 & 0.01 & 11.21 & 0.01 &  9.79 & 0.03 &  7.94 & 0.19 \\
1430+25  & 15.90 & 0.11 & 16.02 & 0.32 & 11.65 & 0.19 &  8.20 & 0.27 \\
1447+27  & 10.70 & 0.01 & 10.71 & 0.01 & 10.04 & 0.04 &  9.65 & 0.84 \\
1450+28  & 12.25 & 0.01 & 12.14 & 0.01 & 12.05 & 0.20 & 12.09 & 5.04 \\
1455+28  & 13.40 & 0.03 & 12.95 & 0.04 &  9.46 & 0.04 &  6.17 & 0.07 \\
1457+29  & 13.44 & 0.01 & 13.08 & 0.01 & 11.63 & 0.07 &  9.20 & 0.37 \\
1502+26  & 12.35 & 0.02 & 12.41 & 0.02 & 11.51 & 0.13 &  8.80 & 0.00 \\
1511+26  & 13.60 & 0.03 & 13.37 & 0.03 & 11.07 & 0.10 &  8.47 & 0.27 \\
1512+30  & 11.95 & 0.01 & 11.86 & 0.01 & 11.47 & 0.10 &  9.78 & 0.94 \\
1521+28  & 12.04 & 0.01 & 11.99 & 0.01 & 11.48 & 0.09 &  9.95 & 0.75 \\
1553+24  & 11.33 & 0.01 & 11.33 & 0.01 & 10.69 & 0.08 & 10.14 & 1.38 \\
1557+26  & 11.54 & 0.01 & 11.52 & 0.01 & 10.86 & 0.10 &  9.10 & 0.48 \\
1609+31  & 13.19 & 0.01 & 13.15 & 0.02 & 13.38 & 0.55 & 10.34 & 1.23 \\
1610+29  & 10.04 & 0.01 & 10.06 & 0.01 &  9.64 & 0.07 &  8.59 & 0.43 \\
1613+27  & 12.44 & 0.01 & 12.48 & 0.02 & 12.21 & 0.39 &  9.56 & 0.68 \\
1615+32  & 11.40 & 0.02 & 10.51 & 0.02 &  8.12 & 0.02 &  5.97 & 0.50  \\
1615+35  & 10.46 & 0.01 & 10.48 & 0.01 &  9.73 & 0.07 &  8.11 & 0.38 \\
1621+38  &  9.67 & 0.01 &  9.68 & 0.01 &  9.13 & 0.03 &  8.20 & 0.34 \\
1626+39  &  9.24 & 0.01 &  9.29 & 0.01 &  8.70 & 0.04 &  7.44 & 0.24 \\
1637+29  & 11.74 & 0.01 & 11.68 & 0.01 & 11.27 & 0.11 &  9.20 & 0.48 \\
1638+32  & 12.83 & 0.01 & 12.65 & 0.01 & 11.80 & 0.14 & 10.17 & 1.00 \\
1643+27  & 12.56 & 0.01 & 12.48 & 0.01 & 12.19 & 0.26 & 11.04 & 2.41 \\
1652+39  &  9.52 & 0.01 &  9.15 & 0.01 &  7.45 & 0.01 &  5.46 & 0.03 \\
1657+32  & 12.24 & 0.01 & 12.23 & 0.01 & 11.91 & 0.21 &   -   &  -   \\
1658+32  & 12.89 & 0.01 & 12.82 & 0.02 & 11.90 & 0.20 &  9.28 & 0.50 \\
1658+30A & 11.02 & 0.01 & 11.02 & 0.01 & 10.50 & 0.06 &  9.81 & 0.60 \\
1726+31  & 13.27 & 0.01 & 12.90 & 0.02 & 11.47 & 0.11 &  8.45 & 0.18 \\
3C192    & 12.44 & 0.01 & 12.43 & 0.02 & 11.08 & 0.17 &  8.35 & 0.27 \\
3C219    & 12.82 & 0.01 & 12.13 & 0.01 & 10.02 & 0.04 &  7.58 & 0.13 \\
3C223    & 13.05 & 0.03 & 11.71 & 0.02 &  7.80 & 0.02 &  5.04 & 0.03 \\
3C234    & 11.07 & 0.00 &  9.59 & 0.00 &  6.03 & 0.00 &  3.61 & 0.00  \\
3C264    &  9.68 & 0.01 &  9.66 & 0.01 &  8.56 & 0.04 &  7.15 & 0.17 \\
3C272.1  &  6.32 & 0.01 &  6.37 & 0.01 &  5.69 & 0.02 &  4.66 & 0.14 \\
3C274    &  5.81 & 0.01 &  5.88 & 0.01 &  5.08 & 0.01 &  3.78 & 0.07 \\
3C285    & 12.64 & 0.01 & 12.00 & 0.01 &  8.83 & 0.01 &  6.15 & 0.03 \\
3C296    &  8.85 & 0.01 &  8.90 & 0.01 &  8.21 & 0.02 &  6.94 & 0.12 \\
3C303    & 12.77 & 0.01 & 11.99 & 0.01 &  9.53 & 0.01 &  7.02 & 0.04 \\
3C305    & 10.99 & 0.01 & 10.75 & 0.01 &  7.98 & 0.01 &  5.60 & 0.03 \\
3C321    & 12.24 & 0.01 & 11.37 & 0.01 &  7.46 & 0.01 &  4.33 & 0.01 \\
3C346    & 13.01 & 0.01 & 12.56 & 0.02 & 10.55 & 0.05 &  7.82 & 0.15 \\
\end{longtable}

}

\longtab{3}{
\begin{longtable}{lrrrrrrrrrr}
\caption{\label{tab:phot} GALEX, SDSS, and 2MASS magnitudes of B2 and 3C radio galaxies}\\
\hline\hline
 Source & $ FUV$ & $NUV$ & $u$ & $g$ & $r$ & $i$ & $z$ & $J$ & $H$ & $K$ \\
             & $\pm$ & $\pm$ & & & & & & $\pm$ & $\pm$ & $\pm$ \\
\hline
\endfirsthead
\caption{continued.}\\
\hline\hline
 Source & $ FUV$ & $NUV$ & $u$ & $g$ & $r$ & $i$ & $z$ & $J$ & $H$ & $K$ \\
             & $\pm$ & $\pm$ & & & & & & $\pm$ & $\pm$ & $\pm$ \\
\hline
\endhead
\hline
\endfoot
B2\ 0034+25 & 21.20 & 19.63 & 16.14 & 14.19 & 13.29 & 12.84 & 12.54 & 10.84 & 10.11 &  9.73 \\
            &  0.38 &  0.16 &       &       &       &       &       &  0.04 &  0.06 &  0.06 \\
B2\ 0648+27 &    .. &    .. & 16.03 & 14.32 & 13.74 & 13.42 & 13.17 & 11.82 & 11.16 & 10.59 \\
            &    .. &    .. &       &       &       &       &       &  0.04 &  0.04 &  0.04 \\
B2\ 0755+37 & 20.27 & 19.23 & 15.93 & 13.96 & 13.05 & 12.60 & 12.28 & 10.89 & 10.22 &  9.87 \\
            &  0.13 &  0.07 &       &       &       &       &       &  0.02 &  0.03 &  0.03 \\
B2\ 0800+24 &    .. &    .. & 17.51 & 15.50 & 14.61 & 14.16 & 13.82 & 12.70 & 11.95 & 11.75 \\
            &    .. &    .. &       &       &       &       &       &  0.04 &  0.06 &  0.06 \\
B2\ 0828+32 &    .. & 21.42 & 17.53 & 15.60 & 14.68 & 14.25 & 13.90 & 12.85 & 11.94 & 11.71 \\
            &    .. &  0.31 &       &       &       &       &       &  0.05 &  0.05 &  0.07 \\
B2\ 0836+29 & 21.05 & 21.54 & 17.52 & 15.28 & 14.28 & 13.78 & 13.44 & 12.27 & 11.16 & 10.60 \\
            &  0.34 &  0.35 &       &       &       &       &       &  0.10 &  0.07 &  0.07 \\
B2\ 0836+29A & 20.26 & 19.61 & 17.42 & 15.69 & 14.81 & 14.32 & 14.12 & 12.68 & 11.91 & 11.62 \\
             &  0.19 &  0.11 &       &       &       &       &       &  0.04 &  0.04 &  0.06 \\
B2\ 0838+32 & 20.77 & 19.91 & 17.47 & 15.50 & 14.58 & 14.12 & 13.77 & 12.28 & 11.38 & 11.22 \\
            &  0.18 &  0.10 &       &       &       &       &       &  0.04 &  0.05 &  0.05 \\
B2\ 0844+31 & 20.30 & 19.80 & 16.69 & 14.67 & 13.74 & 13.25 & 12.93 & 11.76 & 10.97 & 10.59 \\
            &  0.24 &  0.18 &       &       &       &       &       &  0.04 &  0.04 &  0.05 \\
B2\ 0908+37 & 21.29 & 21.57 & 17.95 & 15.91 & 14.91 & 14.45 & 14.13 & 12.71 & 12.19 & 11.64 \\
            &  0.44 &  0.42 &       &       &       &       &       &  0.05 &  0.07 &  0.08 \\
B2\ 0913+38 &    .. & 21.55 & 18.34 & 16.54 & 15.69 & 15.26 & 14.96 & 13.73 & 13.15 & 12.72 \\
            &    .. &  0.43 &       &       &       &       &       &  0.06 &  0.08 &  0.10 \\
B2\ 0915+32 &    .. & 20.58 & 17.51 & 15.56 & 14.63 & 14.20 & 13.89 & 12.52 & 11.81 & 11.58 \\
            &    .. &  0.27 &       &       &       &       &       &  0.05 &  0.07 &  0.09 \\
B2\ 0924+30 & 19.95 & 17.92 & 15.57 & 13.64 & 12.78 & 12.37 & 12.09 & 11.03 & 10.35 & 10.03 \\
            &  0.03 &  0.02 &       &       &       &       &       &  0.03 &  0.03 &  0.04 \\
B2\ 1003+26 &    .. & 21.67 & 17.95 & 15.86 & 14.79 & 14.32 & 14.00 & 12.57 & 12.43 & 11.66 \\
            &    .. &  0.44 &       &       &       &       &       &  0.04 &  0.10 &  0.08 \\
B2\ 1005+28 & 22.84 & 21.98 & 19.08 & 17.03 & 15.86 & 15.38 & 15.02 & 13.62 & 12.83 & 12.70 \\
            &  0.11 &  0.08 &       &       &       &       &       &  0.05 &  0.07 &  0.09 \\
B2\ 1037+30 & 20.66 & 19.36 & 17.89 & 16.39 & 15.67 & 15.24 & 14.96 & 13.79 & 13.04 & 12.54 \\
            &  0.25 &  0.11 &       &       &       &       &       &  0.09 &  0.11 &  0.11 \\
B2\ 1102+30 & 21.36 & 20.69 & 17.20 & 15.27 & 14.33 & 13.91 & 13.62 & 12.43 & 11.63 & 11.35 \\
            &  0.38 &  0.31 &       &       &       &       &       &  0.03 &  0.04 &  0.06 \\
B2\ 1108+27 & 20.77 & 19.28 & 16.32 & 14.31 & 13.32 & 12.83 & 12.46 & 11.19 & 10.38 & 10.13 \\
            &  0.24 &  0.11 &       &       &       &       &       &  0.02 &  0.02 &  0.03 \\
B2\ 1113+24 &    .. &    .. & 17.58 & 15.59 & 14.62 & 14.18 & 13.84 & 12.65 & 11.76 & 11.42 \\
            &    .. &    .. &       &       &       &       &       &  0.06 &  0.06 &  0.08 \\
B2\ 1116+28 & 21.29 & 20.20 & 16.95 & 14.97 & 14.04 & 13.62 & 13.31 & 11.96 & 11.21 & 11.06 \\
            &  0.27 &  0.13 &       &       &       &       &       &  0.03 &  0.03 &  0.04 \\
B2\ 1122+39 & 18.33 & 16.33 & 13.82 & 11.98 & 11.18 & 10.72 & 10.32 &  8.62 &  7.93 &  7.68 \\
            &  0.07 &  0.02 &       &       &       &       &       &  0.01 &  0.01 &  0.01 \\
B2\ 1141+37 & 22.49 & 22.17 & 19.53 & 17.78 & 16.81 & 16.35 & 15.99 & 14.21 & 13.21 & 12.82 \\
            &  0.43 &  0.30 &       &       &       &       &       &  0.07 &  0.07 &  0.08 \\
B2\ 1144+35 & 20.17 & 19.29 & 17.14 & 15.46 & 14.53 & 14.08 & 13.81 & 12.25 & 11.20 & 11.14 \\
            &  0.21 &  0.09 &       &       &       &       &       &  0.04 &  0.04 &  0.07 \\
B2\ 1204+24 &    .. & 22.15 & 18.22 & 16.23 & 15.27 & 14.84 & 14.52 & 13.20 & 12.80 & 12.20 \\
            &    .. &  0.32 &       &       &       &       &       &  0.05 &  0.08 &  0.08 \\
B2\ 1204+34 & 19.78 & 19.52 & 17.52 & 15.97 & 15.17 & 14.72 & 14.43 & 13.35 & 12.78 & 12.25 \\
            &  0.13 &  0.06 &       &       &       &       &       &  0.05 &  0.08 &  0.07 \\
B2\ 1217+29 & 16.60 & 15.47 & 13.38 & 11.52 & 10.70 & 10.30 &  9.92 &  8.09 &  7.42 &  7.18 \\
            &  0.01 &  0.01 &       &       &       &       &       &  0.01 &  0.01 &  0.01 \\
B2\ 1243+26 & 21.94 & 21.08 & 17.79 & 15.81 & 14.80 & 14.37 & 14.03 & 12.72 & 11.89 & 11.67 \\
            &  0.49 &  0.26 &       &       &       &       &       &  0.04 &  0.04 &  0.06 \\
B2\ 1254+27 & 18.87 & 17.42 & 15.38 & 13.33 & 12.50 & 12.07 & 11.80 & 10.09 &  9.33 &  9.20 \\
            &  0.04 &  0.02 &       &       &       &       &       &  0.03 &  0.03 &  0.04 \\
B2\ 1256+28 & 20.77 & 19.26 & 16.37 & 14.43 & 13.62 & 13.20 & 12.91 & 11.64 & 10.84 & 10.61 \\
            &  0.26 &  0.11 &       &       &       &       &       &  0.03 &  0.03 &  0.04 \\
B2\ 1257+28 & 19.92 & 18.10 & 14.84 & 12.93 & 12.10 & 11.66 & 11.39 &  9.85 &  9.02 &  8.86 \\
            &  0.22 &  0.08 &       &       &       &       &       &  0.02 &  0.02 &  0.03 \\
B2\ 1300+32 &    .. & 22.39 & 20.23 & 18.15 & 16.97 & 16.44 & 16.07 & 14.54 & 14.21 & 13.45 \\
            &    .. &  0.50 &       &       &       &       &       &  0.10 &  0.17 &  0.14 \\
B2\ 1303+31 &    .. & 22.22 & 19.62 & 17.40 & 16.18 & 15.68 & 15.31 & 14.33 & 13.63 & 13.17 \\
            &    .. &  0.76 &       &       &       &       &       &  0.08 &  0.09 &  0.11 \\
B2\ 1316+29 & 22.46 & 20.97 & 17.75 & 15.92 & 15.04 & 14.63 & 14.36 & 13.07 & 12.48 & 12.23 \\
            &  0.60 &  0.27 &       &       &       &       &       &  0.06 &  0.06 &  0.12 \\
B2\ 1317+33 & 20.82 & 19.69 & 16.64 & 14.63 & 13.75 & 13.34 & 13.04 & 11.70 & 10.97 & 10.76 \\
            &  0.29 &  0.16 &       &       &       &       &       &  0.03 &  0.03 &  0.05 \\
B2\ 1321+31 & 19.76 & 17.90 & 15.30 & 13.32 & 12.50 & 12.07 & 11.75 & 10.34 &  9.72 &  9.36 \\
            &  0.04 &  0.01 &       &       &       &       &       &  0.02 &  0.03 &  0.03 \\
B2\ 1322+36 & 19.74 & 18.18 & 15.40 & 13.49 & 12.66 & 12.25 & 11.96 & 10.58 &  9.90 &  9.64 \\
            &  0.06 &  0.02 &       &       &       &       &       &  0.01 &  0.02 &  0.02 \\
B2\ 1339+26 & 22.01 & 20.33 & 19.06 & 17.27 & 16.32 & 15.93 & 15.56 & 13.96 & 13.22 & 12.99 \\
            &  0.46 &  0.12 &       &       &       &       &       &  0.06 &  0.08 &  0.06 \\
B2\ 1346+26 & 18.38 & 18.08 & 16.08 & 14.50 & 13.70 & 13.26 & 13.03 & 11.69 & 10.92 & 10.60 \\
            &  0.11 &  0.06 &       &       &       &       &       &  0.06 &  0.06 &  0.08 \\
B2\ 1347+28 &    .. & 21.30 & 18.39 & 16.19 & 15.29 & 14.86 & 14.61 & 13.25 & 12.81 & 12.33 \\
            &    .. &  0.27 &       &       &       &       &       &  0.06 &  0.08 &  0.10 \\
B2\ 1350+31 & 20.28 & 18.62 & 17.02 & 15.12 & 14.24 & 13.77 & 13.42 & 11.89 & 11.15 & 10.84 \\
            &  0.26 &  0.05 &       &       &       &       &       &  0.03 &  0.04 &  0.04 \\
B2\ 1357+28 & 20.92 & 21.43 & 17.67 & 15.68 & 14.75 & 14.30 & 13.97 & 12.55 & 11.69 & 11.57 \\
            &  0.42 &  0.40 &       &       &       &       &       &  0.04 &  0.04 &  0.08 \\
B2\ 1422+26 &    .. & 20.04 & 16.72 & 14.84 & 13.99 & 13.60 & 13.29 & 11.99 & 11.39 & 11.00 \\
            &    .. &  0.23 &       &       &       &       &       &  0.04 &  0.06 &  0.06 \\
B2\ 1430+25 & 21.07 & 20.56 & 18.77 & 17.07 & 16.22 & 15.81 & 15.50 & 14.24 & 13.65 & 13.62 \\
            &  0.43 &  0.24 &       &       &       &       &       &  0.07 &  0.09 &  0.16 \\
B2\ 1447+27 & 20.05 & 19.83 & 16.73 & 14.66 & 13.78 & 13.29 & 13.01 & 11.44 & 10.67 & 10.40 \\
            &  0.27 &  0.12 &       &       &       &       &       &  0.04 &  0.06 &  0.07 \\
B2\ 1450+28 & 21.87 &    .. & 18.97 & 17.03 & 15.95 & 15.50 & 15.09 & 13.41 & 12.79 & 12.45 \\
            &  0.42 &    .. &       &       &       &       &       &  0.08 &  0.10 &  0.13 \\
B2\ 1455+28 &    .. &    .. & 19.24 & 17.41 & 16.21 & 15.68 & 15.33 & 14.38 & 13.75 & 13.18 \\
            &    .. &    .. &       &       &       &       &       &  0.12 &  0.16 &  0.17 \\
B2\ 1457+29 &    .. & 21.78 & 19.47 & 17.53 & 16.45 & 16.00 & 15.64 & 14.79 & 13.85 & 13.38 \\
            &    .. &  0.54 &       &       &       &       &       &  0.16 &  0.16 &  0.20 \\
B2\ 1502+26 &    .. & 21.24 & 17.71 & 15.61 & 14.70 & 14.24 & 13.82 & 14.71 & 14.01 & 13.56 \\
            &    .. &  0.26 &       &       &       &       &       &  0.07 &  0.09 &  0.07 \\
B2\ 1511+26 &    .. &    .. & 19.37 & 17.76 & 16.88 & 16.36 & 16.16 & 14.64 & 14.72 & 13.50 \\
            &    .. &    .. &       &       &       &       &       &  0.14 &  0.10 &  0.20 \\
B2\ 1512+30 &    .. & 21.28 & 17.90 & 15.91 & 14.93 & 14.49 & 14.11 & 12.83 & 12.05 & 11.65 \\
            &    .. &  0.34 &       &       &       &       &       &  0.06 &  0.07 &  0.08 \\
B2\ 1521+28 & 22.19 & 21.19 & 17.95 & 15.97 & 15.02 & 14.54 & 14.20 & 12.82 & 12.13 & 11.95 \\
            &  0.50 &  0.37 &       &       &       &       &       &  0.05 &  0.06 &  0.07 \\
B2\ 1553+24 & 21.52 & 20.68 & 16.91 & 14.96 & 14.06 & 13.61 & 13.28 & 11.84 & 11.24 & 11.08 \\
            &  0.32 &  0.17 &       &       &       &       &       &  0.03 &  0.04 &  0.07 \\
B2\ 1557+26 & 22.75 & 20.97 & 17.41 & 15.46 & 14.52 & 14.09 & 13.73 & 12.36 & 11.52 & 11.41 \\
            &  0.36 &  0.09 &       &       &       &       &       &  0.03 &  0.04 &  0.05 \\
B2\ 1609+31 &    .. & 22.63 & 18.94 & 16.91 & 15.91 & 15.48 & 15.12 & 14.06 & 13.06 & 12.71 \\
            &    .. &  0.49 &       &       &       &       &       &  0.10 &  0.10 &  0.11 \\
B2\ 1610+29 & 20.34 & 19.43 & 15.84 & 13.84 & 12.95 & 12.51 & 12.18 & 11.02 & 10.23 &  9.97 \\
            &  0.25 &  0.15 &       &       &       &       &       &  0.02 &  0.02 &  0.03 \\
B2\ 1613+27 & 21.00 & 21.61 & 18.04 & 16.03 & 15.08 & 14.65 & 14.33 & 13.22 & 12.35 & 12.25 \\
            &  0.32 &  0.25 &       &       &       &       &       &  0.05 &  0.05 &  0.09 \\
B2\ 1615+32 & 20.57 & 19.57 & 18.20 & 17.50 & 16.77 & 16.06 & 16.07 & 14.37 & 13.90 & 13.24 \\
            &  0.20 &  0.05 &       &       &       &       &       &  0.07 &  0.13 &  0.10 \\
B2\ 1615+35 & 19.99 & 18.97 & 16.01 & 14.08 & 13.23 & 12.79 & 12.48 & 11.29 & 10.57 & 10.32 \\
            &  0.24 &  0.11 &       &       &       &       &       &  0.03 &  0.03 &  0.04 \\
B2\ 1621+38 & 19.43 & 18.29 & 15.55 & 13.59 & 12.73 & 12.29 & 12.02 & 10.51 &  9.80 &  9.55 \\
            &  0.15 &  0.05 &       &       &       &       &       &  0.02 &  0.02 &  0.03 \\
B2\ 1626+39 & 18.54 & 17.44 & 15.78 & 13.85 & 13.00 & 12.54 & 12.30 & 10.20 &  9.49 &  9.17 \\
            &  0.04 &  0.02 &       &       &       &       &       &  0.02 &  0.03 &  0.03 \\
B2\ 1637+29 & 20.87 & 20.37 & 17.97 & 15.85 & 14.89 & 14.44 & 14.12 & 12.56 & 11.89 & 11.68 \\
            &  0.33 &  0.27 &       &       &       &       &       &  0.04 &  0.05 &  0.07 \\
B2\ 1638+32 &    .. &    .. & 18.67 & 16.92 & 15.84 & 15.37 & 15.04 & 13.69 & 13.43 & 12.85 \\
            &    .. &    .. &       &       &       &       &       &  0.06 &  0.13 &  0.13 \\
B2\ 1643+27 &    .. &    .. & 18.43 & 16.32 & 15.28 & 14.80 & 14.44 & 13.32 & 12.55 & 12.28 \\
            &    .. &    .. &       &       &       &       &       &  0.07 &  0.07 &  0.10 \\
B2\ 1652+39 & 15.79 & 15.38 & 15.35 & 13.84 & 13.03 & 12.62 & 12.33 & 10.67 &  9.87 &  9.57 \\
            &  0.01 &  0.01 &       &       &       &       &       &  0.02 &  0.02 &  0.03 \\
B2\ 1657+32 & 21.76 & 21.42 & 18.05 & 16.03 & 15.09 & 14.64 & 14.30 & 13.01 & 12.40 & 11.80 \\
            &  0.18 &  0.05 &       &       &       &       &       &  0.06 &  0.07 &  0.07 \\
B2\ 1658+32 & 22.60 & 21.59 & 18.58 & 16.74 & 15.73 & 15.31 & 14.95 & 13.92 & 13.26 & 12.66 \\
            &  0.33 &  0.10 &       &       &       &       &       &  0.09 &  0.11 &  0.10 \\
B2\ 1658+30A & 20.15 & 18.32 & 17.30 & 15.26 & 14.48 & 13.99 & 13.61 & 12.15 & 11.35 & 10.84 \\
             &  0.14 &  0.04 &       &       &       &       &       &  0.08 &  0.08 &  0.10 \\
B2\ 1726+31 &    .. & 22.62 & 19.71 & 17.58 & 16.34 & 15.82 & 15.42 & 14.17 & 13.53 & 13.20 \\
            &    .. &  0.53 &       &       &       &       &       &  0.09 &  0.12 &  0.16 \\
3C\ 192     & 21.32 & 20.55 & 18.06 & 16.23 & 15.38 & 14.88 & 14.56 & 13.19 & 12.47 & 12.21 \\
            &  0.43 &  0.25 &       &       &       &       &       &  0.04 &  0.04 &  0.07 \\
3C\ 219     & 20.96 & 21.45 & 19.21 & 17.89 & 16.74 & 16.26 & 15.98 & 14.61 & 13.92 & 13.12 \\
            &  0.26 &  0.30 &       &       &       &       &       &  0.14 &  0.16 &  0.15 \\
3C\ 223     & 21.41 & 20.47 & 19.36 & 17.83 & 16.73 & 16.30 & 16.08 & 15.04 & 14.29 & 13.93 \\
            &  0.38 &  0.22 &       &       &       &       &       &  0.14 &  0.16 &  0.20 \\
3C\ 234     & 18.54 & 19.05 & 18.53 & 17.99 & 16.85 & 16.64 & 16.85 & 14.74 & 13.66 & 12.74 \\
            &  0.07 &  0.06 &       &       &       &       &       &  0.13 &  0.10 &  0.07 \\
3C\ 264     & 18.72 & 18.07 & 15.38 & 13.53 & 12.68 & 12.26 & 11.95 & 10.48 &  9.77 &  9.49 \\
            &  0.10 &  0.06 &       &       &       &       &       &  0.03 &  0.03 &  0.04 \\
3C\ 272.1   & 17.00 & 15.22 & 12.59 & 10.74 &  9.89 &  9.45 &  9.08 &  7.12 &  6.47 &  6.22 \\
            &  0.04 &  0.02 &       &       &       &       &       &  0.02 &  0.02 &  0.02 \\
3C\ 274     & 15.19 & 15.20 & 13.48 & 11.56 & 10.70 & 10.27 &  9.92 &  6.72 &  6.07 &  5.81 \\
            &  0.02 &  0.01 &       &       &       &       &       &  0.02 &  0.02 &  0.02 \\
3C\ 285     & 19.92 & 19.50 & 18.21 & 16.67 & 15.88 & 15.45 & 15.11 & 13.78 & 13.24 & 12.65 \\
            &  0.17 &  0.08 &       &       &       &       &       &  0.07 &  0.12 &  0.09 \\
3C\ 296     & 19.16 & 17.93 & 14.90 & 12.92 & 12.06 & 11.64 & 11.30 &  9.75 &  9.04 &  8.76 \\
            &  0.15 &  0.04 &       &       &       &       &       &  0.01 &  0.01 &  0.02 \\
3C\ 303     & 19.49 & 19.49 & 18.45 & 17.56 & 16.66 & 16.09 & 15.88 & 14.61 & 13.75 & 13.17 \\
            &  0.11 &  0.08 &       &       &       &       &       &  0.14 &  0.16 &  0.13 \\
3C\ 305     & 19.82 & 18.35 & 16.16 & 14.27 & 13.54 & 13.16 & 12.88 & 11.70 & 11.05 & 10.64 \\
            &  0.18 &  0.07 &       &       &       &       &       &  0.03 &  0.04 &  0.04 \\
3C\ 321     & 19.40 & 18.86 & 17.36 & 16.05 & 15.30 & 14.84 & 14.58 & 13.54 & 12.70 & 12.34 \\
            &  0.11 &  0.06 &       &       &       &       &       &  0.06 &  0.07 &  0.09 \\
3C\ 326     &    .. & 21.47 & 18.89 & 16.91 & 15.90 & 15.35 & 15.03 & 14.41 & 13.73 & 13.41 \\
            &    .. &  0.22 &       &       &       &       &       &  0.09 &  0.11 &  0.13 \\
3C\ 346     &    .. & 21.08 & 18.90 & 17.32 & 16.14 & 15.66 & 15.29 & 14.49 & 13.61 & 12.74 \\
            &    .. &  0.33 &       &       &       &       &       &  0.15 &  0.17 &  0.12 \\
\end{longtable}

}
\longtab{4}{
\begin{longtable}{l@{\hspace{0.5mm}}c@{\hspace{0.5mm}}rrrc@{\hspace{0.9mm}}rrrrl@{\hspace{0.9mm}}rlrc}
\caption{\label{tab:results}Radio and optical parameters}\\
\hline\hline
             & \multicolumn{2}{c}{Galaxy} &  \multicolumn{3}{c}{2nd component} \\ 
 Source & $\log M/M_{\sun}$ & age &  C1\tablefootmark{a}  & C2\tablefootmark{b} &  & $\chi^2_{min}$  & $\log P$ & FR & 
\multicolumn{2}{c}{Spec.} & $L_{H\alpha}$  & XUV & $\log R_W$ & $\log M/M_{\sun}$ \\
             &                              &  $10^9$yr &   &  & &  & & &  & & $10^{33}$ W & mag  & & (B.H.)   \\
\hline
\endfirsthead
\caption{continued.}\\
\hline\hline
             & \multicolumn{2}{c}{Galaxy} &  \multicolumn{3}{c}{2nd component} \\ 
 Source & $\log M/M_{\sun}$ & age &  C1\tablefootmark{a}  & C2\tablefootmark{b} &  & $\chi^2_{min}$ & $\log P$ & FR & \multicolumn{2}{c}{Spec.} & 
$L_{H\alpha}$  & XUV & $\log R_W$ & $\log M/M_{\sun}$ \\
             &                              &  $10^9$yr & & &  & &  & & & & $10^{33}$ W & mag  & & (B.H.) \\
\hline
\endhead
\hline
\endfoot
B2\ 0034+25  & 11.799 & 13.0 &   ..   &   ..   &   &   14.4 & 23.52 &  1.0 & 0 & .. &     ..  & $+0.03\pm .22$ & 1.95 & 8.87\rlap{$^{\mathrm e}$} \\
B2\ 0648+27  & 11.690 &  13.0 &  9.620   &  2.9E8   & g & 6.8  & 23.94 &  3.0 & 3 & S &     ..  &    ..  & 0.25 & 7.49 \\
B2\ 0755+37  & 12.102 & 13.0 &   ..   &   ..   &   &    7.5 & 24.81 &  1.5 &  2 & H:  &   4.65  & $+0.34\pm .09$ & 1.79 & 8.79 \\
B2\ 0800+24  & 11.451 & 13.0 &   ..   &   ..   &   &    6.2 & 23.83 &  1.0 &  1 & L: &   0.29  &    ..  & .. & .. \\
B2\ 0828+32  & 11.505 & 10.0 &   ..   &   ..   &   &    8.2 & 25.04 &  2.0 &  3 & L  &   1.37  & $-0.3\pm .3$ & 1.56 & .. \\
B2\ 0836+29  & 12.036 & 13.0 &   ..   &   ..   &   &   12.9 & 24.86 &  1.5 &  1 & H  &   2.88  & $-0.3\pm .4$ & 1.66 & .. \\
B2\ 0836+29A & 11.975 & 13.0 &  9.640 &  2.9E8 & g &    5.9 & 25.08 &  1.5 &  3 & S  &  25.30  & $+1.77\pm .14$ & 0.51 & .. \\
B2\ 0838+32  & 11.920 & 13.0 &  9.700 &  5.7E8 & g  & 11.7 & 24.81 &  1.0 & .. & .. &     ..  & $+1.06\pm .12$ & 1.80 & .. \\
B2\ 0844+31  & 12.151 & 10.0 &   ..   &   ..   &   &    7.6 & 25.15 &  1.5 &  1 & L: &   2.52  & $+0.52\pm .21$ & 1.91 & .. \\
B2\ 0908+37  & 12.087 & 10.0 &   ..   &   ..   &   &    3.3 & 25.20 &  1.5 &  1 & L  &   3.26  & $-0.3\pm .5$ & 1.91 & 9.29 \\
B2\ 0913+38  & 11.174 &  5.0 &   ..   &   ..   &   &    2.3 & 24.62 &  1.0 &  1 & ?  &   0.40  & $+0.4\pm .5$ & 1.50 & .. \\
B2\ 0915+32  & 11.814 & 13.0 &   ..   &   ..   &   &    1.9 & 24.34 &  1.0 &  1 & L: &   1.00  & $+0.4\pm .3$ & .. & 8.54 \\ 
B2\ 0924+30  & 11.330 & 5.0 &  9.750 &  5.7E8 & g &   38.3  & 23.84 &  1.0 &  1 & A  &   0.12  & $+0.88\pm .02$ & 1.04 & 8.49 \\
B2\ 1003+26  & 12.011 &  5.0 &   ..   &   ..   &   &    8.0 & 24.38 &  ..  &  1 & H  &   7.36  & $-0.3\pm .5$ & 1.98 & 9.51 \\
B2\ 1005+28  & 12.000 & 10.0 &  9.820 &  5.7E8 & g & 13.0 & 24.64 &  1.5 &  0 & -  & $<$0.48 & $+0.34\pm .09$ & 1.73 & .. \\
B2\ 1037+30  & 11.660 & 13.0 &  9.650 &  2.9E8 & g &   7.9  & 24.87 &  ..  &  3 & L: &  11.70  & $+2.53\pm .15$ & 1.22 & 7.76\rlap{$^{\mathrm f}$}  \\
B2\ 1102+30  & 11.941 & 10.0 &   ..   &   ..   &   &    7.7 & 24.62 &  1.0 &  1 & A  &   1.53  & $+0.1\pm .4$ & 1.87 & .. \\
B2\ 1108+27  & 11.775 & 13.0 &   ..   &   ..   &   &   15.2 & 23.62 &  1.0 &  1 & A  &   1.45  & $+0.43\pm.15$ & 1.56 & .. \\
B2\ 1113+24  & 12.168 & 10.0 &   ..   &   ..   &   &    5.2 & 24.21 &  1.0 &  0 & -  & $<$0.10 &    .. & .. & 9.25 \\
B2\ 1116+28  & 11.998 & 10.0 &   ..   &   ..   &   &    6.7 & 24.73 &  1.0 & .. & .. &     ..  & $+0.16\pm .17$ & 2.12 & .. \\
B2\ 1122+39  & 11.100 & 13.0 &  9.600 &  1.0E9 & g &  162.0\rlap{$^{\mathrm c}$} & 22.11 &  1.0 &  2 & .. &  ..  & $+0.74\pm .03$ & 1.42 & 
7.99\rlap{$^{\mathrm e}$} \\
B2\ 1141+37  & 11.580 & 13.0 &  8.600 &  2.9E8 & g &   18.7\rlap{$^{\mathrm c}$} & 25.84 &  2.0 &  0 & .. & $<$0.06 & $+1.1\pm .4$ & 1.16 & .. \\
B2\ 1144+35  & 11.880 & 13.0 &  9.250 &  2.9E8 & g &   22.2 & 24.85 &  1.0 &  4 & S1 &   16.74 & $+1.68\pm .12$ & 1.07 & 8.16\rlap{$^{\mathrm g}$} \\
B2\ 1204+24  & 11.641 & 10.0 &   ..   &   ..   &   &    4.9 & 24.26 &  1.0 & .. & .. &      .. & $-0.4\pm .3$ & 1.91 & .. \\
B2\ 1204+34  & 11.485 &  5.0 &  6.800 &  1.0E6 & g &    9.5 & 24.82 &  2.0 &  3 & S  &   25.42 & $+2.00\pm .08$ & 0.89 & 8.15 \\
B2\ 1217+29  & 10.400 & 13.0 &  8.400 &  5.7E8 & g &  134.0\rlap{$^{\mathrm c}$} & 21.54 &  3.0 &  2 & L &      .. & $+1.64\pm .01$ & 1.83 & 
8.44\rlap{$^{\mathrm e}$} \\
B2\ 1243+26  & 11.971 & 10.0 &   ..   &   ..   &   &    4.1 & 24.58 &  1.0 & 2 & L &      .. & $+0.1\pm .3$ & 0.43 & .. \\
B2\ 1254+27  & 11.800 & 13.0 &  9.700 &  5.7E8 & g &   78.0\rlap{$^{\mathrm c}$} & 22.95 &  1.0 &  0 & -  & $<$0.02 & $+1.15\pm .03$ & 1.38 & .. \\
B2\ 1256+28  & 11.100 & 10.0 &  9.300 &  1.0E9 & g &   11.9 & 23.36 &  1.0 & .. & .. &      .. & $+0.43\pm .15$ & 1.91 & 8.17\rlap{$^{\mathrm e}$} \\
B2\ 1257+28  & 11.982 & 13.0 &   ..   &   ..   &   &   24.5 & 23.39 &  1.0 &  0 & -  & $<$0.01 & $+0.08\pm .12$ & 1.66 & 8.72\rlap{$^{\mathrm e}$} \\
B2\ 1300+32  & 11.755 & 10.0 &   ..   &   ..   &   &    3.4 & 25.35 &  1.0 & .. & .. &      .. & $+0.9\pm .5$ & .. & .. \\
B2\ 1303+31  & 11.878 &  5.0 &   ..   &   ..   &   &   10.3 & 24.80 &  1.0 & .. & .. &      .. & $+0.3\pm .8$ & 1.34 & .. \\
B2\ 1316+29  & 11.436 &  5.0 &   ..   &   ..   &   &    3.4 & 25.17 &  1.0 &  1 & L  &    1.28 & $+0.2\pm .4$ & 1.25 & .. \\
B2\ 1317+33  & 11.700 & 13.0 &   ..   &   ..   &   &    3.6 & 23.40 &  1.0 &  1 & L: &    1.20 & $+0.34\pm .20$ & .. & .. \\
B2\ 1321+31  & 11.250 & 10.0 &  9.750 &  1.0E9 & g &   43.9 & 24.15 &  1.0 &  1 & L: &    0.22 & $+0.61\pm .02$ & 1.86 & 8.08\rlap{$^{\mathrm e}$} \\
B2\ 1322+36  & 11.250 & 10.0 &  9.750 &  1.0E9 & g &   21.0 & 24.86 &  1.0 &  1 & L: &    0.56 & $+0.57\pm .03$ & 1.73 & 8.59 \\
B2\ 1339+26  & 11.250 & 13.0 &  9.000 &  2.9E8 & g &   13.0\rlap{$^{\mathrm c}$} & 24.65 &  1.0 &  0 & -  & $<$0.08 & $+2.26\pm .21$ & 1.91 & 9.26 \\
B2\ 1346+26  & 12.150 & 13.0 &  9.900 &  2.9E8 & g &    4.9 & 24.89 &  1.0 & 2 & L &      .. & $+1.95\pm .08$ & 1.87 & 8.86\rlap{$^{\mathrm e}$} \\
B2\ 1347+28  & 11.561 & 10.0 &   ..   &   ..   &   &    8.5 & 24.40 &  1.5 &  1 & L: &    0.58 & $+0.3\pm .3$ & 1.76 & 8.37 \\
B2\ 1350+31  & 11.700 & 13.0 &  9.100 &  2.9E8 & g &   31.8 & 25.36 &  1.5 &  2 & L  &    6.63 & $+1.86\pm .11$ & 0.96 & 8.15\rlap{$^{\mathrm e}$} \\
B2\ 1357+28  & 11.789 & 13.0 &   ..   &   ..   &   &    5.3 & 24.37 &  1.0 & .. & .. &      .. & $-0.1\pm .4$ & 1.92 & 8.89 \\
B2\ 1422+26  & 11.475 & 10.0 &   ..   &   ..   &   &    2.4 & 24.32 &  1.0 &  3 & S  &    1.72 & $+0.30\pm .24$ & 1.45 & 8.63 \\
B2\ 1430+25  & 11.100 &  5.0 &  6.700 &  6.0E6 & g &    2.6 & 24.55 &  1.0 &  1 & L  &    0.36 & $+2.1\pm .3$ & -0.07 & 8.17 \\
B2\ 1447+27  & 11.552 & 13.0 &   ..   &   ..   &   &   12.2 & 23.10 &  3.0 & .. & .. &      .. & $+0.46\pm .16$ & 2.06 & 8.95 \\
B2\ 1450+28  & 11.909 & 10.0 &   ..   &   ..   &   &    4.5 & 24.88 &  1.0 & .. & .. &      .. &    .. & 2.35 & 9.38 \\
B2\ 1455+28  & 11.852 & 10.0 &   ..   &   ..   &   &    5.2 & 25.61 &  2.0 &  3 & S  &   35.48 &   .. & 0.25 & ..  \\
B2\ 1457+29  & 11.579 &  5.0 &  6.700 &  6.0E6 & g &    3.1 & 25.28 &  1.0 & .. & .. &      .. & $+1.1\pm .6$ & 1.25 & .. \\
B2\ 1502+26  & 11.690 & 13.0 &   ..   &   ..   &   &  327.0\rlap{$^{\mathrm d}$} & 25.69 &  1.0 &  3 & S  &    3.98 & $+0.0\pm .3$ & 1.51 & 9.36 \\
B2\ 1511+26  & 11.072 &  5.0 &   ..   &   ..   &   &   37.8\rlap{$^{\mathrm c}$} & 25.71 &  1.0 &  3 & .. &      .. &    .. & 0.93 & .. \\ 
B2\ 1512+30  & 11.988 & 10.0 &   ..   &   ..   &   &    2.8 & 24.18 &  ..  &  0 & -  & $<$0.11 & $+0.1\pm .4$ & 1.89 & 9.01 \\
B2\ 1521+28  & 11.838 & 10.0 &   ..   &   ..   &   &    3.5 & 24.93 &  1.0 &  1 & L: &    2.24 & $+0.2\pm .4$ & 1.89 & 8.98 \\
B2\ 1553+24  & 11.625 & 10.0 &   ..   &   ..   &   &    2.4 & 23.90 &  1.0 &  1 & L: &    0.77 & $-0.06\pm .21$ & 2.03 & 8.66 \\
B2\ 1557+26  & 11.480 & 10.0 &   ..   &   ..   &   &    7.0 & 23.14 &  1.0 &  1 & A  &    0.61 & $+0.06\pm .16$ & 1.79 & 8.91 \\
B2\ 1609+31  & 11.606 & 10.0 &   ..   &   ..   &   &    4.5 & 24.50 &  1.5 &  0 & -  & $<$0.06 & $-0.2\pm .5$ & 1.88 & .. \\
B2\ 1610+29  & 11.745 & 10.0 &   ..   &   ..   &   &    6.2 & 23.35 &  1.0 & .. & .. &      .. & $-0.01\pm .18$ & 2.04 & 9.04\rlap{$^{\mathrm e}$} \\
B2\ 1613+27  & 11.320 &  5.0 &  5.900 &  1.3E5 & g &   14.7 & 24.37 &  1.0 &  1 & L: &    0.58 & $+0.0\pm .3$ & 1.77 & 8.51 \\
B2\ 1615+32  & 11.330 &  5.0 &  6.850 &  0.40  & p &   15.3 & 26.18 &  2.0 &  4 & S1 &   27.14 & $+3.42\pm .09$ & 0.84 & .. \\
                         & 11.800 & 13.0 & 9.400 & 1.0E8 & g & 66.2 \\
B2\ 1615+35  & 11.586 & 10.0 &   ..   &   ..   &   &   11.7 & 24.62 &  1.0 &  1 & C  &    1.03 & $+0.57\pm .15$ & 1.79 & .. \\ 
B2\ 1621+38  & 11.900 & 13.0 &   9.600   &   5.7E8   & g  &   39.5 & 23.97 &  1.0 & .. & .. &      .. & $+0.70\pm .08$ & 2.00 & 8.85\rlap{$^{\mathrm e}$} \\
B2\ 1626+39  & 11.750 & 13.0 &  9.350 &  2.9E8 & g &  268.0\rlap{$^{\mathrm c}$} & 24.81 &  1.0 & 2 & L: &      .. & $+1.93\pm .03$ & 1.93 & 
8.89\rlap{$^{\mathrm e}$} \\
B2\ 1637+29  & 12.050 & 13.0 &  8.920 &  2.9E8 & g &    5.7 & 24.75 &  1.0 &  0 & .. &      .. & $+1.0\pm .3$ & 1.81 & .. \\
B2\ 1638+32  & 11.750 &  5.0 &   ..   &   ..   &   &    2.3 & 25.19 &  1.0 &  1 & L: &    2.55 & .. & 1.70 & .. \\
B2\ 1643+27  & 12.001 & 13.0 &   ..   &   ..   &   &    8.4 & 24.41 &  1.5 &  1 & L: &    0.94 &    .. & 2.03 & .. \\
B2\ 1652+39  & 11.850 & 13.0 &  7.050 &  2.0   & p &   90.8\rlap{$^{\mathrm c}$} & 24.55 &  3.0 & .. & .. &      .. & $+4.05\pm .01$ & 1.26 & .. \\
                         & 11.850 & 13.0 & 7.900 & 6.0E6 & g & 97.9 \\
B2\ 1657+32  & 11.661 & 13.0 &   ..   &   ..   &   &    3.5 & 23.99 &  1.0 &  2 & L  &    2.40 & $+0.34\pm .08$ & 1.73 & ..  \\
B2\ 1658+32  & 11.700 & 10.0 &  8.500 &  2.9E8 & g &   14.5 & 24.24 &  1.0 &  1 & L: &    0.96 & $+0.57\pm .16$ & 1.50 & .. \\
B2\ 1658+30A & 11.330 & 13.0 & 9.000 &  2.9E8 & g &   52.7\rlap{$^{\mathrm c}$} & 24.20 &  1.5 & 2 & L &      .. & $+2.22\pm .06$ & 1.86 & 8.94 \\
B2\ 1726+31  & 11.786 &  5.0 &   ..   &   ..   &   &    3.9 & 26.29 &  2.0 & .. & .. &      .. & $+0.3\pm .6$ & 1.16 & 8.82 \\
3C\ 192      & 11.410 & 10.0 &  5.550 &  0.50  & p &    3.1 & 25.59 &  2.0 &  3 & S  &   15.32 & $+1.3\pm .3$ & 1.32 & 8.07\rlap{$^{\mathrm e}$} \\
                    & 11.500 & 13.0 & 9.000 & 2.9E8 & g & 2.0 \\
3C\ 219      & 11.975 & 13.0 &  6.400 &  0.60  & p &    6.2 & 26.86 &  2.0 &  3 & S  &   54.94 & $+1.7\pm .3$ & 0.93 & .. \\
                    & 12.000 & 13.0 & 8.700 & 5.7E7 & g & 9.9 \\
3C\ 223      & 11.270 &  5.0 &  6.240 &  0.50  & p &    7.8 & 26.25 &  2.0 &  3 & S  &   60.34 & $+2.69\pm .27$ & 0.02 & 8.15\rlap{$^{\mathrm e}$} \\
                    & 11.250 & 5.0 & 9.500 & 2.9E8 & g & 7.4 \\
3C\ 234      & 11.900 & 13.0 &  7.200 &  2.00  & p &   80.0\rlap{$^{\mathrm c}$} & 26.81 &  2.0 &  3 & S  &  235.86 & $+4.18\pm .07$ & 0.20 & .. \\
                    & 11.900 & 13.0 & 8.100 & 1.3E5 & g & 1060 \\
3C\ 264      & 11.620 & 13.0 &  8.650 &  2.9E8 & g &    7.4 & 24.73 &  1.0 &  1 & H: &    1.21 & $+1.03\pm .07$ & 1.68 & 8.67\rlap{$^{\mathrm e}$} \\
3C\ 272.1    & 10.900 & 13.0 &  9.500 &  1.0E9 & g &  294.0\rlap{$^{\mathrm c}$} & 23.54 &  1.0 & .. & .. & .. & $+0.79\pm .03$ & 1.94 & 
8.75\rlap{$^{\mathrm e}$} \\
3C\ 274      & 11.000 & 13.0 &  6.200 &  1.3E5 & g & 1380.0\rlap{$^{\mathrm c}$} & 25.08 &  1.0 & 2 & L &      .. & $+2.16\pm .01$ & 1.84 &
9.04\rlap{$^{\mathrm e}$} \\
3C\ 285      & 11.550 & 13.0 &  8.520 &  5.7E7 & g &    4.7 & 25.53 &  2.0 &  3 & S: &    6.59 & $+2.64\pm .10$ & 0.47 & 7.97\rlap{$^{\mathrm e}$} \\
3C\ 296      & 11.900 & 13.0 & 10.000 &  1.0E9 & g &   47.6\rlap{$^{\mathrm c}$} & 24.73 &  1.0 &  1 & L: &    0.91 & $+0.37\pm .07$ & 1.90 & 
8.75\rlap{$^{\mathrm e}$} \\
3C\ 303      & 11.400 &  5.0 &  6.800 &  1.30  & p &    3.9 & 26.14 &  2.0 &  4 & S1 &   50.24 & $+3.35\pm .09$ & 0.76 & .. \\
                    & 11.700 & 13.0 & 9.200 & 5.7E7 & g & 11.9 \\
3C\ 305      & 11.400 &  5.0 & 10.000 &  5.7E8 & g &   17.1  & 25.05 &  1.5 &  3 & S  &   17.88 & $+1.38\pm .10$ & 0.93 & 8.08\rlap{$^{\mathrm e}$} \\
3C\ 321      & 11.500 &  5.0 & 10.100 &  2.9E8 & g &   10.3 & 25.88 &  2.0 &  2 & .. &      .. & $+2.81\pm .08$ & 0.07 & .. \\
3C\ 326      & 11.320 &  5.0 &  6.400 &  1.3E5 & g &   23.7 & 26.37 &  2.0 &  2 & C  &   18.60 & $+0.97\pm .23$ & .. & .. \\
3C\ 346      & 11.800 &  5.0 &  6.300 &  0.30  & p &    4.5 & 26.37 &  3.0 &  3 & S &      .. & $+1.7\pm .5$ & 0.97 & .. \\
                    & 11.800 & 5.0 & 8.700 & 5.7E7 & g & 7.6 \\
\end{longtable}
\tablefoottext{a}{If the second component is a (young) galaxy model, then $C1=log(M/M_{\sun})$; if it is a power law, then
$C1=\log(L^{2000\AA}_{p.l.}/L_{\sun})$.}

\tablefoottext{b}{For a second galaxy component, $C2$ is the age (in years); for a power law, $C2$ is the spectral index $\alpha$ ($\lambda^{-\alpha}$).}

\tablefoottext{c}{$J$, $H$, $K$  systematically higher than the other bands, in many cases resulting in high $\chi^2$ values.}

\tablefoottext{d}{$J$, $H$, $K$  systematically lower than the other bands, resulting in high $\chi^2$ value.}

\tablefoottext{e}{Hyperleda.}
\tablefoottext{f}{Mezcua et al. (\cite{mezcua11}).}
\tablefoottext{g}{Snellen et al. (\cite{snellen03}).}
}

\end{document}